\documentclass[a4paper,11pt]{article}
\usepackage{amssymb}
\usepackage{jheppub} 

\usepackage[T1]{fontenc} 
\usepackage[dvipsnames]{xcolor}
\usepackage{dsfont}

\newcommand{\la}[1]{\label{#1}}
\newcommand{\eq}[1]{(\ref{#1})}

\newcommand*{\PO}{\text{P}}

\def\[{\left[}
\def\]{\right]}
\def\({\left(}
\def\){\right)}
\def\d{\partial}
\newcommand{\beq}{\begin{equation}}
\newcommand{\eeq}{\end{equation}}
\newcommand\beqa{\begin{eqnarray}}
\newcommand\eeqa{\end{eqnarray}}

\newcommand\baln{\begin{align}}
\newcommand\ealn{\end{align}}

\usepackage{amsmath}

\usepackage{amssymb}

\DeclareMathOperator{\tr}{tr}

\usepackage{graphicx}

\usepackage{subfig}

\title{\boldmath Open Fishchain in N=4 Supersymmetric Yang-Mills Theory}

\author[a,b]{Nikolay Gromov,}
\author[a]{Julius Julius,}
\author[a]{Nicol\`{o} Primi}

\affiliation[a]{Department of Mathematics, King's College London,
The Strand, London WC2R 2LS, UK}
\affiliation[b]{St.Petersburg INP, Gatchina, 188 300, St.Petersburg,
  Russia}

\emailAdd{nikolay.gromov$\bullet$kcl.ac.uk, julius.julius$\bullet$kcl.ac.uk, nicolo.primi$\bullet$kcl.ac.uk}

\abstract{
We consider a cusped Wilson line with $J$ insertions of  scalar fields in ${\cal N}=4$ SYM and prove that in a certain limit the Feynman graphs are integrable to all loop orders.
We identify the integrable system as a quantum fishchain with open boundary conditions. The existence of the boundary degrees of freedom results in the boundary reflection operator acting non-trivially on the physical space.
We derive the Baxter equation for Q-functions and provide the quantisation condition for the spectrum. 
This allows us to find the non-perturbative spectrum numerically.
}

\begin{document} 

\maketitle
\flushbottom

\section{Introduction}

Integrability in $D>2$ quantum field theories takes its roots in quantum chromodynamics where it was first observed that in the BFKL limit the evolution kernel admits integrability~\cite{Lipatov:1993yb,Faddeev:1994zg}.
Later on it was found in $\mathcal{N}=4$ Supersymmetric Yang-Mills Theory (SYM) in~\cite{Minahan:2002ve} in a different regime, where the one-loop mixing matrix of a single trace of scalars was identified with the Hamiltonian of a closed integrable $SO(6)$ Heisenberg spin-chain in the large-$N$ limit (see
\cite{Beisert:2010jr,Dorey:2019gkd,Gromov:2017blm}
for recent reviews). Even though this observation was further generalised to two loops and a number of tests was done at higher loop orders and also non-perturbatively, there is still no direct proof of integrability of ${\cal N}=4$ SYM. Even if there is very little doubt about the integrability of this theory, having a rigorous proof of it would give us new tools and may also allow us to go beyond the spectrum in applications of integrability to non-perturbative gauge theories\footnote{For long operators the integrability based {\it Hexagon} approach~\cite{Basso:2015he} works very well in some regimes.}.

Recently, the so-called {\it fishnet limit} of ${\cal N}=4$ SYM attracted much attention~\cite{Gurdogan:2015csr}. In its simplest version this is a limit where only two scalar fields remain coupled and have a Yukawa-type interaction.
This theory has much simpler Feynman diagrams in the planar limit and thus provides perfect playground to test various non-perturbative techniques including integrability.
In \cite{Zamolodchikov:1980mb,Chicherin:2012yn,Gurdogan:2015csr,Gromov:2017cja,Chicherin:2017frs,Grabner:2017pgm,Gromov:2018hut,Gromov:2019jfh} it was shown how the integrability emerges directly from the Feynman graphs and the connection with the integrability structure of ${\cal N}=4$ SYM such as Quantum Spectral Curve was established. This gives a number of clues of how the integrability realises itself in more complicated theories such as ${\cal N}=4$ SYM. The main drawback of the fishnet theory is that it is not a unitary CFT and it is not known whether the conformal symmetry persists beyond the planar limit.

In this paper we consider a Wilson loop with local operator insertions in
undeformed ${\cal N}=4$ SYM and then take the so-called ladders limit.
We will use the methods developed for the fishnet theories~\cite{Gromov:2017cja,Gromov:2019jfh,Cavaglia:2020hdb} to develop an integrability based description of these observables and obtain a solution for the spectrum. The solution takes the form of a Baxter finite difference equation supplemented with a particular quantisation condition.

\begin{figure}[ht]%
 \centering
{\includegraphics[]{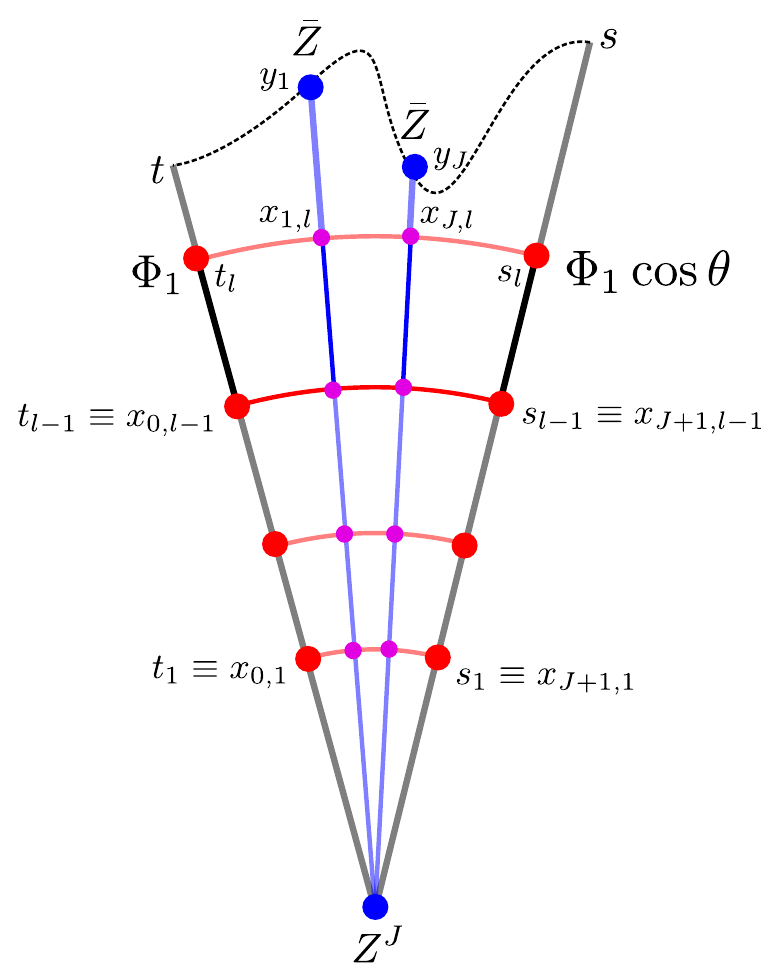}}
 \caption{The CFT wavefunction for $J=2$
 is a sum of the fishnet diagrams with any number of bridges. This figure shows one such diagram with $l=4$
 bridges. The graph building operator is highlighted.}%
 \label{fig:CFTWF}
 \end{figure}

The setup that we will consider in this paper is the following: we have a cusped Maldacena-Wilson line \cite{Maldacena:1998im,Erickson:2000af} with internal cusp angle $\varphi$, as in figure~\ref{fig:CFTWF}. The scalar $\Phi_1$ couples to the left ray and $\Phi_1 \cos\theta + \Phi_2 \sin\theta$ couples to the right ray.  $J$  scalars $Z =\frac{1}{\sqrt{2}}\left(\Phi_5 + i\,\Phi_6\right)$, orthogonal to the ones that couple to the rays, are inserted at the cusp. Here $\Phi_i$, $i = 1,\cdots, 6$ are the six scalars of $\mathcal{N}=4$ SYM theory. In addition, we can include in our description the excited states, in analogy with \cite{Cavaglia:2018lxi,Grabner:2020nis,excitedstates}, which corresponds to insertions of linear combinations of the scalars coupled to the lines.
This observable has a well defined anomalous dimension, which was studied in  \cite{Drukker:2012de,Correa:2012hh,Gromov:2015dfa} by means of TBA and then QSC methods. 
In this paper we will start from scratch, carrying out a first principles derivation in the so-called {\it ladders limit}, which we describe below.
The only insight we borrow from the QSC approach is a simple quantisation condition, which would require further efforts to derive from first principles. 

Another motivation behind the work we present in this paper is due to the recent study of structure constants by the separation of variables (SoV) method \cite{Cavaglia:2018lxi,Cavaglia:2020hdb,Gromov:2019wmz,Gromov:2020fwh}, where the explicit form of the Baxter equation was shown to be at the heart of the SoV approach.
An alternative to the approach of this paper would be to derive the Baxter equation 
from the QSC, which has a number of technical complications. Whereas at least numerically QSC~\cite{Gromov:2015wca} would give us a full control over this observable for a very wide range of parameters, extracting the closed system of equations in the ladders limit analytically has proven to be quite a challenging task (which was performed for $J=0$ case in~\cite{Gromov:2016rrp}).

The {\it ladders limit} which we study in this paper was first introduced
for the case $J=0$
in~\cite{Erickson:1999qv} and then used in~\cite{Correa:2012nk}.
This is obtained by taking the coupling $g\rightarrow 0$ and $\theta\rightarrow i\,\infty$, in such a way that $\hat g \equiv g \,\left(\frac{\exp\left(-i \theta/2\right)}{2}\right)^{\frac{1}{J+1}}$ is kept constant.
For the case $J=0$ it was noticed
in~\cite{Erickson:1999qv,Correa:2012nk} that only the ladder graphs contribute to the anomalous dimensions and the correlation functions. In this paper we show that for the general $J>0$ case the diagrams which survive are those of the fishnet type with a boundary corresponding to the two rays of the Wilson line~(see figure~\ref{fig:CFTWF}).
This drastic simplification in Feynman graphs allows us to construct the resummation procedure involving a {\it graph-building} operator. Such an operator was first constructed in the case of a Wilson-Maldacena loop with no scalar insertions in \cite{Cavaglia:2018lxi} and for the fishnet theory in \cite{Gromov:2017cja}. A new ingredient in the construction 
is the boundary of the fishnet, which itself carries a $1$D dynamics.
We had to adapt the boundary integrability methods for spin chains, developed by Sklyanin in~\cite{Sklyanin_1988}. In our case, however, the boundary reflection matrix itself is a nontrivial operator in the physical space.\footnote{Similar situation can be found e.g. in \cite{Gombor:2020kgu}.}

In this paper we first explore the integrability in the classical (strong coupling) limit $\hat g\to\infty$ and then quantise this system and develop the full quantum integrability.
Like in the case of the fishnet theory the integrability description comes from a chain of particles living on $AdS_5$ (with radius going to zero at strong coupling) also known as ``fish-chain''~\cite{Gromov:2019aku,Gromov:2019bsj,Gromov:2019jfh}. This time, however, we have two particles
with zero conformal weight at the ends of the chain whose motion is additionally restricted to the 
Wilson lines. 
In the quantum construction we
identify explicitly the conserved charges of the system in the commuting family of operators, and prove that the graph-building operator of the Feynman graph in the perturbation theory is one of them. In this way we obtain a full quantum non-perturbative description for the spectrum.

We also briefly discuss an interesting limit when the cusp becomes a straight line. In this limit the insertion becomes an operator in 1D defect CFT,
which has been intensively studied in recent years~\cite{Giombi:2017cqn,Mazac:2018mdx,Mazac:2018ycv,Dolan:2011dv,Mazac:2016qev,Beccaria:2017rbe,Kim:2017sju,Cooke:2017qgm,Grabner:2020nis,excitedstates}. Furthermore, in this limit, we can make a connection with the bootstrap methods of~\cite{Liendo:2016ymz,Liendo:2018ukf}.

This paper is organised as follows. In section~\ref{sec:derive}, we derive the graph building operator starting from Feynman diagrams in the ladders limit. In section~\ref{sec:dual}, we construct the Lagrangian of the open fishchain and solve the equations of motion. In section~\ref{sec:integrablity} we show that our model is classically integrable and construct the Lax and the boundary reflection operators. Then in section~\ref{sec:quantum} we show that integrability carries forward to the quantum case. In section~\ref{Baxtersection} we construct the Baxter equation for arbitrary number of scalar insertions of $Z$ at the cusp. In section~\ref{sec:numerics} we present the numerical non-perturbative spectrum for various  insertions and geometric parameters. Finally we conclude in section~\ref{sec:conclusion}.

\section{Ladders limit and graph building operator}\label{sec:derive}
In this section we will describe the Feynman diagrams contributing to the expectation value of the cusped Wilson line. We show that in the ladders limit it gets an iterative Dyson-Schwinger structure, governed by a graph building operator. The graph building operator is a hybrid between that obtained for $J=0$ in \cite{Erickson:1999qv,Correa:2012nk} for the cusp without insertion and the one for the fishnet theory \cite{Zamolodchikov:1980mb,Gurdogan:2015csr}.
In the rest of the paper we  develop the integrability based method to diagonalise this operator.

\subsection{Graph building operator}\label{sec:graph}

 \begin{figure}[ht]%
 \centering
 {\includegraphics[]{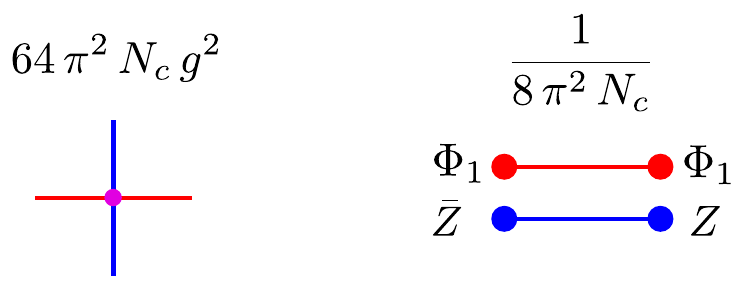}}
 \caption{We only need a subset of all Feynman diagrams.
 Above are the conventions for the scalar propagators
 and the interaction vertex between $\Phi_1$ and $Z=\frac{\Phi_5+i\Phi_6}{\sqrt 2}$.
 We use the standard definition $g=\frac{\sqrt\lambda}{4\pi}$ with the `t Hooft coupling $\lambda=g_{YM}^2N$.
 }%
 \label{fig:Feynmanrules}
\end{figure}

The Maldacena-Wilson Loop with $J$ insertions of scalar fields is given by:
\begin{multline}
\label{def:WL}
    W =\frac{1}{N} \tr\PO\exp \int_{0}^{\infty} dt\, 4\,\pi g \left(i\,A\cdot x'(t) + \,\Phi_{1}|x'(t)|\right) \\ \times  Z(0)^{J} \times \PO\exp \int_{0}^{\infty} ds\, 4\,\pi g \left[-i\,A\cdot x'(s) +\,(\Phi_{1}\cos\theta + \Phi_{2}\sin\theta)|x'(s)|\,\right]\; , 
\end{multline}
where $x'(t)\equiv \frac{\partial x(t)}{\partial t}$ and $x'(s)\equiv \frac{\partial x(s)}{\partial s}$. The two scalars that couple to the individual Wilson rays form an angle $\theta$ between each other. The expectation value of this quantity is divergent in both the IR and the UV, with the divergence controlled by the dimension $\Delta$\footnote{Strictly speaking for $J>0$ it is only divergent for large enough coupling as at tree level we have $\Delta=J$. For $J=0$ it is divergent for any $g>0$.}:
\begin{align}
    \langle W\rangle \sim\left(\frac{R^{IR}}{\epsilon^{UV}}\right)^{-\Delta} \;,
\end{align}
where $\Delta$ corresponds to the overall scaling dimension of $W$ and is also known as the cusp anomalous dimension in the $J=0$ case. The cusp anomalous dimension was studied intensively in perturbation theory and integrability \cite{Correa:2012at,Fiol:2012sg,Gromov:2012eu,Gromov:2013qga,Sizov:2013joa,Beccaria:2013lca,Dekel:2013dy,Janik:2012ws,Bajnok:2013sya,Drukker:2011za,Henn:2013wfa,Drukker:2006xg,Makeenko:2006ds,Drukker:2000rr,Erickson:2000af,Erickson:1999qv,Drukker:2012de,Correa:2018fgz,Correa:2012hh,Correa:2012nk,Alday:2007he,Bruser:2018jnc}. 
In this paper we will study a more general observable
which is the expectation value of $W$ with $J$
additional insertions (under the trace) of complex scalar fields $\bar{Z}=\frac{1}{\sqrt{2}}(\Phi_5-i\Phi_6)$ at points $y_i$ which lie outside of the countour,
and also truncate the upper limit in 
Wilson lines at some finite $t$ and $s$.
In analogy with~\cite{Gromov:2019jfh}
we call this object the CFT wavefunction $\psi(t,s,y_i)$.
At first sight this object is not gauge invariant, however in the  {\it ladders limit} it is well defined.
In fact, it can be made gauge invariant by closing the Wilson loop by introducing additional segments of non-supersymmetric Wilson lines running through the $\bar{Z}$ insertions, which will decouple in the ladders limit, as in fig.~\ref{fig:CFTWF}. As we will see, the role of the effective coupling in the ladders limit is played by
 \beq\la{ghat}
\hat g \equiv g \,\left(\frac{\exp\left(-i \theta/2\right)}{2}\right)^{\frac{1}{J+1}}\;,
\eeq
which we will assume finite while $g\to 0$ and $\theta\to i\,\infty$. 
In this limit we will get the following simplifications:
\begin{itemize}
    \item First of all, since we are taking the 't Hooft coupling to zero, the gluons and fermions decouple, and we are left with a theory of interacting scalar fields. Hence, we can drop out the gauge field $A$ from the definition \eqref{def:WL}.
    \item In a Feynman diagram expansion, only the contributions with the highest power of $\cos\theta$ will survive. For $J=0$, the only diagrams at $l$-loop order correspond to ladder diagrams, that is, diagrams that contain $l$ scalar propagators beginning on one of the Wilson lines and ending on the other \cite{Correa:2012nk}.
    \item For $J>0$, the scalars at the cusp $Z$ can only contract with the external insertions of $\bar{Z}$. This means that only one type of vertex allowed, i.e. the one in figure~\ref{fig:Feynmanrules}. This is analogous to what one finds in the simplest fishnet CFT. Consequently, only ``fishnet'' diagrams contribute.
\end{itemize}
Using these simplifications, we can define the CFT wavefunction in the ladders limit as:
\begin{multline}
\label{def: CFTWF}
    \psi(t,y_{1},\dots,y_{J},s) \equiv\frac{1}{N} \Bigg\langle \tr \prod_{j = 1}^{J}\bar{Z}(y_{j}) \times\PO\exp \int_{0}^{t} dt'\,(4 \pi g )\,|x'(t')|\Phi_1 \\ \times  Z(0)^{J} \times  \PO\exp\int_{0}^{s} ds'\,(4 \pi g )\,|x'(s')|\Phi_1\cos\theta\Bigg\rangle 
   \; .
\end{multline}
The CFT wavefunction is obtained expanding the path-ordered exponentials
\begin{equation}
\begin{split}
    &\psi(t,y_{1},\dots,y_{J},s) = \\
  & \sum_{ l =0}^\infty\psi_{ l }(t,y_{1},\dots,y_{J},s)=\sum_{ l =0}^{\infty}\tr \int_{0}^{t}dt_{ l }|x'(t_{ l })|\int_{0}^{t_{ l }}dt_{ l -1}|x'(t_{ l -1})|\,\dots\int_{0}^{t_{2}}dt_{1}|x'(t_{1})|\\&\int_{0}^{s}ds_{ l }|x'(s_{ l })|\int_{0}^{s_ l }ds_{l - 1}|x'(s_{ l -1})|\,\dots\int_{0}^{s_{2}}ds_{1}|x'(s_{1})|
    F_{ l }(y_j,t_i,s_i)\;.
\end{split}
\end{equation}
Here, 
$\psi_{ l }(y_j,t_i,s_i)$ represents the contribution of the $ l $-bridge fishnet Feynman graph, where a bridge is defined as a series of $J+1$ propagators connecting the left and right Wilson rays, as can be seen in figure~\ref{fig:CFTWF}. Note that the sum goes in the number of bridges
$ l $. The integrand in $s_i,\;t_i$ is given by:
\begin{multline}\la{Fal}
    F_{ l }(y_j,t_i,s_i)
    =\frac{1}{N} \left(\frac{1}{8\,\pi^2\,N}\right)^{ l (J + 1)+ J( l  + 1)}(64\,\pi^2\,N\,g^2)^{ l \,J}(16\,\pi^2\,g^2\cos\theta)^{ l } N^{( l +1)(J+1) + 1}\\
    \int  \left(\prod_{i=1}^J\prod_{j=1}^{ l } d^4x_{i,j}\right) \left(\prod_{r=0}^{J}\prod_{k=1}^{ l }\dfrac{1}{(x_{r+1,k}-x_{r,k})^2}\right)\left(\prod_{m=1}^J\prod_{n=0}^{ l }\dfrac{1}{(x_{m,n+1}-x_{m,n})^2} \right)\;.
\end{multline}

Here we have defined  $x_{k,0}\equiv y_0\equiv\left(e^{t_k},0,0,0\right)$, 
$x_{k,J+1}\equiv  y_{J+1}\equiv\left(e^{s_k} \cos\varphi, e^{s_k}\sin\varphi,0,0\right)$ $\forall k=1\dots l $, and
$ x_{ l +1,j}\equiv y_j ,\, x_{0,j}\equiv 0$ $\forall j=1\dots J$.
In the formula~\eq{Fal}, the second factor in the first line of the r.h.s contains the contribution from the propagators, the third the one from the vertices, the fourth comes from the expansion of the path-ordered exponentials, while the fifth represents the contribution from the closed index loops
of the planar diagram. In the second line we first integrate over all
positions of the vertices, the second term is a collection of all vertical propagators, while the third contains that of the horizontal ones (see~figure~\ref{fig:CFTWF} for the case of $J = 2$ and $ l  = 4$). Notice that at any loop order these graphs have the same order in  $N$, coherent with the fact that we are using a planar diagram expansion. 
Instead of computing this integral we notice that we can define it recursively in terms of the inverse of a graph building operator as we illustrate below. First, notice that $\square_{y_i}$ acts on scalar propagators as:
\begin{align}
    \square_{y_j}\frac{1}{(y_{j}-x_{j,l})^{2}} = -4\pi^2\,\delta(y_{j}-x_{j,l})\;.
\end{align} 
Moreover, acting with $\partial_{t}\partial_{s}$ on the contour of a Wilson line brings down the expansion of the path ordered exponential by one step, at the cost of a factor $|y'_0(t)||y'_{J+1}(s)|$.
Therefore acting on $\psi$ with a string of $\square_{y_j}$, followed by $\partial_{t}\partial_{s}$, we get back $\psi$ expanded to one less bridge, up to a multiplicative factor: 
\begin{align}
\label{graphbuilding1}
    \partial_{t}\partial_{s}\prod_{j = 1}^{J}\square_{y_j} \psi_{ l } = 
  {(-1)^{J}}(4\hat{g}^2)^{J+1} \frac{|y_{0}'||y_{J+1}'|}{\prod_{i = 0}^{J}(y_{i}-y_{i+1})^2}\psi_{ l -1}\;,
\end{align}
where we use the definition of $\hat g$ from \eq{ghat}. From this we can extract an operator annihilating the CFT wavefunction:
\begin{equation}\la{Bm1}
    (\hat B^{-1}-1)\psi=0\;\;,\;\;\hat{B}^{-1}\equiv\frac{(-1)^{J}}{(4\hat{g}^2)^{J+1}}\frac{\prod_{i = 0}^{J}(y_{i}-y_{i+1})^2}{|y_{0}'||y_{J+1}'|} \,\partial_{t}\partial_{s}\prod_{j = 1}^{J}\square_{y_j}\;.
\end{equation}
We refer to $\hat{B}^{-1}$ as an inverted graph-building operator.
The role of $\hat{B}^{-1}-1$ was realised in~\cite{Gromov:2019aku}
to be the analogue of the world-sheet Hamiltonian of a string theory. We will explore this further in the next section.

The Wilson loop with insertion $W$ is invariant under dilatations, which stretches the space around the origin (which we take to be the position of the cusp). Thus we can use the following dilatation operator, acting on the CFT wavefunction
\beq
\hat D =-i\(\d_t+\d_s+\sum_{i=1}^{J}( y_i \cdot\d_{y_i}+1)\)\;,
\eeq
to measure the dimension $\Delta$ of the initial cusped Wilson line. 
More precisely, the eigenvalue of $\hat D$ is $i\,\Delta$. This operator commutes with $\hat B$ as it is easy to see. Another operator which commutes with $\hat{B}$ is the generator of rotations in the orthogonal plane to the Wilson line:
\beq
\hat S=i \sum_{i=1}^{J}\(y_{i}^3\, \d_{y_i^4}-y_{i}^4\, \d_{y_i^3}\)\;.
\eeq
This operator measures the spin of $W$. For $Z^J$ scalar insertions $ S=0$, but one can also study more general insertions with derivatives in the orthogonal plane, corresponding to $S\neq 0$,
which are also described by our construction.

\begin{figure}[ht]%
 \centering
{\includegraphics[scale = 1]{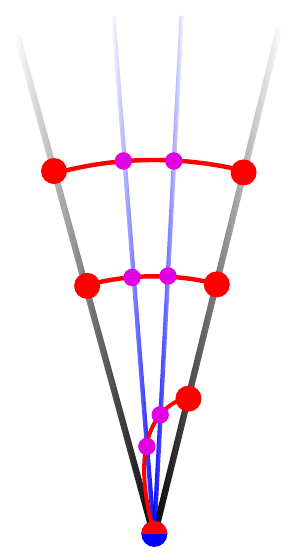}}
 \caption{An example of an ``excited state'' for $J=2$. Here, propagators from the extra insertion of $\Phi_1$ at the cusp contract with the Wilson line without crossing any other propagators of $\Phi_1$ (shown in red), as such diagrams would be subleading in the ladders limit.
 }%
 \label{fig:excited}
 \end{figure}

In analogy with the fishnet \cite{Gromov:2019bsj}
one should diagonalise both ${\hat S}$ and ${\hat D}$. After doing so, 
the equation $(\hat B^{-1}-1)\,\psi =0$ should restrict us to the discrete spectrum of eigenvalues of the dilatation operator, which would give us all the anomalous dimensions of the operators with given quantum numbers.
Indeed we will find that there are infinitely many (but a discrete set) of such $\psi$'s diagonalising  all the $3$ operators.
In analogy with~\cite{Cavaglia:2018lxi,Grabner:2020nis,excitedstates}
we expect 
each of them to correspond to a particular insertion of operators,
which could include derivatives and extra $\Phi_1,\Phi_2$ fields in addition to $Z^J$, whose number is fixed by the R-charge. These type of insertions at the cusp will not modify the iterative structure of the diagrams, instead just adding a finite number of propagators close to the cusp (cf.~figure~\ref{fig:excited}). Therefore,  all these states should be governed by the same equation~\eq{graphbuilding1}.

In the next sections we will explore how the integrability arises explicitly in this system. In particular, we will show first in the classical (strong coupling) limit and then in general that the operators $\hat{B},\hat D$ and $\hat S$
are part of a bigger commuting family of operators.

\section{Classical open fishchain}\label{sec:dual}

In this section, following~\cite{Gromov:2019aku}, we interpret the inverse of the graph-building operator as a Hamiltonian of a quantum system of particles. Then we take the quasi-classical strong coupling limit of the system, deriving the classical fishchain with specific open boundary conditions.
We analyse in detail the classical system and find some of the solutions of the equations of motion. 

\subsection{Strong coupling limit}  The starting point for the strong coupling $\hat g\to\infty$ analysis is equation \eqref{graphbuilding1}.
By re-writing \eq{Bm1} in terms of the conjugate momenta:
\begin{equation}
    \label{classical: momenta}
    p_i=-i\partial_{y_i},\qquad \pi_t=-i\partial_{t},\qquad \pi_s=-i\partial_{s},
\end{equation}
we obtain the Hamiltonian $\hat{H}$,
governing a system with $4J+2$ degrees of freedom, given by:
\begin{align}
    \label{eq:ham1}
    \hat{H} =\pi_{t}\,\pi_{s}\prod_{i = 1}^{J}p^2_{i} +(4\hat{g}^2)^{J+1} \dfrac{|y_{0}^{'}(t)||y_{J+1}^{'}(s)|}{\prod_{i = 0}^{J}(y_{i}-y_{i+1})^2}\,.
\end{align}
In this section we will be treating this Hamiltonian as the one of a classical system. In analogy with \cite{Gromov:2019aku}
we will see that the classical limit
 corresponds to the strong coupling $\hat g\to\infty$ limit of the original quantum system~\eq{Bm1}.
We will now demonstrate the classical integrability of this system and then describe its quantisation in section~\ref{sec:quantum}.

We remark that $y_i,\, i=1\dots J$ are 4D vectors with four bulk degrees of freedom for each one, while $y_0$ and $y_{J+1}$ are 4D vectors having one boundary degree of freedom each. Therefore, without loss of generality, we parametrise the latter as:
\beq\la{xb4D}
y_0(t)=\left(e^t,\,0,\,0,\,0\right)\,,\qquad y_{J+1}(s)=\left(e^s \cos{\varphi},\,e^s\sin{\varphi},\,0,\,0\right)\,.
\eeq

We will find it beneficial to embed the system in $6D$ space, which will allow to make the conformal symmetry of the system manifest, but will also result in a local action with nearest neighbour interaction. 

In the rest of this section we will deduce the classical equations of motion of this system, using the Lagrangian formalism.
First, by performing a Legendre transformation on \eq{eq:ham1}, we find the Lagrangian to be:
\begin{align}
    \label{eq: Lagrangian0}
    L=
    2^{\frac{-2J}{2J+1}}(2J+1)\left(\dot{t}\,\dot{s}\prod_{i=1}^J\dot{y}_i^2\right)^{\frac{1}{2J+1}}-(4\hat{g}^2)^{J+1}\,\dfrac{|y_{0}^{'}(t)||y_{J+1}^{'}(s)|}{\prod_{i = 0}^{J}(y_{i}-y_{i+1})^2}\,,
\end{align}
where $\dot f{\,}\equiv\frac{d}{d\tau} f$, with $\tau$ being a ``world-sheet'' time variable (conjugate to the Hamiltonian \eq{eq:ham1}).
The action $S=\int L \,d\tau$ is not invariant under  time reparametrisation symmetry $\tau\rightarrow f(\tau)$, which is needed to ensure $\hat H\psi=0$. In order to enforce this symmetry we introduce an auxiliary field $\gamma$ transforming as $\gamma\rightarrow \dfrac{\gamma}{\dot{f}}$
when $\tau\to f(\tau)$
which gives
\begin{align}
    \label{eq: Lagrangian1}
    L=2^{\frac{-2J}{2J+1}}(2J+1)\left(\frac{1}{\gamma}\dot{t}\,\dot{s}\prod_{i=1}^J\dot{y}_i^2\right)^{\frac{1}{2J+1}}-\gamma\,(4\hat{g}^2)^{J+1}\,\dfrac{|y_{0}^{'}(t)||y_{J+1}^{'}(s)|}{\prod_{i = 0}^{J}(y_{i}-y_{i+1})^2}\,.
\end{align}
This is now time-reparametrisation invariant.
We then eliminate the auxiliary field setting it to its extremum (by a suitable time reparametrisation).
We have to remember that 
$y_0$ (and $y_{J+1}$) is not itself a canonical coordinate, but depends on the world-sheet time through $t(\tau)$ (and $s(\tau)$ respectively).
Thus we can use 
$\dot{y}_0=y'_0 \dot{t}$ and similarly $\dot{y}_{J+1}={y'_{J+1}}{\dot{s}}$. After that we get:
\begin{align}
    L= (2J+2)(2i)^{\frac{1}{J+1}}\,\hat{g}\left[\dfrac{|\dot{y}_0||\dot{y}_{J+1}|\prod_{i=1}^J\dot{x}_i^2}{\prod_{i=0}^J|y_i-y_{i+1}|^2} \right]^{\frac{1}{2(J+1)}}\,.
\end{align}
We now embed the system in $6D$ Minkowksi spacetime, using lightcone coordinates in the Poincare' slice:
\begin{equation}
    y_i^{\mu}=\frac{X_i^{\mu}}{X_i^+},\qquad X_i^2=0,\qquad X_i^+=X_i^0+X_i^{-1}\,.
\end{equation}
Hence we get:
\begin{align}\label{eqn:NGLag}
L=(2J+2)(2{i})^{\frac{1}{J+1}}\,\hat{g}\left[\dfrac{|\dot{X}_0||\dot{X}_{J+1}|\prod_{i=1}^J\dot{X}_i^2}{\prod_{i=0}^J(-2X_i.X_{i+1})} \right]^{\frac{1}{2(J+1)}}\;.
\end{align}
Furthermore, we can disentangle this action to bring it to a Polyakov-like form, by introducing auxiliary fields $\alpha_i$, getting:
\begin{equation}
    \label{Lagrangianfinal}
    L=\xi\(\alpha_0\frac{|\dot{X}_0||\dot{X}_{J+1}|}2+\sum_{i=1}^J\left(\alpha_i\frac{\dot{X}^2_i}{2}+\eta_iX_i^2\right)+(J+1)\prod_{k=0}^J(-\alpha_k X_k.X_{k+1})^{-\frac{1}{J+1}}\)\,,
\end{equation}
where
\beq\la{xitog}
\boxed{\xi\equiv (2{i})^{\frac{1}{J+1}}\hat g}\;.
\eeq
In \eq{Lagrangianfinal}
 we also introduced the light-cone
constraint $X_i^2=0$ via the Lagrange multiplier $\eta_i$.
In order to get back the Nambu-Goto-like form~\eq{eqn:NGLag}, we have to extremise the fields $\alpha_i$ and plug these values back into \eqref{Lagrangianfinal}. It is possible to do this due to the new re-scaling symmetry of $X_i$.
More precisely, the Lagrangian \eqref{Lagrangianfinal} has $J+3$ gauge symmetries: time-dependent rescaling $X_i\to g_i(\tau) X_i,\;\alpha_i\to \alpha_i g^{-1/2}_i(\tau),\;\eta_i\to \eta_i g^{-1/2}_i(\tau),\,i=0\dots J+1$ and time reparametrisation $\tau\rightarrow f(\tau)$, under which fields transform as ${X}_i\rightarrow \frac{{X}_i}{\dot{f}},\,\alpha_i\rightarrow \dot{f}\alpha_i,\,\eta_i\rightarrow\frac{\eta_i}{\dot{f}}$. 
Instead of setting $\alpha_i$'s to their extreme values we can use
the symmetries to impose $\alpha_i=1,\,\forall i=0,\dots,J$. 
This would lead to the following constraints (the same way as one gets Virasoro constraints):
\beq
\la{dXX}\dot{X}_{k}^2  =\mathcal{L}\;,
\eeq
where
\beq
\la{def:calL}
\mathcal{L}\equiv 2\prod_{i = 0}^{J}(-X_{i}\cdot X_{i+1})^{-\frac{1}{J+1}}\,,
\eeq
with $k=1,\dots,J$ in the first equation.
Furthermore, from the equation of motion for $\alpha_0$ we get
$|\dot{X}_{0}| |\dot{X}_{J+1}| = \mathcal{L}
$: this still leaves us with the freedom to rescale $X_0\to h(\tau) X_0$ and simultaneously $X_{J+1}\to \frac{1}{h(\tau)}\, X_{J+1}$, which we can fix by imposing additionally
$|\dot{X}_{0}| =|\dot{X}_{J+1}| $. Hence, we can just extend the range of $k$ in \eq{dXX} to $k=0,\dots,J+1$. Finally,
to fix the remaining time-reparametrisation gauge freedom we can set:
\beq\la{con2}{\cal L}=1\,,
\eeq
 which is a convenient gauge to work with. We have imposed $J+3$ conditions, so all gauge degrees of freedom are fixed. The gauge fixed Lagrangian is then:
 \begin{equation}
    \label{Lagrangianfixed}
    L=\xi\(\frac{|\dot{X}_0||\dot{X}_{J+1}|}2+\sum_{i=1}^J
    \frac{\dot{X}^2_i}{2}
    +(J+1)\prod_{k=0}^J(-X_k.X_{k+1})^{-\frac{1}{J+1}}\)\,,
\end{equation}
Finally, by noticing that $
2|\dot X_0||\dot X_{J+1}|
=\dot X_0^2+\dot X_{J+1}^2-(|\dot X_0|-|\dot X_{J+1}|)^2$ we can 
replace $|\dot X_0||\dot X_{J+1}|\to
\frac{\dot X_0^2}{2}+\frac{\dot X_{J+1}^2}{2}
$
in  \eq{Lagrangianfixed},
modulo terms quadratic in constraints.
Similarly, defining $y=2\prod_{i=0}^J(-X_i\cdot X_{i+1})^{-\frac{1}{J+1}}\simeq 1$ on constraints, we have $y=e^{\log y}=1+\log y+{\cal O}(\log^2 y)$, which allows us to replace the potential term by $\sum_{k=0}^J\frac{1}{2}\log\frac{- X_{k}.X_{k+1}}{2e}$. Therefore we get the gauge fixed Lagrangian:
\beqa
    \label{Lagrangianfixed2}
    L&=&\xi\(
    \frac{\dot{X}^2_0}{4}+
    \sum_{i=1}^{J}
    \frac{\dot{X}^2_i}{2}
    +\frac{\dot{X}^2_{J+1}}{4}
    -\sum_{i=0}^J\frac{1}{2}\log\frac{- X_{i}.X_{i+1}}{2e}\)\;,\eeqa
    with constraints given by:
    \beqa\la{eqH0}
    &&\prod_{i=0}^J
    \frac{-X_i. X_{i+1}}{2}=1\;,\\
    &&X_i^2=0\;\;,\;\;\dot X_i^2=1\;\;,\;\;i=0,\dots,J+1\;.
\eeqa
Note that on the constraints we also have $L\simeq \xi (J+1) $.
In this form the Lagrangian is explicitly local and the interaction is only between the nearest neighbours.
It may appear a bit strange that the boundary particles has mass $1/2$ w.r.t. to the particles in the bulk, however,  we will see in the next section that in this way the equations of motion are more uniform. The reason is that the bulk particles has to be split in two and reflected, unlike those at the boundaries.
In \eq{Lagrangianfixed} the $6D$ variables $X_i$,
$i=1,\dots,J$ are independent canonical coordinates, constrained by \eq{dXX} and \eq{con2}.
At the same time the boundary particles $X_0$
and $X_{J+1}$ are encoded in terms of one variable each $t(\tau)$ and $s(
\tau)$, due to \eq{xb4D}. Explicitly:
\begin{align}
\label{def:X0}
    X_{i}(\tau) =r_i(\tau)\left(\cosh w_i(\tau), -\sinh w_i(\tau), \cos\phi_i,\sin\phi_i,0, 0\right)\;\;i=0,J+1,
\end{align}
where $\phi_0=0,\;\phi_{J+1}=\varphi$,
 $w_0(\tau)=t(\tau)$ and $w_{J+1}(\tau)=s(\tau)$.
On the constraint $\dot X_i^2=1$ we also have $r_i(t)=\frac{1}{\dot{w}_i(\tau)}$.
Apart from this,
the Lagrangian~\eq{Lagrangianfixed} is very similar to the one found in the classical limit of the fishnet graphs in \cite{Gromov:2019aku}.
It can be interpreted as the one of a discretised string with string-bits having a nearest neighbour interaction. 
However, due to the difference in the boundary DOFs it still remains to see whether the system is classically integrable, as it was in the original case  \cite{Gromov:2019aku}.

\subsection{Equations of motion}
We now compute the Euler-Lagrange equations starting from \eq{Lagrangianfixed2}. The $J$ equations for bulk variables are interpreted as equations of motions for $J$ bulk particles, while the equations for $X_0$ and $X_{J+1}$ are interpreted as equations of motion for two particles constrained on the two Wilson lines.
Since the Lagrangian \eqref{Lagrangianfixed2} has nearest neighbour interactions, only the first and last bulk particles in the spin chain will feel the presence of the particles on the Wilson lines. For example, for the bulk particle $j$ we have~\cite{Gromov:2019aku}:
\begin{equation}
    \label{EOMS: pcle1}
    \ddot{X}_j=2\eta_jX_j-\frac{1}{2}
    \left(\frac{X_{j+1}}{X_{j+1}.X_j}+\frac{X_{j-1}}{X_j.X_{j-1}} \right)\;,\quad\, j=1,\dots,J
\end{equation}
For the particles on the Wilson lines, we only have one physical degree of freedom for each, $t(\tau)$ and $s(\tau)$. The relative equations of motion are given by:
\beq
\label{eom0}
\frac{\ddot{t}}{\dot{t}^2}=
\frac{X_1.\partial_{t(\tau)}X_0}{X_1.X_0}\;\;,\;\;
\frac{\ddot{s}}{\dot{s}^2}=
\frac{X_{J}.\partial_{s(\tau)}X_{J+1}}{X_J.X_{J+1}}\;
.
\eeq
These two equations can be written in the form
\eq{EOMS: pcle1} by introducing
the reflected particles $X_{-1}$
and $X_{J+2}$ as the reflection of the particles $X_1$
and $X_{J}$ w.r.t. the ray parametrised by $t$ and $s$ respectively. More precisely we introduce the reflection matrix and rotation matrices:
\begin{align}
\label{def:Cmatrix}
    C^{M}_{\,\,N} = \left(\begin{array}{cccccc}
         1 & 0 & 0 & 0 & 0 & 0 \\
         0 & 1 & 0 & 0 & 0 & 0 \\
         0 & 0 & 1 & 0 & 0 & 0 \\
         0 & 0 & 0 & -1 & 0 & 0 \\
         0 & 0 & 0 & 0 & -1 & 0 \\
         0 & 0 & 0 & 0 & 0 & -1 
    \end{array}\right)_{MN}
\;\;,\;\;
    G^{\,M}_{\quad N}=\begin{pmatrix}
1 & 0 & 0 &0&0&0\\
0 & 1 & 0 &0&0&0\\
0&0&\cos{\varphi}&-\sin{\varphi}&0&0\\
0&0&\sin{\varphi}&\cos{\varphi}&0&0\\
0&0&0&0&1&0\\
0&0&0&0&0&1
\end{pmatrix}_{MN}\;.
\end{align}
Then we define the images of the particles $1$ and $J$ by the reflection about the ray parametrised by $t$ and $s$ respectively as $X_{-1}=C.X_1$
and $X_{J+2}=G .C .G^{-1}. X_{J}=
G^2 .C. X_{J}$.
With these definitions the equations \eq{eom0}
coincide with \eq{EOMS: pcle1} for $j=0$ and $j=J+1$ correspondingly.

\begin{figure}[ht]%
\centering
\includegraphics[scale = 1]{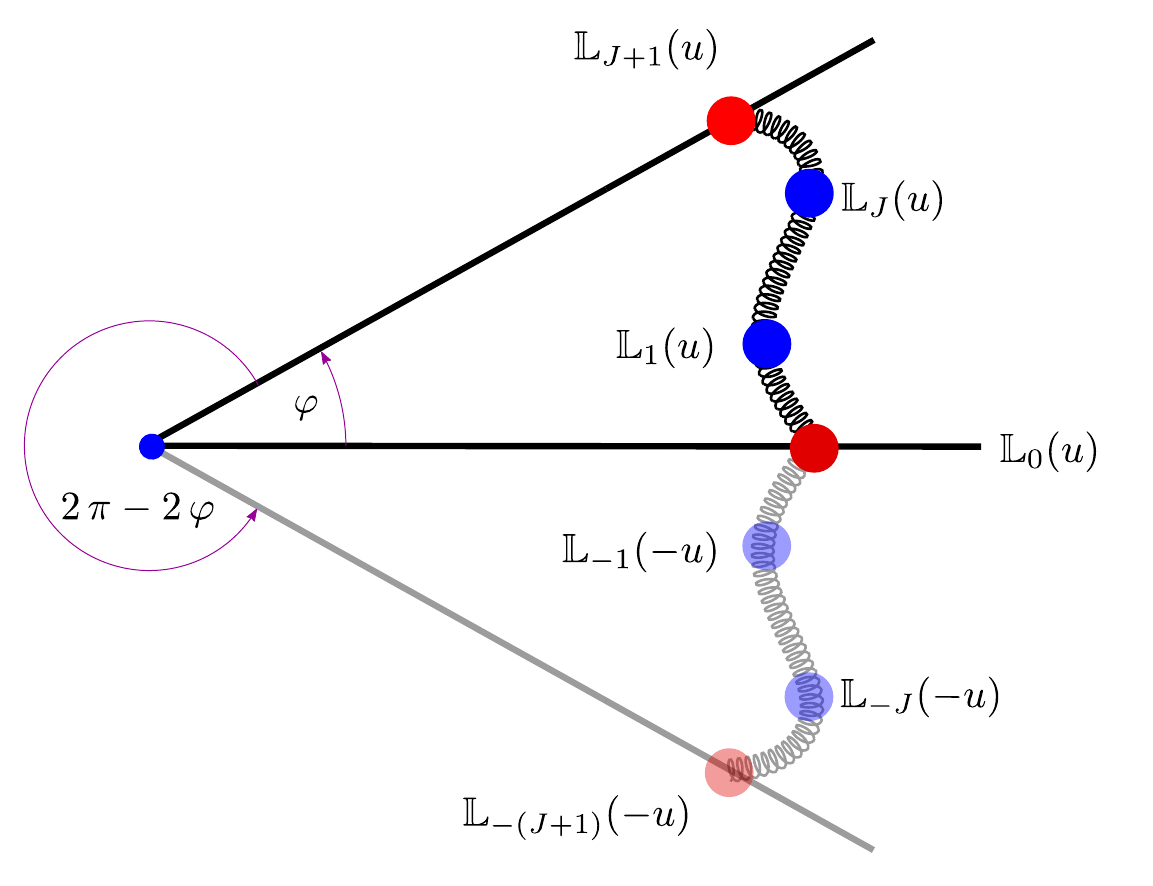}
\caption{Schematic representation of our method of images construction }%
\label{fig:doubletwist}%
\end{figure}

Thus, we conclude that at the level of the
classical equations of motion the open version of the fishchain we consider here is identical to the double-size closed fishchain of \cite{Gromov:2019aku}, with length $2J+2$  and quasi-periodic boundary condition
twisted by a $2\varphi$ rotation
(see figure~\ref{fig:doubletwist}).
However, there are some important differences in the Poisson structure and consequently the quantisation is different.

\subsection{Conserved charges} 
The presence of boundaries in the open fishchain breaks the $SO(1,5)$ symmetry that its closed counterpart enjoyed to the subgroup $SO(2)\times SO(1,1)$. Nevertheless it is useful to define 
\begin{align}
    q_{j}^{M N} \equiv {\dot X_j^{M}}X_j^{N} - {\dot X_j^{N}}X_j^{N} = 2 {\dot X}_{j}^{[M}X_j^{N]} \;,
\end{align}
for $j = 0, \dots J + 1$, 
which are the local $SO(1,5)$
generators for $j=1,\dots, J$.
We also define the total charge 
\begin{align}
    Q^{MN} 
    = \xi\,\left(\frac{q_{0}^{MN}}{2} +\sum_{i = 0}^{J} q_j^{MN}+\frac{q_{J + 1}^{MN}}{2}\right)\;.
\end{align}
As the $SO(1,5)$ symmetry is broken, only the components of $Q^{MN}$ corresponding to the unbroken symmetry subgroup will remain conserved in time. Thus we only have two Noether charges, corresponding to the $SO(2)$ angular momentum and to the scaling dimension:
\beq\la{clcha}
S=Q_{3,4}\;\;,\;\;D=Q_{-1,0}=i\,\Delta\;.
\eeq

\subsection{Example of solution of the classical equations of motion}
Now we proceed to the numerical solution of the system~\eqref{EOMS: pcle1}. To do so, we introduce the following parametrisation for the bulk particles, which is similar to the one used for the boundary particles~\eqref{def:X0}:
\begin{align}\label{eqn:Bulkparam}
    X_{a}(\tau) = \frac{1}{\sqrt{\dot{w}^2_a(\tau)+\dot{\phi}_a^2(\tau)}}
    \left(\begin{array}{c c c c c c}
        \cosh w_a(\tau), & -\sinh w_a(\tau), & \cos\phi_a(\tau), &   \sin\phi_a(\tau),  & 0, & 0
    \end{array}\right)\;.
\end{align}
where $a = 1,\dots,J$. 
This resolves the $X^2=0$ and $\dot X^2=1$ constraints. For the ansatz
\eq{eqn:Bulkparam} the particles are all in the same plane. 
The boundary particles are constrained to move on the Wilson rays, so their angular position is fixed
\begin{align}
    \phi_{0}(\tau) &= 0 \;, \\
    \phi_{J+1}(\tau) &=  \varphi \;.
\end{align}
We can imagine the simplest solution would be when these particles move along straight lines.
For that we need to compensate the interaction with neighbours, which could otherwise bend the trajectory, so we require
\beq
\phi_k(\tau)=\frac{k}{J+1}\varphi\;.
\eeq
To simplify our ansatz further we can assume that $w_k(\tau)=W(\tau)$. Plugging this ansatz into the EOMs \eq{EOMS: pcle1}
we obtain
\beq
w_{k}(\tau) =  
    \beta
    \,\tau\;.
\eeq
Finally, constraint \eq{eqH0} 
gives
\beq
\(\frac{\sin\(\frac{\varphi}{2J+2}\)}{\beta}\)^{2J+2}=1\,,
\eeq
which has $2J+2$ different solutions
\begin{align}
    \beta = e^{2\pi i \frac{n}{2J+2}} \sin\left(\frac{\varphi}{2\,J + 2}\right)\;\;,\;\;n=1\dots 2J+2\;.
\end{align} 
To get an interpretation of this, we also compute the anomalous dimension, using
\eq{clcha}
\beq\la{delta}
\Delta=-\frac{(J+1)\,i}{\beta}\xi
\;.
\eeq
We see that the ambiguous factor can be absorbed into $\xi$. In fact, the initial graph building operator only depends on $\xi^{2J+2}$, thus this type of ambiguity is expected.
In fact this is the same as in the case of the closed fishchain~\cite{Gromov:2019aku}, where the solutions were found to multiply in a similar way and were responsible for the different asymptotics of a $4$ point correlator.
We can check our classical solution by comparing with some known results for $J=0$ case. From \eq{delta} for $J=0$ we obtain:
\beq
\Delta = \pm\frac{2\hat g}{\sin\frac{\varphi}{2}}
\eeq
which agrees perfectly with the equation (E.6) in \cite{Cavaglia:2018lxi}.
We note that for $\hat g>0$ only the minus sign solution appears in the spectrum whereas the plus sign solution corresponds to large and negative $\hat g$.

\begin{figure}[ht]%
 \centering
 \subfloat[$J = 2$ --- Notice that the boundary particles run away to infinity in a finite amount of time, while the particles in the bulk have only moved a finite amount.]{\includegraphics[scale = 0.8]{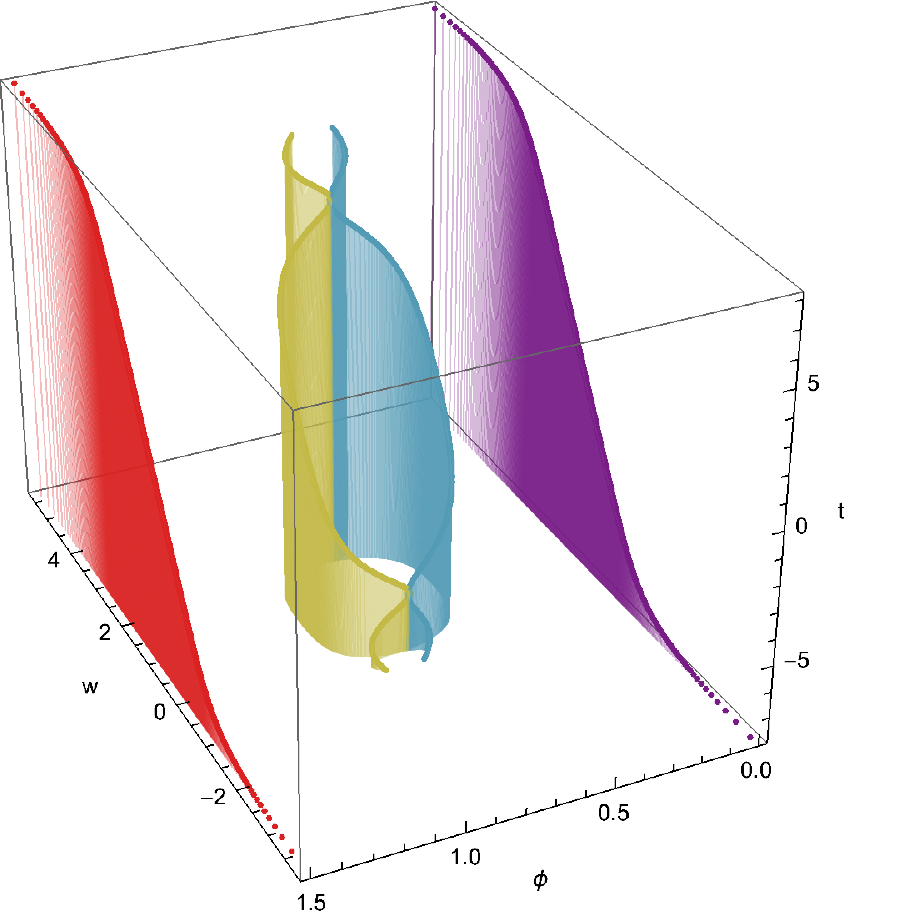}\label{fig:J2EoM}}%
 \hfill
 \subfloat[$J=3$ --- Notice that one of the bulk particles proceeds to make a complete loop.]{\includegraphics[scale = 0.8]{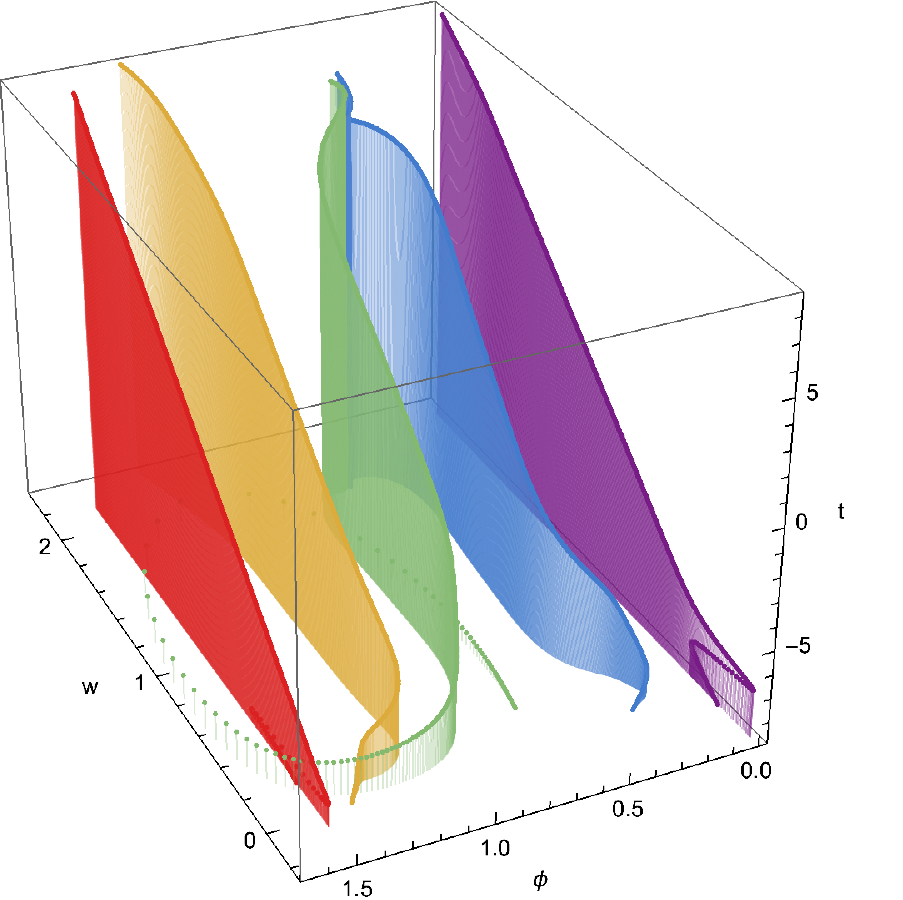}\label{fig:J4EoM}}%
 \hfill
\caption{Plot of the motion of particles obtained by a numerical solution of the classical equations of motion. In these solutions, motion is restricted to the plane of the Wilson loop. As expected, the boundary particles are confined to fixed rays whereas the bulk particles are free to move anywhere in the plane. }
\end{figure}

More general solutions can be obtained numerically.
We generated a couple of solutions, obtained by perturbing the analytic solution we just presented. These can be found in figure~\ref{fig:J2EoM} and figure~\ref{fig:J4EoM}.

\section{Classical integrability}\label{sec:integrablity}
In this section we prove that the dual model is integrable at the classical level
by studying its Poisson structure. 
We will construct the Lax matrices, corresponding to the particles in the bulk,
and the dynamical reflection matrix will represent the boundary particles. Using these building blocks we will construct a family of mutually Poisson-commuting objects.

The main purpose of this section is to establish the grounds for quantisation. For this reason we will only build here a subset of all commuting integrals of motion, as they will anyway appear in the quantum case in full generality.

\subsection{Poisson brackets}
In this section we discuss the Poisson structure following from the Lagrangian \eq{Lagrangianfixed2}. For the bulk DOFs 
the Poisson structure is identical to the closed fishchain case already studied in~\cite{Gromov:2019bsj}.
One can find the conjugate momenta and define the Poisson bracket
in the standard way. In particular, for the bulk particles the momentum conjugate to $X_{i,M}$ is
\beq
P_i^M=\xi \dot X_i^M\;,
\eeq
and the Poisson bracket is defined as $\{X_{i,M},P_j^N\}=\delta_{ij}
\delta_M^N$. Due to the constraints the Poisson brackets is ambiguous, and we could define a Dirac bracket. Alternatively, one can work with gauge invariant quantities. The gauge invariant combination of phase space coordinates for the bulk particles are the local symmetry generators
\beq\label{qdef}
q_i^{MN}=
\frac{1}{\xi}\(X_i^N P_i^M-X_i^M P_i^N\)
=X_i^N \dot X_i^M-X_i^M \dot X_i^N\;,
\eeq
which form the $SO(1,5)$ algebra under the Poisson bracket:
\begin{equation}\la{PBqq}
     \{q_k^{MN},q_k^{KL}\}=\frac{1}{\xi}\(-\eta^{MK}q_k^{NL}+\eta^{NK}q_k^{ML}+\eta^{ML}q_k^{NK}-\eta^{NL}q_k^{MK}\)\;,\;\;k=1,\dots,J\,.
\end{equation}
Similarly, one can proceed with the boundary degrees of freedom $t$ and $s$.
The canonically conjugate momenta to $t(\tau)$ and $s(\tau)$ are
\beq
\Pi_t=\frac{\xi}{2 t'(\tau)}\;\;,\;\;
\Pi_s=\frac{\xi}{2 s'(\tau)}\;.
\eeq
Even though the boundaries explicitly break down the $SO(1,5)$ symmetry,
it is still useful to define $q_0^{MN}$
and $q_{J+1}^{MN}$ in a similar to \eq{qdef} way
\beq
q_0^{NM}=\frac{2}{\xi}\(Y_0^{M}Y_0^{'N}-Y_0^{N}Y_0^{'M}\)\Pi_t\,,\;\;
q_{J+1}^{NM}=\frac{2}{\xi}\(Y_{J+1}^{M}Y_{J+1}^{'N}-Y_{J+1}^{N}Y_{J+1}^{'M}\)\Pi_s\,,
\eeq
where
\beq\la{Ys}
Y_0=\{\cosh t,-\sinh t,1,0,0,0\}\;\;,\;\;
Y_{J+1}=\{\cosh s,-\sinh s,\cos\varphi,\sin\varphi,0,0\}\;.
\eeq
Since the Wilson lines explicitly break conformal symmetry, the Poisson bracket of $q_0$  is modified to
\begin{equation}\la{PBq0q0}
     \{q_0^{MN},q_0^{KL}\}=\frac{1}{\xi}\(-\tilde\eta^{MK}q_0^{NL}+\tilde\eta^{NK}q_0^{ML}+\tilde\eta^{ML}q_0^{NK}-\tilde\eta^{NL}q_0^{MK}\)\,,\;\;
     \tilde\eta = \eta\(\mathds{1}+C\)\,, 
\end{equation}
where $C$ is the reflection matrix defined in \eq{def:Cmatrix}.
Similarly for the right boundary we get
\begin{equation}\la{PBqJ+1qJ+1}
     \{q_{J+1}^{MN},q_{J+1}^{KL}\}=\frac{1}{\xi}\(-\tilde\eta_\phi^{MK}q_{J+1}^{NL}+\tilde\eta_\phi^{NK}q_{J+1}^{ML}+\tilde\eta_\phi^{ML}q_{J+1}^{NK}-\tilde\eta_\phi^{NL}q_{J+1}^{MK}\)\,,\;\;
     \tilde\eta_\phi = \eta\(\mathds{1}+G.G.C\)\,, 
\end{equation}
where $G$ is the rotation matrix defined in \eq{def:Cmatrix}.

Finally, let us write the Hamiltonian $H$, corresponding to the lagrangian \eq{Lagrangianfixed2} in terms of the local symmetry generators $q_i$. 
For this we introduce
\beq\la{Hqdef}
H_q\equiv \frac{1}{2^{2J+2}}
\tr\(
q_0^2.q_1^2.\dots q_{J}^2.q_{J+1}^2.G.G.C.q_J^2.\dots.q_1^2.C
\)-1\,.
\eeq
Then we find that $H_q$ is proportional to the Hamiltonian $H$ up to a constant multiplier and up to second order in constraints
\beq
H_q=\exp\(\tfrac{4}{\xi}H\)-1\simeq \tfrac{4}{\xi}H+{\cal O}(H^2)\;.
\eeq
As our constraint implies $H=0$ we can equivalently use $\tfrac 4\xi H_q$ instead. The advantage of $H_q$ over $H$ is that it is written explicitly in terms of $q_i$'s. At it is explained in \cite{Gromov:2019bsj}
in the case of $q_i$'s there is no difference between the Poisson and Dirac brackets and so they are more convenient for the quantization.

Next, we will build the Lax representation based on the Poisson structure explained here and develop the integrability construction.

\subsection{Lax representation}
In this section we will build the classical integrals of motion. As at the classical level the equations of motion mostly coincide with the closed fishchain case of~\cite{Gromov:2019aku, Gromov:2019bsj}, we will review the construction from there adapting for our notations. 

In order to build the Lax representation, it is useful to introduce the local current
$j_i^{MN}=-2\dfrac{X_{i-1}^{[M}X_i^{N]}}{X_{i-1}.X_i}$, satisfying
\begin{equation}
    \dot{q}_i^{MN}=\{q_i^{MN},H\}=-\frac{1}{2}\left(j_{i+1}^{MN}-j_i^{MN}\right)\;\;,\;\;i=0,\dots,J+1\;.
\end{equation}
This allows us to define the Lax pair of matrices $\mathbb{L}_i$ and $\mathbb{V}_i$:
\begin{equation}
\label{L4def}
    \mathbb{L}_i=u\,\mathbb{I}_{4x4}+\frac{i}{2}q_i^{MN}\Sigma_{MN}\;\;,\;\;
        \mathbb{V}_i=-\frac{i}{4u}\,j^{MN}_i\Sigma_{MN}
    \,,
\end{equation}
where $\Sigma_{MN}$ are the $6D$ $\sigma$ matrices, giving a $4D$ representation of $SO(1,5)$. The explicit form we are using can be found in \cite{Gromov:2019bsj}.
One can show~\cite{Gromov:2019aku} 
from \eq{EOMS: pcle1}
that $\mathbb{L}_i$ and $\mathbb{V}_i$ satisfy the flatness condition
\begin{equation}
\label{eqn:flatness}
    \dot{\mathbb{L}}_i=\mathbb{L}_i.\mathbb{V}_{i+1}-\mathbb{V}_{i}.\mathbb{L}_i=\mathbb{V}_{i+1}.\mathbb{L}_i-\mathbb{L}_i.\mathbb{V}_{i}
    \,.
\end{equation}
From that it follows immediately that the combination
\begin{align}\la{Tfund}
    \mathbb{T}(u)=\tr \mathbb{L}_{-J}(u).\mathbb{L}_{-1}(u).\mathbb{L}_0(u).\mathbb{L}_{1}(u)\cdots\mathbb{L}_{J}(u). \mathbb{L}_{J+1}(u). G^{\bf 4}. G^{\bf 4} \;,
\end{align}
is conserved in time for any value of $u$, i.e. $\{ \mathbb{T}(u),H\}=0$. 
In the above expression we have defined
\beq
\mathbb{L}_{-i}(u)=C^{\bf 4}.\mathbb{L}^t_{i}(-u).C^{\bf 4}\;,
\eeq
and also the reflection matrix 
and the twist matrix in irrep. $\bf 4$:
\begin{equation}
\la{C4andG4}
{C^{\bf 4}}^{ab}=C^{\bf 4}_{ab}=\left(
\begin{array}{cccc}
 0 & 0 & 0 & 1 \\
 0 & 0 & 1 & 0 \\
 0 & 1 & 0 & 0 \\
 1 & 0 & 0 & 0 \\
\end{array}
\right)_{ab}\;\;,\;\;
    {G^{\bf 4}}^{\,a}_{\;\;\,\,f}=\begin{pmatrix}
e^{i \frac{\varphi}{2}} & 0 & 0 &0\\
0 & e^{-i \frac{\varphi}{2}} & 0 &0\\
0&0&e^{-i \frac{\varphi}{2}}&0\\
0&0&0&e^{i \frac{\varphi}{2}}
\end{pmatrix}^a_{\,\,f}\,.
\end{equation}

As each coefficient in the polynomial in $u$,
${\mathbb T}(u)$, is an integral of motion we get $\leq 2J+2$ integrals of motion. As we have $4J+2$ degrees of freedom in our system, some integrals of motion are still missing. The remaining ones are hiding in ${\mathbb T}^{\bf 6}(u)$ and ${\mathbb T}^{\bf \bar{4}}(u)$ -- the transfer matrices in vector and anti-fundamental representations. We will discuss them in detail in the quantum case in the next section. The classical construction for ${\mathbb T}^{\bf 6}(u)$ can be also deduced from \cite{Gromov:2019bsj}.
To prove that the model is classically integrable, we also need to show that the integrals of motion are in convolution,
i.e. that $\{{\mathbb T}(u),{\mathbb T}(v)\}=0$.
This in turn is less trivial and cannot be obtained from the closed fishchain case immediately, as the Poisson 
structure is modified due to the presence of the boundary particles.

In order to prove the convolution property of integrals of motion one can use \eq{PBqq} and \eqref{L4def} 
to show that, for $1\leq n,m \leq J$:
 \begin{equation}
 \label{PB1}
    \xi\; \{(\mathbb{L}_n)^{a}_{\,\,b}(u),(\mathbb{L}_m)^{c}_{\,\,d}(v)\}=\dfrac{(\mathbb{L}_n)^{a}_{\,\,d}(u)(\mathbb{L}_n)^{c}_{\,\,b}(v)-(\mathbb{L}_n)^{c}_{\,\,b}(u)(\mathbb{L}_n)^{a}_{\,\,d}(v)}{u-v}\,\delta_{nm}\,
 \end{equation}
 and analogously for $-J\leq n,m \leq -1$.
This relation can also be written by defining the dynamical $r$-matrix $r(u,v)=\frac{\mathbb{P}}{u-v}$, where $\mathbb{P}$ is the $16\times16$ permutation matrix:
 \begin{equation}\label{eqn:YBcl}
\xi\;     \{\mathbb{L}_n(u),\mathbb{L}_m(v)\}=[r(u,v),\mathbb{L}_n(u)\otimes\mathbb{L}_m(v)]\delta_{nm}\,.
 \end{equation}
For the boundary particles we have a different relation due to the modifications in the Poisson brackets \eq{PBq0q0} and \eq{PBqJ+1qJ+1}.
Denoting
\beq\la{Kdef}
\mathbb{K}(u)\equiv C. \mathbb{L}_0(u)
\;\;,\;\;
\bar{\mathbb{K}}(u)\equiv G^{-1}.\mathbb{L}_{J+1}(u).G.C\,,
\eeq
we have:
\begin{align}
\label{KPB}
\xi\;    \{\mathbb{K}_{ab}(u),\mathbb{K}_{cd}(v)\}= \frac{\mathbb{K}_{ad}(u)
   \mathbb{K}_{cb}(v)-\mathbb{K}_{ad}(v)
   \mathbb{K}_{cb}(u)}{u-v}-\frac{\mathbb{K}_{db}(u)
   \mathbb{K}_{ca}(v)-\mathbb{K}_{bd}(v)
   \mathbb{K}_{ac}(u)}{u+v}\;.
\end{align}
and the same for $\bar{\mathbb{K}}$.
In Appendix~\ref{Appendixcommutes}
we use these identities to show that
indeed
\begin{equation}
    \{\mathbb{T}(u),\mathbb{T}(v)\}=0\,.
\end{equation}
In the next section we show how this consideration extend to the quantum case.

\section{Quantum integrability}\label{sec:quantum}
In order to demonstrate the integrability at the quantum level we will have to embed the graph building operator into a family of commuting operators. To first approximation, one can replace the local $SO(1,5)$ generators $q_i$ by the operators $\hat q_i$. However, there are some quantum corrections to work out due to non-commutativity of various components of $\hat q_i^{MN}$, and this is what we will do in this section.

We will define the $\mathbb{\hat{L}}$ and $\mathbb{\hat{K}}$ operators as a quantum version of the classical ones. They will continue to be $4\times 4$ matrices, but now each component will become a differential operator.
Thus we will treat them as tensors acting on a tensor product of a $4D$ vector space and a functional space. We will refer to these spaces as auxiliary and physical spaces as usual.

Our strategy to fix the quantisation ambiguities is to make sure that $\mathbb{\hat{L}}$ and $\mathbb{\hat{K}}$
satisfy the Yang-Baxter equation and the Boundary Yang-Baxter equation correspondingly, which are generalisations of the classical Poisson brackets \eqref{PB1} and \eqref{KPB}.
After that we will build explicitly the quantum integrals of motion,  demonstrate that the graph building operator \eq{Bm1} is one of them
and prove that they mutually commute with each other. In order to get the complete set of integrals of motion, we will have to construct the transfer matrices in all antisymmetric representations of ${\mathfrak {sl}}(4)$.
We do this via the fusion procedure \cite{Lipan:1997bs}.

Next we will use integrability to compute the quantum spectrum. For that we will construct the Baxter equation~\cite{Baxter:1982zz} and use it to determine the Q-functions. 
Then imposing a suitable quantisation condition on the Q-functions we will demonstrate how to obtain the spectrum non-perturbatively.

\subsection{Quantisation of the integrability relations}
We need to build the quantum analogue of \eqref{eqn:YBcl}, which is the Yang-Baxter equation, and of \eqref{KPB}, which is given by the boundary Yang-Baxter equation. 

\paragraph{Quantum Lax matrix.}
The quantum version of the Lax matrix is\footnote{Note that our conventions differ by sign in comparison with \cite{Gromov:2019bsj}.} \begin{equation}
\label{def:Lax}
   \mathbb{\hat{L}}^{\;a}_{i\;\;b}(u)= u\, \delta^a_{b}+\frac{i}{2}\hat{q}^{MN}_i\Sigma^{\quad \;\;a}_{MN\;\;b}\;,
\end{equation}
where $\hat q_i^{MN}$ is the local generator of $SO(1,5)$, obtained as a quantisation
of \eq{qdef}, i.e. by replacing $P_j^K\to
\hat P_j^K =-i\d_{X_{j,K}}$:
\beq
\hat q_j^{MN}=-\frac{i}{\xi}\(X_j^N \frac{\partial}{\partial X_{j,M}}-
X_j^M \frac{\partial}{\partial X_{j,N}}\)\;.
\eeq
It satisfies the $SO(1,5)$ commutation relation:
\begin{equation}\la{CMqq}
\[\hat q_k^{MN},\hat q_k^{KL}\]=\frac{i}{\xi}\(-\eta^{MK}\hat q_k^{NL}+\eta^{NK}\hat q_k^{ML}+\eta^{ML}\hat q_k^{NK}-\eta^{NL}\hat q_k^{MK}\)\;\;,\;\;k=1,\dots,J\,.
\end{equation}
As explained in \cite{Gromov:2019bsj} $\hat q_i$ can be understood as acting on the functions of $4$-dimensional variables $y_i$ (e.g. CFT wave function) as if it was the corresponding conformal generator in $4D$. In other words one can use the following map between the functions 
of $4D$ coordinates $y_m$ and functions of $6D$ coordinates $X^M$
\beq\la{f6f4}
f(y_1,\dots,y_m)\to \frac{1}{X^{-1}+X^{0}}
f\(\frac{X^1}{X^{-1}+X^{0}},\dots,\frac{X^4}{X^{-1}+X^{0}}\)
\eeq
as $q_i$ preserves the $6$ interval  $X^M X_M$ we can set it to zero consistently. Note the action on the $4D$ is only well defined for observables build out of $q_i$'s. In particular $\hat P_j$ and $\hat X_j$
themselves are operators living in $AdS_5$~\cite{Gromov:2019bsj}.

The Yang-Baxter (YB) equation
can also be obtained by replacing $\{\cdot,\cdot\}$ by $\frac{1}{i}[\cdot,\cdot]$
in \eq{PB1}. One should, however, pay attention to the order of terms in the r.h.s. of \eq{PB1}
as at the quantum level the order does matter.
So the correct generalisation of \eq{PB1}, following from \eq{def:Lax} is:
\beq\la{YB1}
\hat{\mathbb L}^{b}_{\; e}(u)\hat{\mathbb L}^{d}_{\; f}(v)R_{ac}^{ef}(v-u) = R^{bd}_{f e}(v - u)\hat{\mathbb L}^{e}_{\;c}(v)\hat{\mathbb L}^{f}_{\;a}(u)\;,
\eeq
where we introduced the $R$ matrix, a quantum version of the $r$ matrix seen in~\eq{eqn:YBcl}, which acts on two copies of the $4D$ auxiliary space and is defined as:
\begin{align}
    \label{eqn:Rdef}
    R^{b\,c}_{a\,d}(u) = \mathbb{I}_{(\text{aux}\times\text{aux})} + \frac{i}{\xi\, u} \mathbb{P}\;=\delta^b_a\, \delta^c_d+\frac{i}{\xi\, u}\delta^b_d\,\delta^c_a\;.
\end{align}
Here $\mathbb{P}$ is the permutation operator, acting on vectors in the direct product of two spaces by interchanging them.

It will be also convenient to introduce
the Lax operator for the reflected particles:
\begin{equation}
\label{def:Laxbar}
   \hat{\bar{\mathbb{L}}}^{\;\;\,\,a}_{i\,b}(u)=- \mathbb{\hat{L}}^{\;a}_{i\;\;b}(u)\;,
\end{equation}
and the corresponding $\bar R$-matrix:
\begin{align}
\label{eqn:Rbardef}
    \bar{R}^{b\,c}_{a\,d}(u)={R}^{b\,c}_{a\,d}(-u)\;.
\end{align}

\begin{figure}[h]
    \centering
    \includegraphics[scale=0.5]{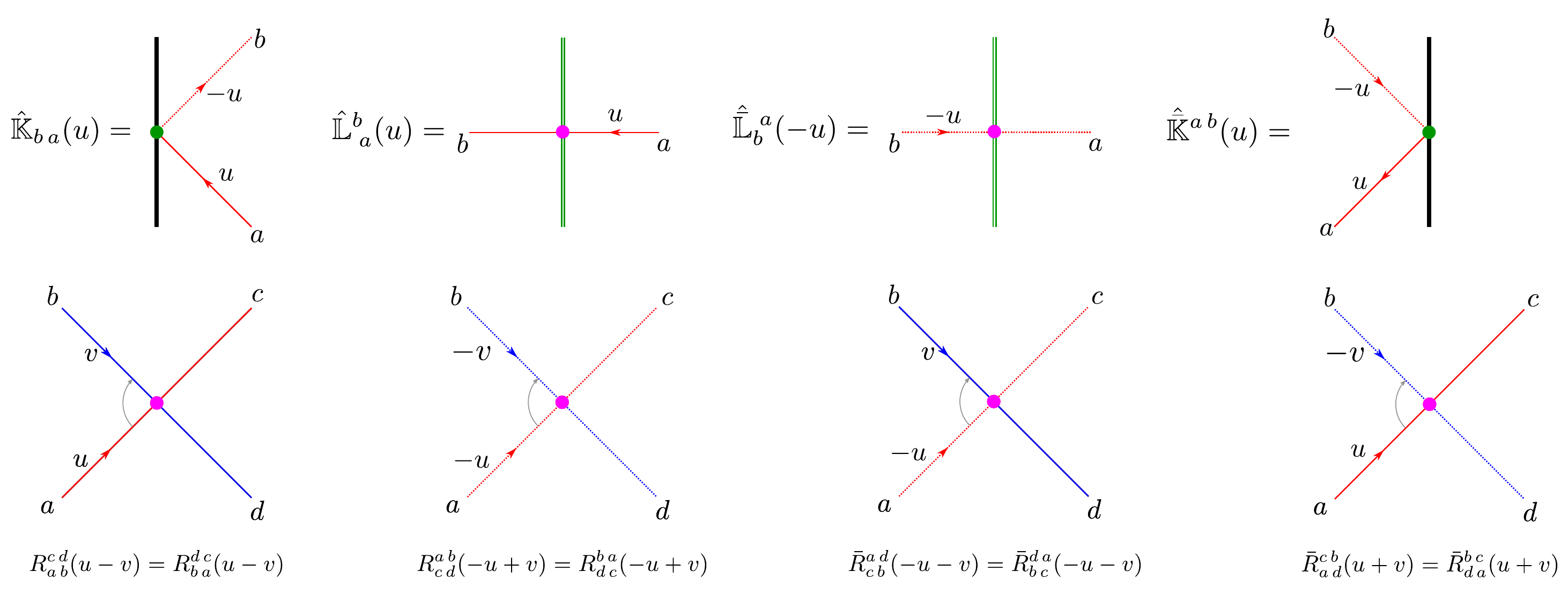}
    \caption{Diagrammatic rules}
    \label{fig:alldiags}
\end{figure}
\paragraph{Diagrammatic notation.}\la{diagrammatic}
In what follows it is extremely convenient to use the following notations: we denote the physical space by a double green line, the boundary spaces by thick black lines;
the auxiliary space will be denoted by a solid line,
and for the reflected auxiliary space we use a dotted line;
the auxiliary space is equipped with a direction and a spectral parameter; then various tensors correspond to intersection vertices following the rules depicted in figure~\ref{fig:alldiags}.

For example, the YB equation~\eq{YB1}  can be easily expressed using this notation in the following way:\\
\begin{center}
\includegraphics[scale = 0.5]{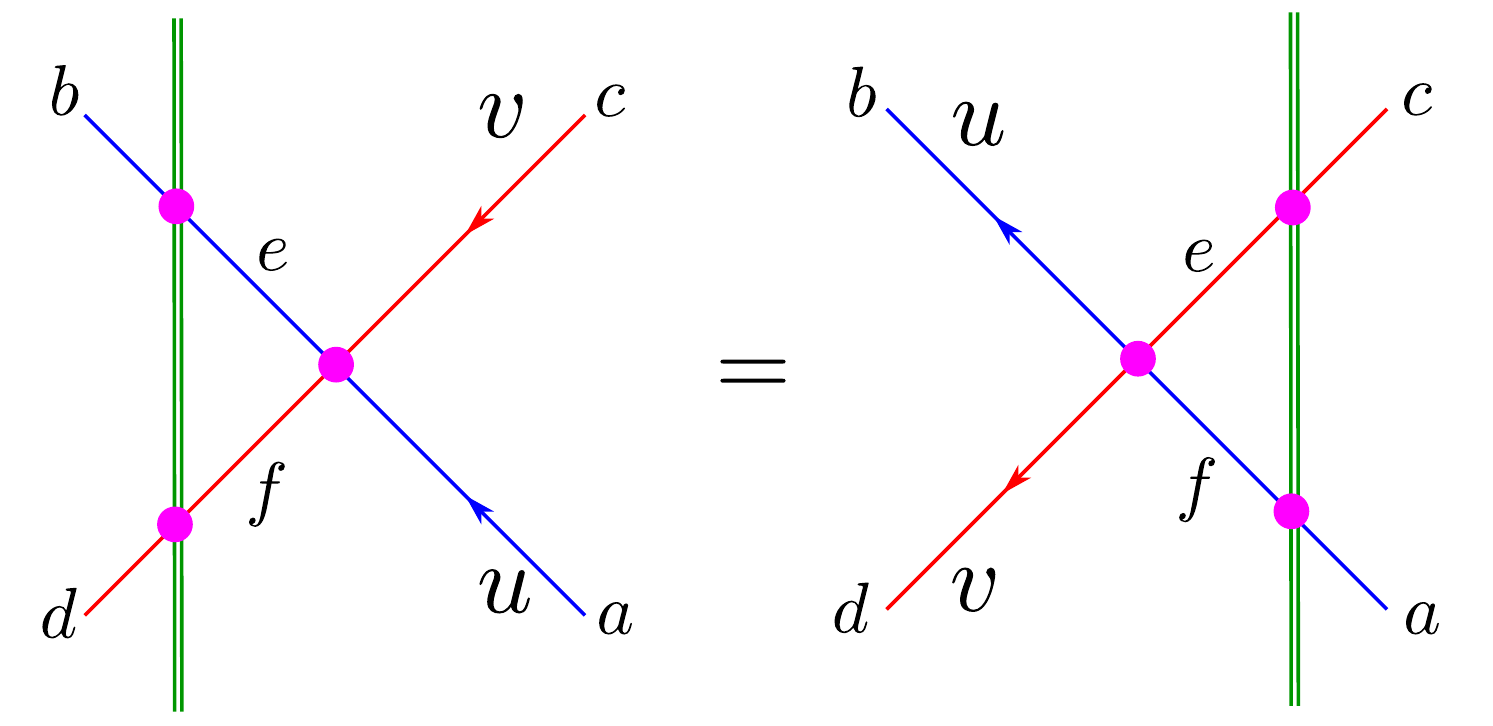}\label{fig:RLL}
\end{center}
In addition we have two other YB equations involving reflected auxiliary space (see figure~\ref{fig:RLLM}).
\begin{figure}[h]
\centering
\includegraphics[scale = 0.5]{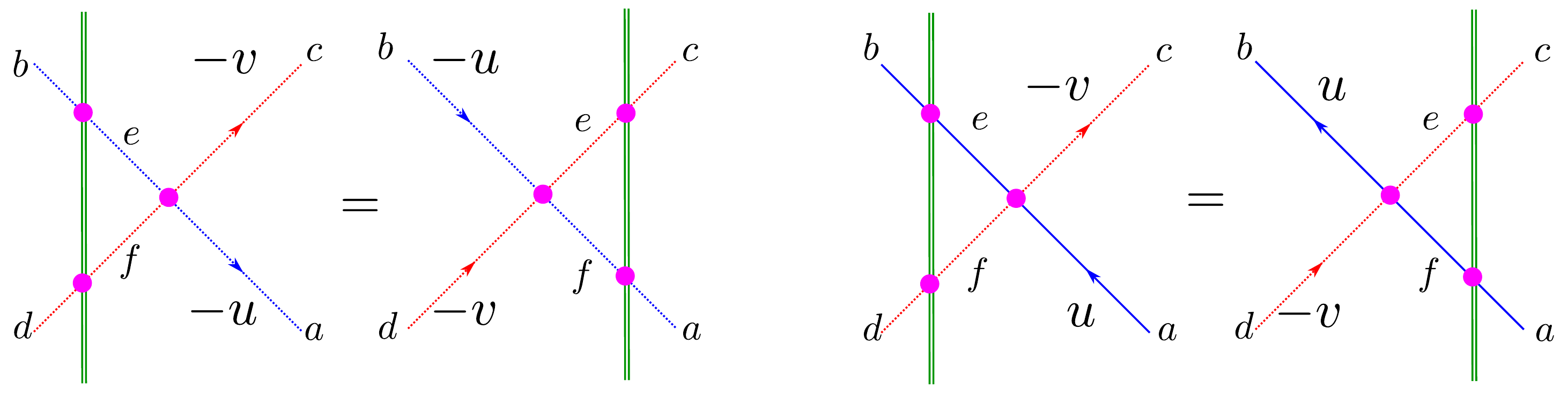}
\caption{Additional Yang-Baxter equations, which follow from \eq{YB1}, but look a slightly differently diagrammatically.}
\la{fig:RLLM}
\end{figure}

\subsection{Boundary reflection operator}
In the classical case at the boundary 
we found that $q_0$ and $q_{J+1}$ satisfied the modified Poisson brackets \eq{PBq0q0}.
This in turn results in a different relation~\eq{KPB} for the boundary Lax-type operator
denoted by ${\mathbb K}$.
In order for integrability to persist at the quantum level \eq{KPB} should become the boundary Yang-Baxter equation (BYBE)~\cite{Sklyanin_1988}.

The quantum version of $q_0$ and $q_{J+1}$
are again obtained by replacing $\Pi_t\to -i\partial_t$
and $\Pi_s\to -i\partial_s$,
and
read:
\begin{align}
\label{eqn:defq0}
\hat{q}_{0}^{NM}\equiv -i\frac{2}{\xi}(Y_0^M\dot Y_0^N-Y_0^N\dot Y_0^M)\partial_t\;\;,\;\;
\hat{q}_{J+1}^{NM}\equiv -i\frac{2}{\xi}(Y_{J+1}^M\dot Y_{J+1}^N-Y_{J+1}^N\dot Y_{J+1}^M)\partial_s\;,
\end{align}
where $Y's$ are explicit functions of $s$ and $t$, parameterising the Wilson rays defined in \eq{Ys}. Following the classical case, we also introduce:
\begin{align}
    \mathbb{\hat{L}}^{\;\;\,a}_{0\;\;\;b}(u)= u\, \delta^a_{b}+\frac{i}{2}\hat{q}^{MN}_0\Sigma^{\quad \;\;a}_{MN\;\;b}\;\;,\;\;
    \mathbb{\hat{L}}^{\;\;\quad a}_{J+1\;\;b}(u)= u\, \delta^a_{b}+\frac{i}{2}\hat{q}^{MN}_{J+1}\Sigma^{\quad \;\;a}_{MN\;\;b}\;.
\end{align}
Next we need to identify the quantisation of $\mathbb K$ \eq{Kdef}, such that 
\eq{KPB} becomes the BYBE, which 
for the left boundary is:
\begin{align}
\label{BYBEleft}
   \mathbb{\hat{K}}_{d_{2}c_{2}}(v)\bar{R}_{a_{2}d_{1}}^{c_{2}c_{1}}(u+v)\mathbb{\hat{K}}_{c_{1}a_{1}}(u)R_{b_{1}b_{2}}^{d_{1}d_{2}}(v-u)=R_{a_{2}a_{1}}^{c_{2}c_{1}}(v-u)
   \mathbb{\hat{K}}_{b_{1}d_{1}}(u)
   \bar{R}_{c_{1}b_{2}}^{d_{1}d_{2}}(u+v)
   \mathbb{\hat{K}}_{d_{2}c_{2}}(v)
   \;.
\end{align}
Diagrammatically this equation becomes
\begin{center}
\raisebox{-0.5\height}{
    \includegraphics[scale = 0.6]{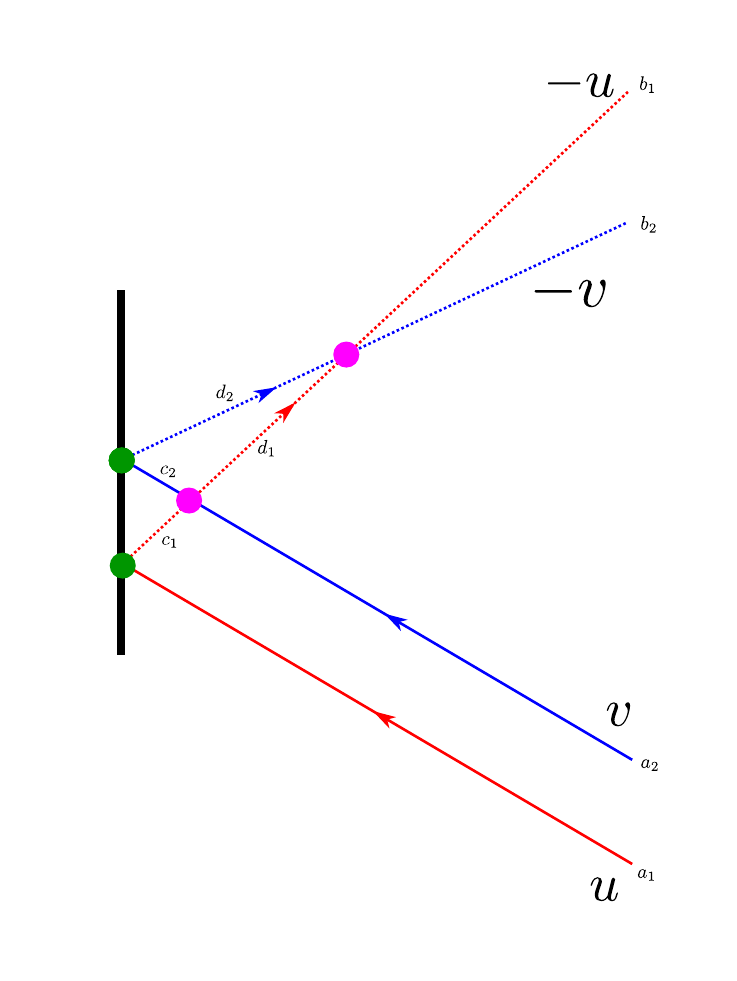}
}
=
\raisebox{-0.5\height}{
    \includegraphics[scale = 0.6]{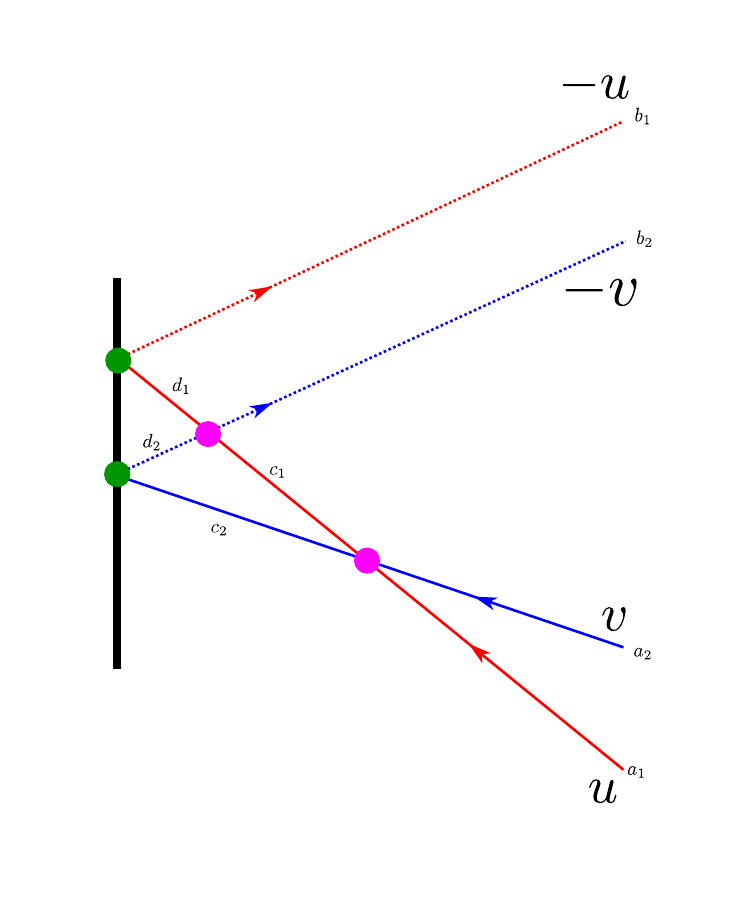}
}
\end{center}
We find that at the quantum level
there is a quantum correction to the spectral parameter, invisible in the classical $\xi\to\infty$ limit.
Namely the equation \eq{BYBEleft} is solved by
\beqa
\label{def:K}
\hat{\mathbb{K}}(u) = C.\hat{\mathbb L}_0(u-\tfrac{i}{2\xi})\;,
\eeqa
where $C$ is the same reflection matrix as the classical case \eqref{C4andG4}.
Similarly for the right boundary 
we have the following BYBE:
\begin{align}\label{BYBEright}
    R^{b_{1}b_{2}}_{c_{1}c_{2}}(u-v)\hat{\bar{\mathbb{K}}}^{d_{2}c_{2}}(v)\bar{R}_{d_{2}d_{1}}^{a_{2}c_{1}}(-u-v)\hat{\bar{\mathbb{K}}}^{a_{1}d_{1}}(u)  = \hat{\bar{\mathbb{K}}}^{c_{1}b_{1}}(u)\bar{R}_{c_{1}c_{2}}^{d_{1}b_{2}}(-u-v)\hat{\bar{\mathbb{K}}}^{d_{2}c_{2}}(v)R_{d_{1}d_{2}}^{a_{1}a_{2}}(u-v)\;,
\end{align}
which can be expressed diagrammatically as follows
\begin{center}
\raisebox{-0.5\height}{
    \includegraphics[scale = 0.6]{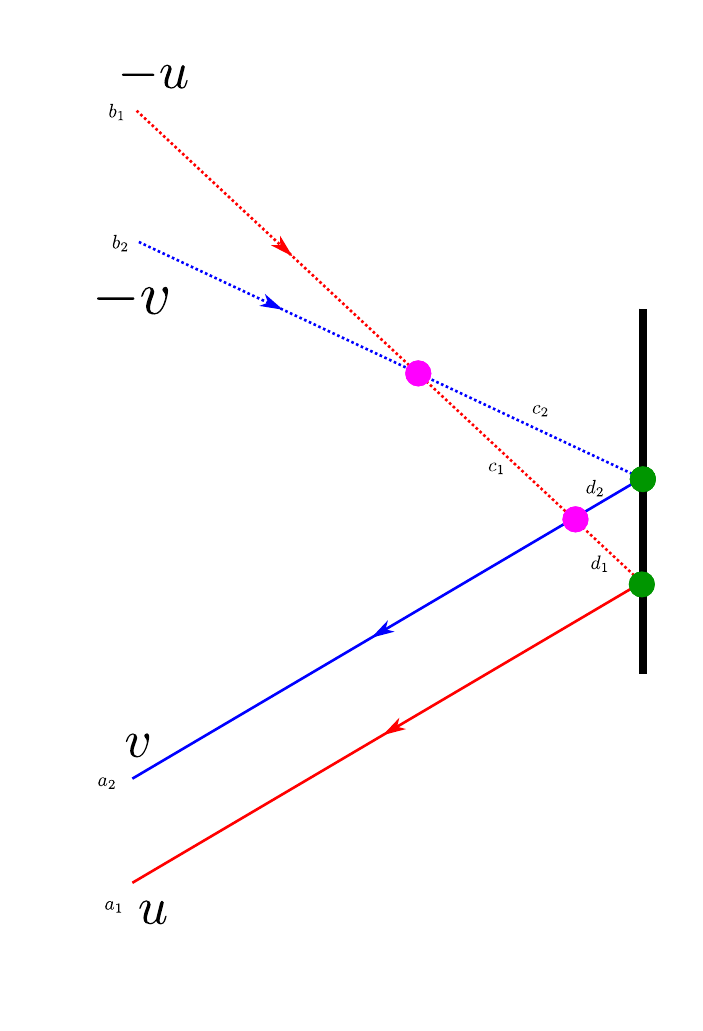}
}
=
\raisebox{-0.5\height}{
    \includegraphics[scale = 0.6]{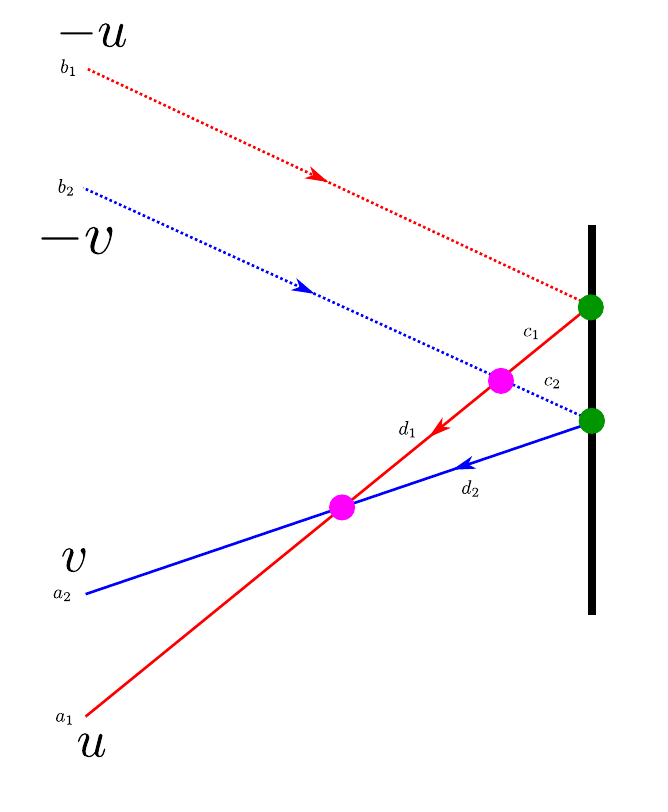}
}
\end{center}
One can easily verify that the solution to this equation has the following form:
\beqa
\label{def:Kb}
\hat{\bar{\mathbb{K}}}(u) =  G^{-1}.\mathbb{L}_{J+1}(u+\tfrac{i}{2\xi}).G.C\;,
\eeqa
where $G$ is the twist matrix defined in \eqref{C4andG4}. This expression is again identical to the classical expression up to a quantum correction in the spectral parameter.

In the rest of this section we will use the building blocks
${\mathbb K}$, ${\mathbb L}$ and $R$ to build a complete system of 
conserved, mutually commuting operators and also show that the inverse graph-building operator is  part of this family.

\subsection{Normalisation of the $R$-matrix}\la{Rmatrixnorm}
The $R$-matrix itself is defined up to an arbitrary scalar factor, which does not affect any of the previous relations.
However, in the next sections we will be using the fusion procedure for the boundary reflection matrix which is sensitive to the normalisation.

In order to fix the normalisation one can think about
the $R$-matrix as an S-matrix and impose unitarity. 
Therefore we denote the normalised R-matrix by $S$:
\begin{equation}
\label{eqn: Smatrix}
    S(u)=a(u) R(u)\;\;,\;\;    \bar S(u)=a(-u) \bar R(u)\;,
\end{equation}
where $a(u)$ is the normalisation factor which we fix by the unitarity condition
\beq
\label{unitarity}
S(u)S(-u)=I\;,
\eeq
which diagrammatically an be expressed as follows:
\begin{center}
\raisebox{-0.5\height}{\includegraphics[scale = 0.35]{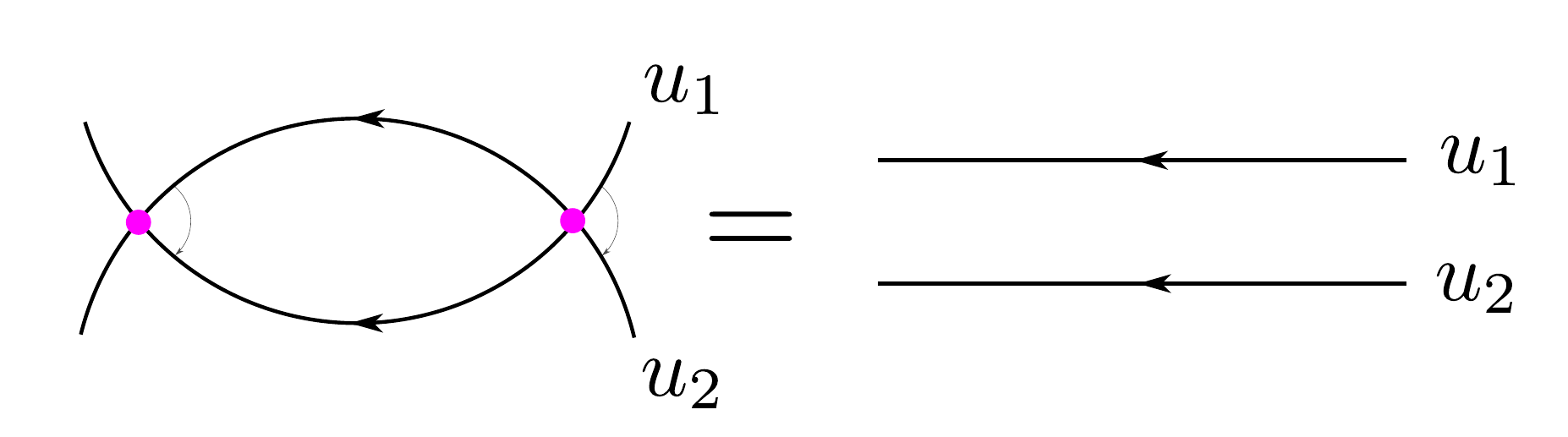}\label{fig:uni}}\;\;\;\;
\raisebox{-0.5\height}{\includegraphics[scale = 0.35]{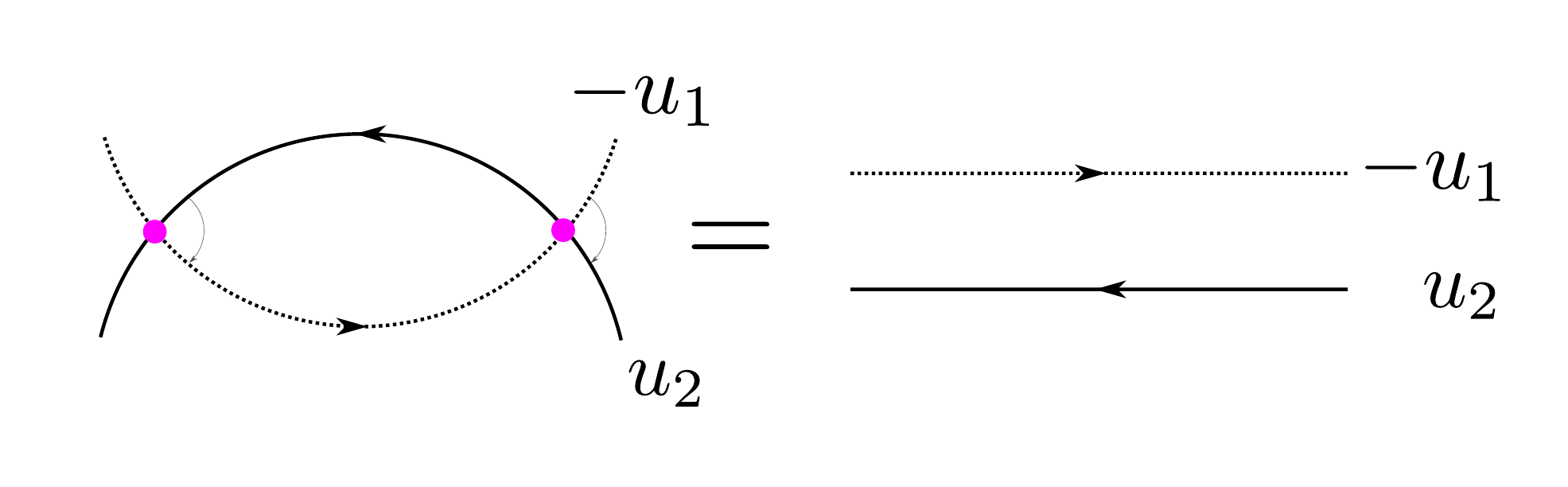}\label{fig:unibar}}
\end{center}
and takes the following form when written explicitly with all indices:
\begin{equation}
    \label{eqn:unitarity}
    S^{a\,\,d}_{\,\,b\,\,e}(u)S^{b\,\,e}_{\,\,c\,\,f}(-u)=\bar{S}^{a\,\,d}_{\,\,b\,\,e}(-u)\bar{S}^{b\,\,e}_{\,\,c\,\,f}(u)=\delta^a_c\delta^d_f\,.
\end{equation}
To satisfy this, we must have that:
\begin{equation}
\label{def:alpha}
    A(u)\equiv a(u)a(-u)=\frac{u^2\,\xi^2}{1-u^2\xi^2}\;.
\end{equation}
Below we will only need the combination of the scalar factors $A(u)$, so we do not need to decode individual $a(u)$ by imposing additional analyticity conditions.

\subsection{Transfer matrix}
In this section we will build a family of mutually commuting operators
out of the building blocks discussed above. First we will build the transfer matrix in fundamental representation, following the discussion in the classical case. As we know already, it cannot encode all the integrals of motion. In order to complete our system of IMs
we will also have to build the transfer matrix in vector and anti-fundamental representations. For that we will follow the fusion procedure.

\begin{figure}[ht]
    \centering
    \includegraphics[scale=0.5]{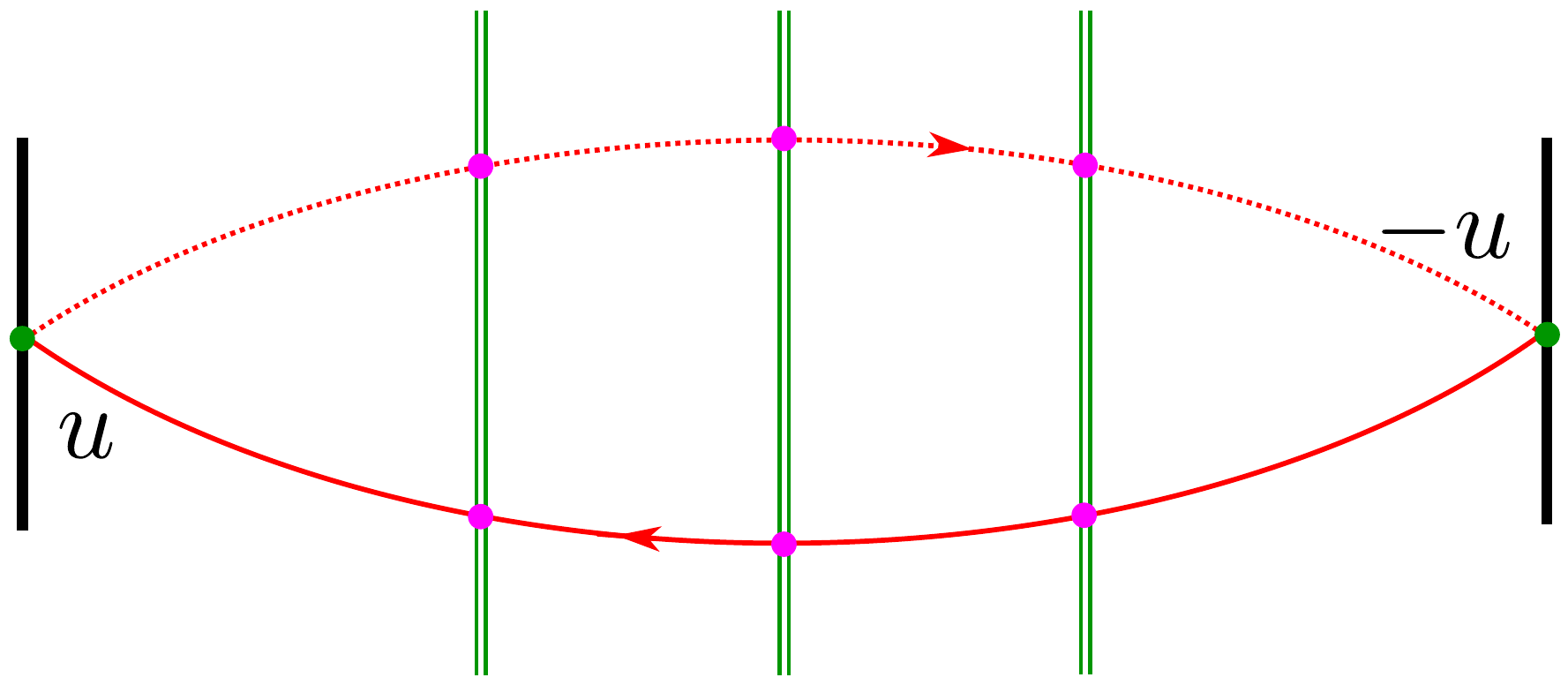}
    \caption{Transfer matrix in fundamental representation for $J=3$.}
    \label{fig:transfer}
\end{figure}
\subsubsection{Transfer matrix in fundamental representation}

\begin{figure}[ht]%
 \centering
 \subfloat[Step 1 --- We start off with $\mathbb{T}(u)\mathbb{T}(v)$, which acts as a differential operator on the quantum space.]{\includegraphics[scale = 0.6]{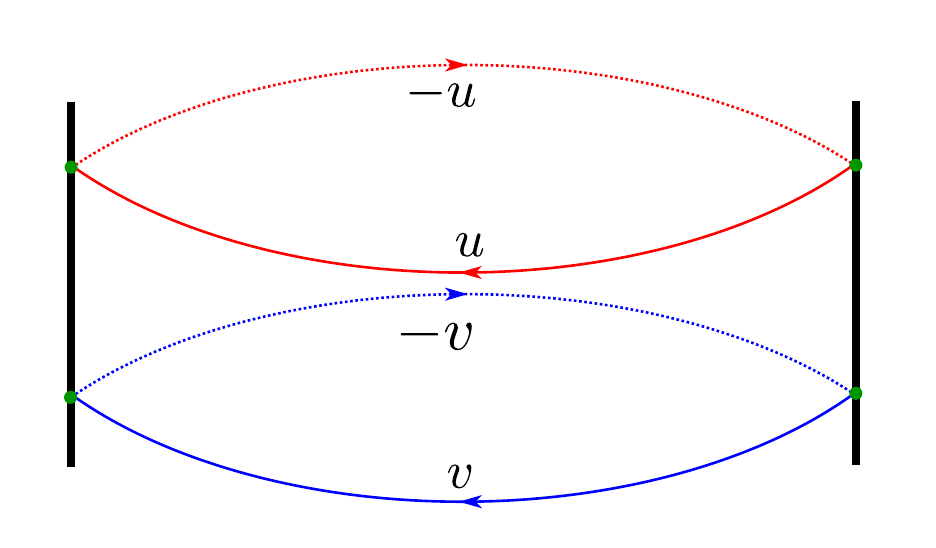}\label{fig:step1}}%
 \hfill
 \subfloat[Step 2 --- By introducing the identity as a product of $S$-matrices from \eq{unitarity}, we can pass the particle line of $\mathbb{T}(u)$ through $\mathbb{T}(v)$. ]{\includegraphics[scale = 0.6]{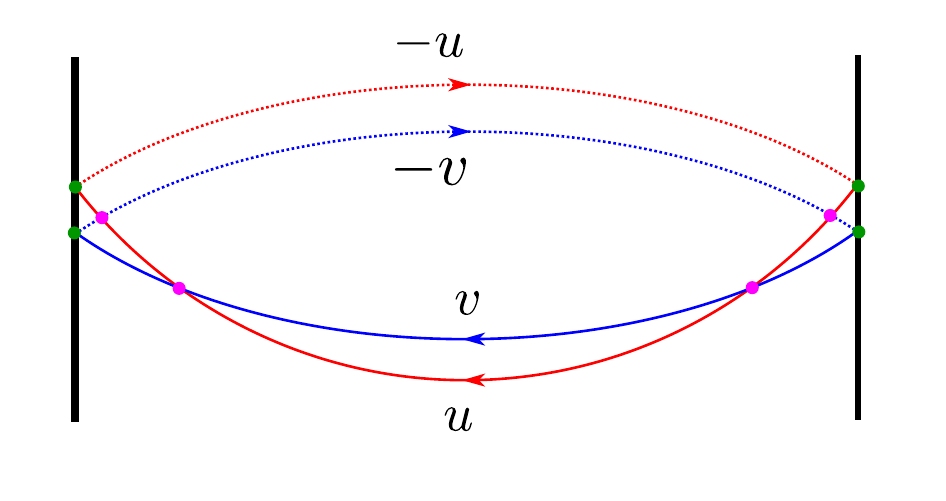}\label{fig:step2}}%
 \hfill \\
 \subfloat[Step 3 --- Now we apply the boundary Yang-Baxter equations \eq{BYBEleft} and \eq{BYBEright}.]{\includegraphics[scale = 0.6]{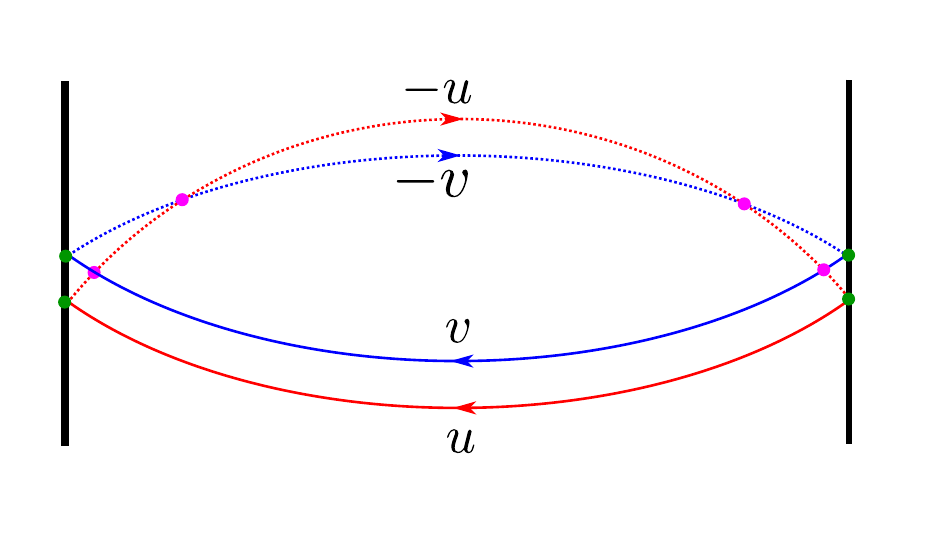}\label{fig:step3}}%
 \hfill
 \subfloat[Step 4 --- Finally we resolve the identity again using \eq{unitarity}, to obtain $\mathbb{T}(v)\mathbb{T}(u)$ hence proving that indeed the transfer matrices commute with each other for arbitrary values of the spectral parameter.]{\includegraphics[scale = 0.6]{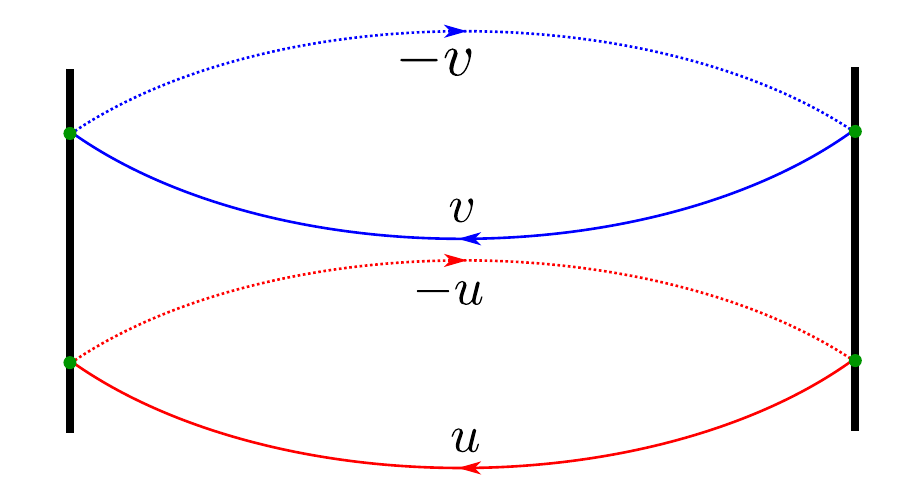}\label{fig:step4}}%
 \hfill
\caption{Diagrammatic proof of $[\mathbb{\hat{T}}^{\bf{4}}(u),\mathbb{\hat{T}}^{\bf{4}}(v)]=0$}%
 \label{T commutes}
\end{figure}

For the transfer matrix in fundamental representation we can mainly mimic
the classical transfer matrix \eq{Tfund}.
From that we can deduce the diagrammatic representation as in figure~\ref{fig:transfer} and follow the rules outlined in section~\ref{diagrammatic}
to deduce the quantum counterpart.
We will use the index $\bf 4$ to indicate the fundamental representation, and define the transfer matrix as:
\begin{align}\la{def:Tq}
    \mathbb{\hat{T}}^{\bf{4}}(u) = &\tr\[\mathbb{{\bar{L}}}_{J}(-u)\cdots\mathbb{{\bar{L}}}_{2}(-u)
    \mathbb{\bar{L}}_{1}(-u)
    {\mathbb{{K}}}(u){\mathbb{{L}}_{1}}(u)\dots
    {\mathbb{{L}}_{J-1}}(u)
    {\mathbb{{L}}_{J}}(u)
        G
    \mathbb{{\bar{K}}}(u)
    G^t\] \;.
\end{align}
We  show now that the transfer matrices form a family of mutually commuting operators:
\beq
[\mathbb{\hat{T}}^{\bf{4}}(u),\mathbb{\hat{T}}^{\bf{4}}(v)]=0\;.
\eeq
This is particularly easy to see using the diagrammatic representation
as we do in figure~\ref{T commutes} for the particular case $J=0$ for simplicity. The step 1 represents $\mathbb{\hat{T}}^{\bf{4}}(u)\mathbb{\hat{T}}^{\bf{4}}(v)$. In step 2, we use unitarity of $S$-matrix \eqref{unitarity} to insert 4 scattering matrices. In step 3 we use BYBE \eqref{BYBEleft} and \eqref{BYBEright}. In step 4 we cancel the $S$-matrices using unitarity again, obtaining $\mathbb{\hat{T}}^{\bf{4}}(v) \mathbb{\hat{T}}^{\bf{4}}(u)$.

In order to be able to conclude that the quantum system is integrable we need to demonstrate that the Hamiltonian is part of the system of commuting operators. For that end in the next section we will build 
the transfer matrix in vector representation and demonstrate that it does contain the Hamiltonian.

\subsubsection{Ingredients of the transfer matrix in vector representation}
In order to build $\mathbb{T}^{\bf 6}$ -- the transfer matrix in
vector representation -- we will need the corresponding building blocks, which are
${\mathbb K}$ and ${\mathbb L}$ in vector representation.
The simplest way to obtain those is by applying the fusion procedure~\cite{Lipan:1997bs}. Roughly speaking, we will need two
copies of ${\mathbb L}$ (or ${\mathbb K}$) in the fundamental representation with spectral parameters $u\pm \frac{i}{2\xi}$ combined together in an antisymmetrised way.
This procedure was already applied to ${\mathbb L}$ in \cite{Gromov:2019bsj} where ${\mathbb L}^{\bf 6},\;{\mathbb L}^{\bar {\bf 4}}$
and ${\mathbb L}^{\bar {\bf 1}}$ -- i.e. ${\mathbb L}$ in all anti-symmetric representations -- were computed.
\begin{figure}[ht]%
 \centering
\subfloat[$\mathbb{\hat{K}}^6$]{\includegraphics[scale = 0.9]{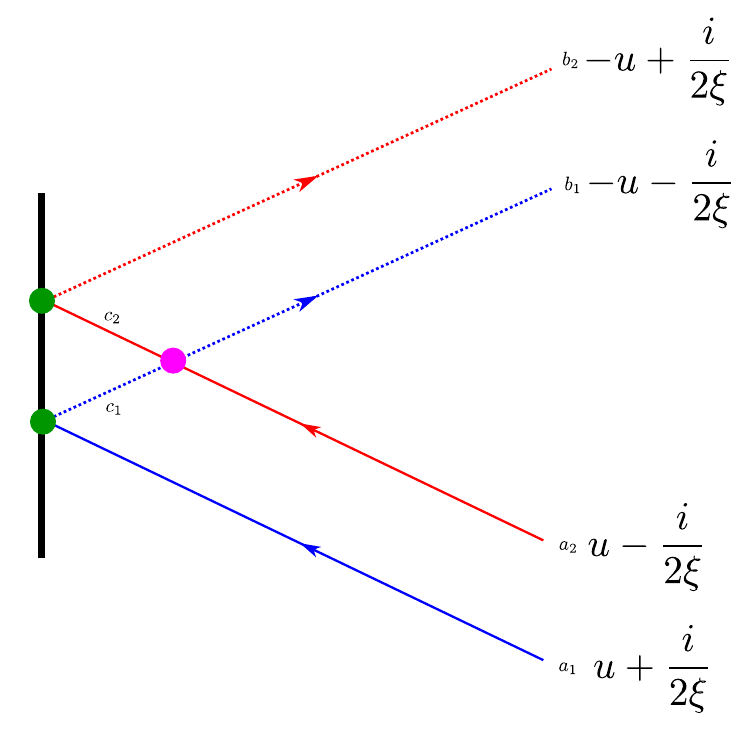}\label{fig:K6left}}
 \hfill
 \subfloat[$\hat{\bar{\mathbb{K}}}^6$]{\includegraphics[scale = 0.9]{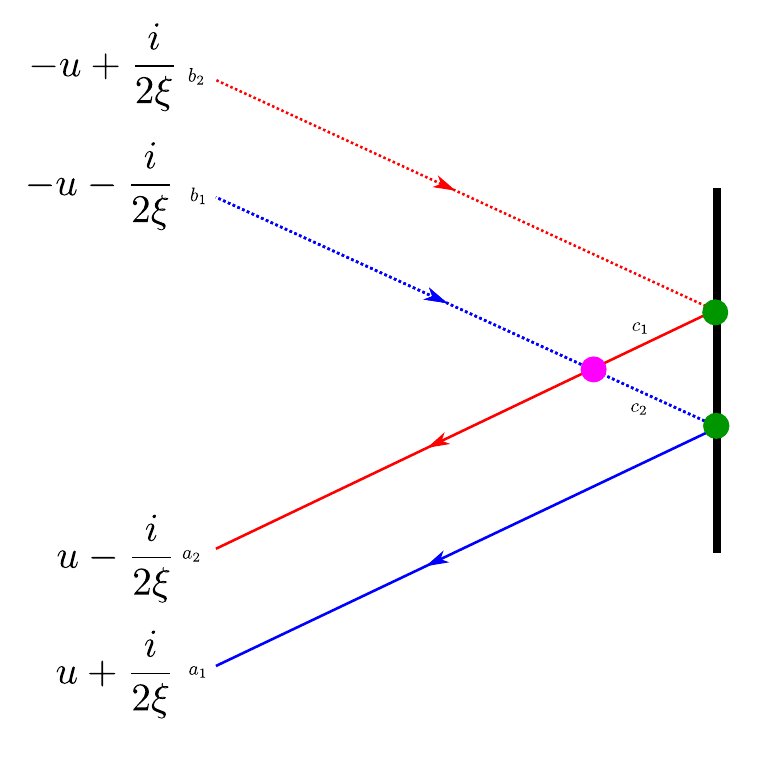}\label{fig:K6right}}
 \hfill
 \caption{Boundary reflection matrices in vector representation via fusion procedure.}%
 \label{fig:fus6}%
\end{figure}

The ${\mathbb L}^{\bf 6}$ needed in this section has auxiliary space being the $6$-dimensional Minkowski space with metric $\eta_{MN}$.
The Lax operator is now a quadratic polynomial with coefficients built out of local charge operators $\hat q_i$ as follows:\footnote{Due to the sign difference in $\hat{\mathbb{L}}_i^{{\bf 4}}(u)$  in comparison with \cite{Gromov:2019bsj}, the linear term in $u$ also has a different sign.} 
\begin{equation}
\label{eqn:L6}
    \hat{\mathbb{L}}_i^{{\bf 6}\;MN}(u)=\left(u^2-\frac{1}{8}\tr\hat{q}_i^2\right)\eta^{MN}-u \hat{q}_i^{MN}+\left(\frac{1}{2}\hat{q}_i^{2\;MN}-\frac{i}{\xi}\hat{q}_i^{MN}+\frac{1}{4\xi^2}\eta^{MN}\right)\;.
\end{equation}
We also define the reflected operators $\hat{\bar{\mathbb{L}}}^{\bf6}_i$ by
\beq
\hat{\bar{\mathbb{L}}}_i^{\bf6}(u)\equiv\mathbb{\hat{L}}^{\bf6}_i(-u)\;.
\eeq
Having 
$\mathbb{\bar{L}}^6$ defined  allows us to maintain the same definition for $\mathbb{T}^6$ as the other $\mathbb{T}$-operators like \eq{def:Tq}.

\paragraph{Boundary reflection operator.}
What remains to be done is fusing the reflection operators. 
In order to keep them covariant 
and ensure the structure of the BYBE,
we have to insert an additional $R$-matrix between the tensor product of two reflection matrices, as shown in figure~\ref{fig:fus6}.
In terms of $Y=Y_0$, defined in \eq{Ys}, we get:
\beqa\la{K6def}
\hat{\mathbb{K}}^{\bf 6}_{MN}(u)&=&C^{\bf 6}_{MN} u\(u-\frac{i}{\xi}\)
+u\frac{2i}{\xi}(Y_N \dot Y_M-Y_M \dot Y_N)\d_t\\
\nonumber&+&\frac{2}{\xi^2}
 Y_N \hat \d_t Y_M
\hat \d_t
-\frac{2i}{\xi^3 u}
 Y_M \hat \d_t Y_N
\hat \d_t\;.
\eeqa
As we can see, it is a second order differential operator in $t$ and a second order polynomial in the spectral parameter $u$.
Similarly for the right boundary we get (replacing $t$ by $s$):
\beqa\la{K6defbar}
 \hat{\bar{\mathbb{K}}}^{\bf 6}_{MN}(u)&=&C^{\bf 6}_{MN} u\(u+\frac{i}{\xi}\)
+u\frac{2i}{\xi}(Y_N \dot Y_M-Y_M \dot Y_N)\d_s\\
\nonumber&+&\frac{2}{\xi^2}
 Y_M \hat \d_s Y_N
\hat \d_s
+\frac{2i}{\xi^3 u}
 Y_N \hat \d_s Y_M
\hat \d_s\;.
\eeqa
In the equations above we are using the reflection matrix $C^{\bf 6}$ in vector representation is the matrix $C$ we introduced in \eq{def:Cmatrix}.
An important property, which follows directly from the definitions \eqref{K6def} and \eqref{K6defbar}, is that
${\mathbb K}^{\bf 6}(+i/\xi)=0$
and
$\bar {\mathbb K}^{\bf 6}(-i/\xi)=0$.

Using $\hat q_0$ \eqref{eqn:defq0}  we can write
\beqa
\label{eqn:K6}
\hat{\mathbb{K}}^{\bf 6}(u)&&=C^{\bf 6} u\left(u-\frac{i}{\xi}\right)-u\hat{q}_{0}+\frac{1}{2}\hat{q}^2_{0}-\frac{i}{2\xi u}(\hat{q}^2_{0})^T\;.
\eeqa
For the right boundary we have very similar expression
\beqa\la{Kb6}
\hat{\bar{\mathbb{K}}}^{\bf 6}(u)&&=
(G^{\bf 6})^{-1}.\(u\left(u+\frac{i}{\xi}\right)-u\hat{q}_{J+1}+\frac{1}{2}(\hat{q}^2_{J+1})^T+\frac{ i}{2\xi u}\hat{q}^2_{J+1}\).G^{{\bf 6}}.C^{\bf 6}\;.
\eeqa
where
$G^{\bf 6}$ in vector representation is the twist matrix $G$ from \eq{def:Cmatrix}. The twist matrix $G^{\bf 6}$
appears in the expression \eq{Kb6} for the right boundary reflection operator, as it is defined in a way that does not depend on $\varphi$. 

Having all the needed ingredients we can compute ${\mathbb T}^{\bf 6}$
by replacing in the r.h.s. of \eq{def:Tq} all the operators and matrices by their vector representation counterpart. In addition, one should multiply the result by $A(2u)$ in order to account for the correct normalisation of the extra $R(2u)$ and $\bar R(2u)$ appearing in the fusion procedure of 
the boundary reflection operators (see figure~\ref{fig:fus6}).

We will present some explicit examples of the transfer matrices later in section~\ref{sec:J0} and Appendix~\ref{sec:J1T}. In the next sections we will first prove that the transfer matrix in the vector representation contains the Hamiltonian, and then proceed with the antifundamental representation.

\subsubsection{Hamiltonian from the transfer matrices}\la{sec:ham}
In this section we will show that the Hamiltonian of the system is a part of the commuting family of operators. For that consider:
\beq
{\mathbb T}^{\bf 6}(0)=
4\lim_{u\to 0}u^2\xi^2
\tr\[\mathbb{{\bar{L}}}_{J}(0)\cdots\mathbb{{\bar{L}}}_{2}(0)
    \mathbb{\bar{L}}_{1}(0)
    {\mathbb{{K}}}(u){\mathbb{{L}}_{1}}(u)\dots
    {\mathbb{{L}}_{J-1}}(0)
    {\mathbb{{L}}_{J}}(0)
        G
    \mathbb{{\bar{K}}}(u)
    G^t\] \;.
\eeq
First we can use that:
\begin{equation}
\label{eqn:L60}
    \hat{\mathbb{L}}_i^{{\bf 6}\;MN}(0)=\frac{\hat{q}_i^{2\;MN}}{2}-\frac{i}{\xi}\hat{q}_i^{MN}
    -\frac{\eta^{MN}}{8}\tr\hat{q}_i^2
    +\frac{\eta^{MN}}{4\xi^2}=
    \frac{:\hat{q}_i^2:^{MN}}{2}=\frac{1}{2\xi^2}X_i^M X_i^N \d^2_{X_i^K}\;.
\end{equation}
where in the last equality we used the identity from
 \cite{Gromov:2019bsj}.
 Also from~\eq{K6def} and~\eq{K6defbar} we have
\beqa
\label{eqn:K60}
\left.u\,\xi\, \hat{\mathbb{K}}^{{\bf 6}\;MN}(u)\right|_{u=0}&=&
-\frac{i}{2}
(\hat{q}^2_{0\;\;\;\;\;})^{NM}=
-\frac{2i}{\xi^2}
 Y^M_0 \hat \d_t Y^N_0
\hat \d_t\;\;,
\\
\left.u \xi \(G^{\bf 6}\hat{\bar{\mathbb{K}}}^{\bf 6}(u)
(G^{\bf 6})^{-1}\)^{MN}
\right|_{u=0}&=&
+\frac{i}{2}
(\hat{q}^2_{J+1})^{MN}=
+\frac{2i}{\xi^2}Y_{J+1}^{N}\hat \d_s Y^M_{J+1} \hat \d_s
\;.
\eeqa
Combining all parts together,
up to sub-leading terms in $1/\xi$ we get the quantum version of $H_q+1$, where $H_q$ is defined in
\eq{Hqdef}. In order to check that this produces
the correct quantisation of $H_q$, i.e. the one related to the graph building operator, we have to analyse the expression \eq{eqn:L60} more carefully. Paying attention to the order of the operators we get 
\beqa
{\mathbb T}^{\bf 6}(0)&=&4\frac{4}{2^{2J}\xi^{4J+4}}
\eta_{NM}X_J^M X_J.X_{J-1}\dots
X_1.Y_0\d_t\prod_{i=1}^J\Box^{(6)}_i\\
\nonumber
&\times&Y_0.X_1
X_1.X_2\dots
X_{J}.Y_{J+1}\d_s Y_{J+1}^N\d_s\d_t\prod_{i=1}^J\Box^{(6)}_i\;,
\eeqa
where all derivatives are understood as operators acting on the CFT wavefunction embedded in the lightcone of $6D$ Minkowski spacetime.
In order to relate the above expression with the graph building operator \eq{Bm1}, which is expressed in terms of derivatives acting on functions in $4D$ Euclidean spacetime, we recall that ${\mathbb T}^{\bf 6}$ is built out of $q_i$'s
and as such we can act with it, in a consistent way, on functions of $4D$ coordinates, following the prescription \eq{f6f4}.
Furthermore, one can just replace the $6D$ d'’Alembertian operator in $4D$ d’'Alembertian due to the identity
\begin{align}
    \square^{(6)} = \square^{(4)} + \partial_{X_+}\partial_{X_-}  \,,
\end{align}
and the fact that there is no dependence on $X^{-}$ in the $4D$ functions, by construction~\eq{f6f4}. Therefore,
\beqa
Y_0.X_1
X_1.X_2\dots
X_{J}.Y_{J+1}\d_s\d_t\prod_{i=1}^J\Box^{(6)}_i&=&\\
\(-\tfrac12\)^{J+1}\frac{\d_s\d_t}{|y_0'||y_{J+1}'|}\prod_{i=0}^J (y_i-y_{i+1})^2
\prod_{i=1}^J\Box^{(4)}_i&=&
\(\tfrac12\)^{J+1}(4\hat g^2)^{J+1}\hat B^{-1}\,,
\eeqa
where we used that $X_1.X_2=-\frac{1}{2}(x_1-x_2)^2$
and $Y_0.X_1=-\frac{e^{-t}}{2}(x_0-x_1)^2$.
We use the expression for the inverse of the graph-building operator  $\hat B^{-1}$ from \eq{Bm1}.
Then for ${\mathbb T}^{\bf 6}(0)$ one gets precisely
\beq\la{BT}
{\mathbb T}^{\bf 6}(0)=4\hat B^{-2}\;.
\eeq
Where we used \eq{xitog} to relate $\xi$ and $\hat g$.
We see that all factors cancel exactly, implying that at the quantum level we also have ${\mathbb T}^{\bf 6}(0)\psi=4\psi$
as it follows from \eq{Bm1}.
At the same time we see that the quantum graph building operator $\hat B$ is indeed a part of the commuting family of operators, which demonstrates the integrability of the initial system of Feynman graphs.

Now in order to find the spectrum $\Delta(\xi)$ we will have to build two remaining transfer matrices in the two sections below.

\subsubsection{Ingredients of the transfer matrix in the anti-fundamental representation}
\label{T4barstuff}
\begin{figure}[ht]%
 \centering
\subfloat[$\mathbb{\hat{K}}^{\bar{4}}$]{\includegraphics[scale = 0.85]{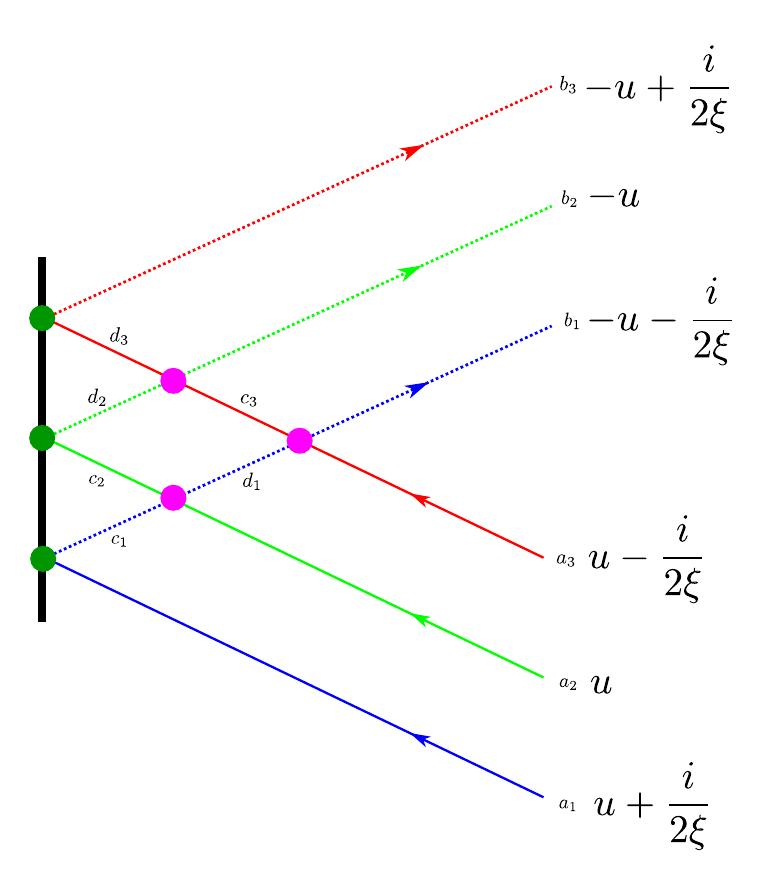}\label{fig:K4barleft}}
 \hfill
 \subfloat[$\hat{\bar{\mathbb{K}}}^{\bar{4}}$]{\includegraphics[scale = 0.85]{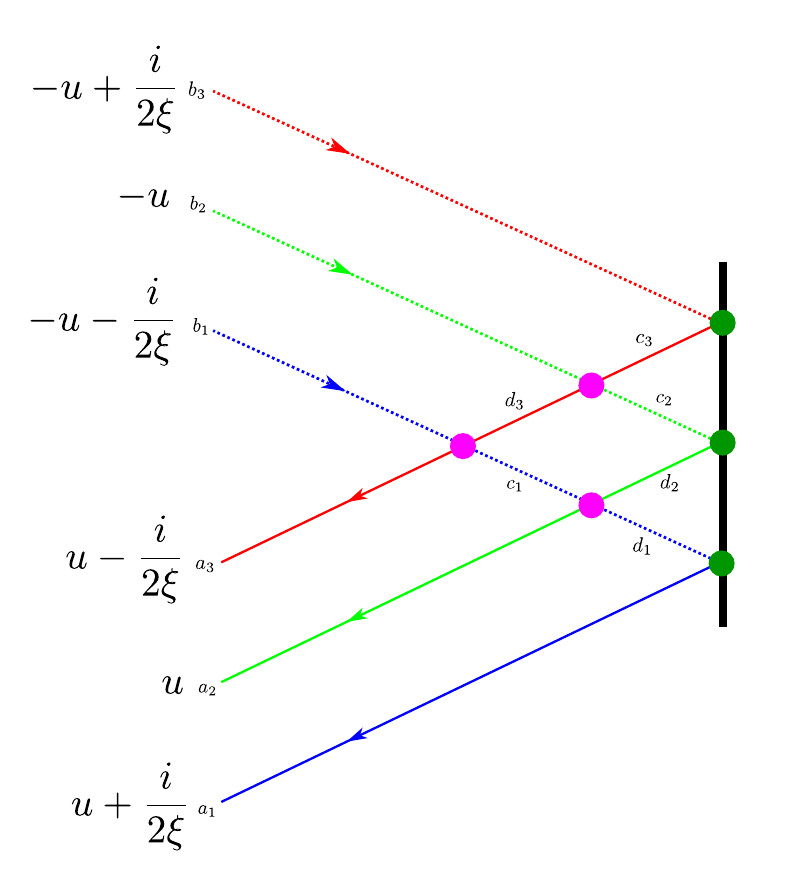}\label{fig:K4barright}}
 \hfill
 \caption{Fusion of the boundary reflection operators to anti-fundamental irrep. ${\bf\bar{4}}$}%
 \label{fig:fus4bar}%
 \label{fig:Fusionto4bar}
\end{figure}

Here we compute the $\bar {\bf 4}$ transfer matrix, corresponding to the antisymmetrisation of the tensor product of three copies of $\bf{4}$ irrep.~ingredients with the corresponding shifts in the spectral parameters, dictated by the fusion procedure. The calculation for ${\mathbb L}^{\bf 4}$ was done in \cite{Gromov:2019bsj}. The result for ${\mathbb L}^{\bar{\bf 4}}$ can be re-expressed in terms of one
${\mathbb L}^{\bf 4}$ times a scalar polynomial factor
\begin{equation}\la{L4bar}
    \mathbb{\hat{L}}_{k\,a}^{\bar{\bf 4}\,\,\,\,\,b}(u)=\left(u^2-\frac{\text{tr}\hat{q}_k^2}{8}+\frac{1}{\xi^2} \right)\hat{\bar{\mathbb{L}}}_{k\,a}^{{\bf 4}\,\,\,\,\,b}(-u)\;,
\end{equation}
and $\hat{\bar{\mathbb{L}}}_{k\,\,\,\,\,b}^{\bar{\bf 4}\,a}(u)=
\mathbb{\hat{{L}}}_{k\,b}^{\bar{\bf 4}\,\,\,\,a}(u)$.
The fusion of the boundary reflection operator
is done in analogy with $\bf 6$ representation. For that one follows the diagram in figure~\ref{fig:Fusionto4bar} to obtain:
\begin{equation}\la{K4bar}
    \mathbb{\hat{K}}^{\bar{\bf 4}\,ab}(u)=-\left(u^2-\frac{iu}{\xi}+\frac{3}{4\xi^2}\right)\hat{\bar{\mathbb{K}}}^{{\bf 4}\,ba}(-u)\;\;,\;\;
    \hat{\bar{\mathbb{K}}}^{\bar{\bf 4}}_{ab}(u)=-\left(u^2+\frac{iu}{\xi}+\frac{3}{4\xi^2}\right)\mathbb{\bar{K}}^{\bf 4}_{ba}(-u)\;.
\end{equation}
Finally, the twist matrix is the inverse of the one for the $\bf4$ irrep. \eqref{C4andG4}.
The polynomial factors in \eq{L4bar} and \eq{K4bar} will play an important role below, when we derive the TQ-relations.

\subsubsection{Ingredients of the quantum determinant}
\label{1barstuff}
\begin{figure}[ht]%
 \centering
\subfloat[$\mathbb{K}^{\bar{1}}$]{\includegraphics[scale = 0.85]{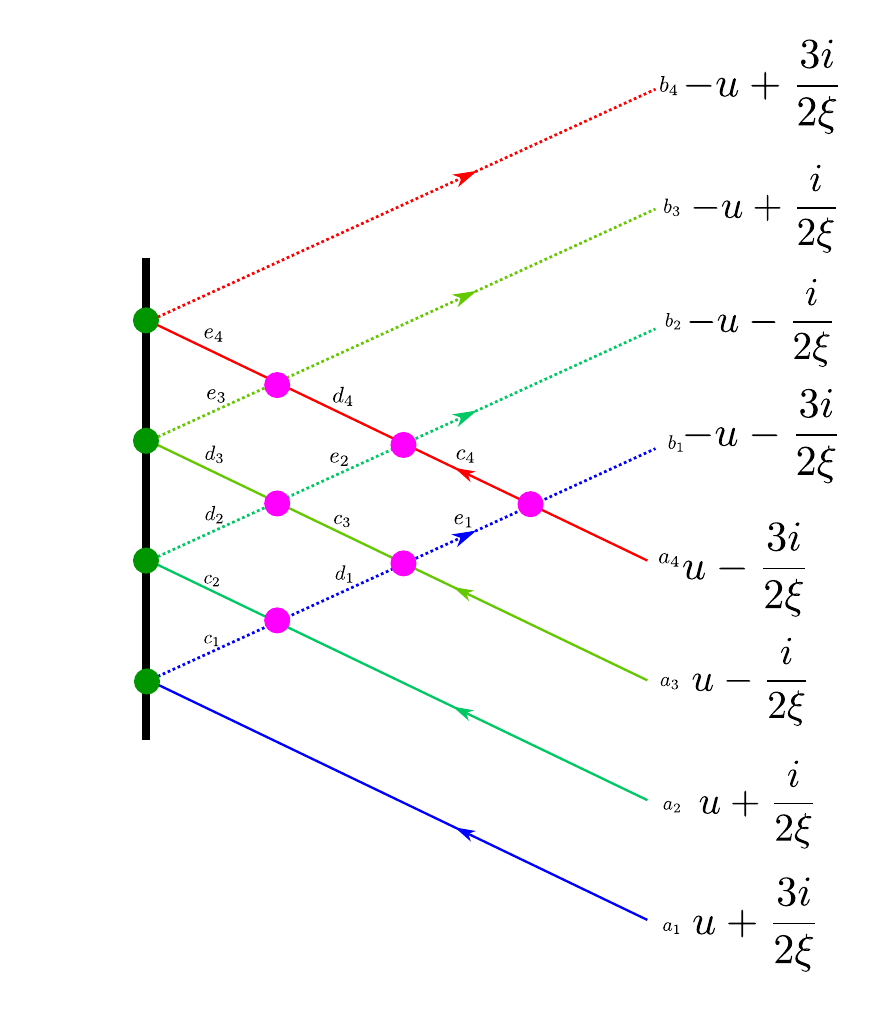}\label{fig:K6barleft}}
 \hfill
 \subfloat[$\mathbb{\bar{K}}^{\bar{1}}$]{\includegraphics[scale = 0.85]{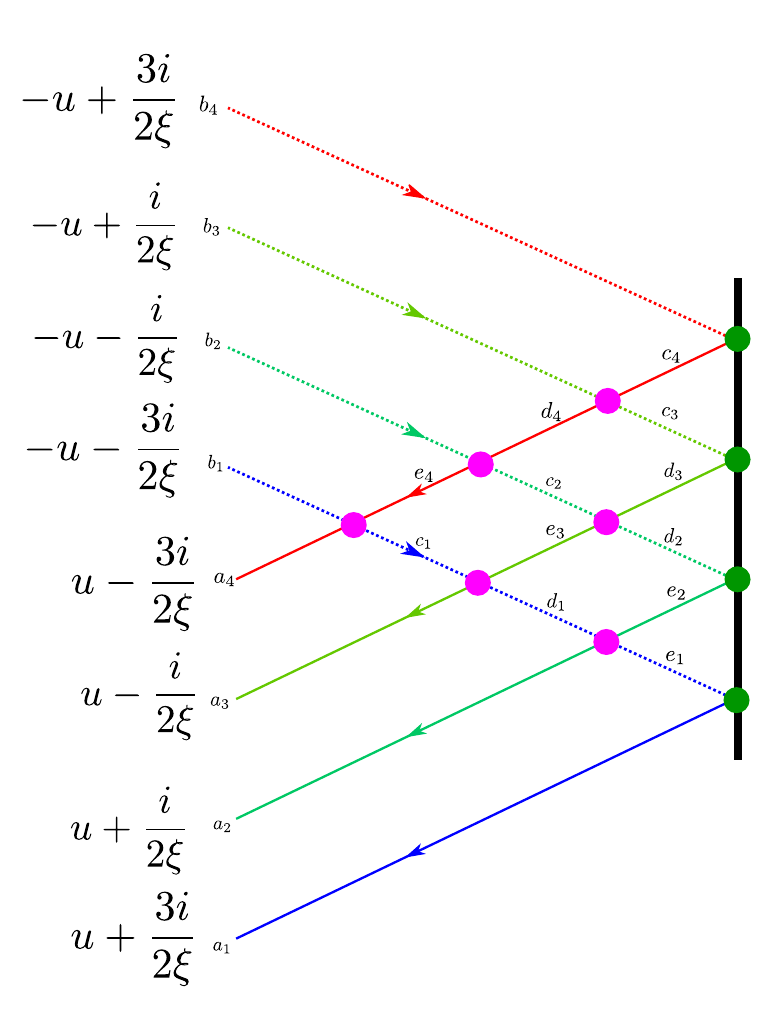}\label{fig:K6barright}}
 \hfill
 \caption{Fusion of the boundary reflection operators to obtain the quantum determinant.}%
 \label{fig:fus1bar}%
 \label{fig:Fusionto1bar}
\end{figure}

Here we compute the ingredients of the transfer matrix in 
the representation $\bf\bar 1$, also known as the quantum determinant. 
Like in the previous sections, this can be computed as an antisymmetrisation of the tensor product of four copies of ${\mathbb L}$ and ${\mathbb K}$ in the $\bf4$ irrep.
For both ${\mathbb L}^{\bar 1}$ and ${\mathbb K}^{\bar 1}$
we find that they are just  fourth order polynomials in $u$ acting trivially on the physical space.
Again the calculation of ${\mathbb L}^{\bar 1}$ was already performed in \cite{Gromov:2019bsj} and the result reads:
\beq
\mathbb{\hat{L}}_i^{\bar{\bf1}}(u)=\left(u^2-\frac{\text{tr}\,\hat{q}^2_i}{8}+\frac{5}{4\xi^2} \right)^2+\frac{\text{tr}\,\hat{q}^2_i}{8\xi^2}-\frac{1}{\xi^4}\;.
\eeq
The expression for $\hat{\bar{\mathbb{L}}}_i^{\bar{\bf1}}$ is the same.
For the boundary reflection operator we follow the diagram on figure~\ref{fig:Fusionto1bar} to obtain:
\begin{equation}
    \mathbb{K}^{\bar{\bf1}}(u)=\left(u-\frac{2i}{\xi}\right)\left(u-\frac{i}{\xi}\right)u\left(u+\frac{i}{\xi}\right)
    \;\;,\;\;
     \bar{\mathbb{K}}^{\bar{\bf1}}(u)=\left(u+\frac{2i}{\xi}\right)\left(u+\frac{i}{\xi}\right)u\left(u-\frac{i}{\xi}\right)\;.
\end{equation}

In the next section we will discuss the implications of these polynomial factors for the analytical properties of the transfer matrices and then derive the TQ-relations.

\subsubsection{$J=0$ example}\label{sec:J0}
Before discussing the general case we first give the explicit result for the simplest case of a chain of zero length. This means that we are only left with the boundary reflection operators. Furthermore, the graph-building operator is a
second order differential operator in $s$ and $t$, as it should commute with the dilatation operator only one variable remains. If we further impose ${\mathbb T}^{\bf 6}(0)=4$, we will automatically diagonalise all transfer matrices obtaining the following results for their eigenvalues:
\begin{align}\label{eqn:J0T}
    \begin{split}
{\mathbb T}^{\bf 4}(v)&=
\frac{1}{\xi^2}\(
4 v^2 \cos \varphi +\cos \varphi+
8 \xi ^2 \)\;,\\
{\mathbb T}^{\bf 6}(v)&=
A(2v)
\frac{v^2+1}{\xi^4 v^2}
\(
 v^4 (2\cos (2 \varphi )+4)+v^2
   \left(16 \xi ^2 \cos \varphi -4 \Delta
   ^2 \sin ^2\varphi \right)
+
16 \xi ^4
\)\;,\\
{\mathbb T}^{\bf {\bar 4}}(v)&=
A(2v)A(2v+i)A(2v-i)
\frac{\left(4 v^2+1\right) \left(4
   v^2+9\right)}{16 \xi ^6}
\(
4 v^2 \cos \varphi +\cos \varphi+
8 \xi ^2 \)\;,\\
{\mathbb T}^{\bar {\bf 1}}(v)
&=
A^2(2v)A(2v+i)A(2v-i)
A(2v+2i)A(2v-2i)
\frac{v^2 \left(v^2+1\right)^2
   \left(v^2+4\right)}{\xi ^8}\;,
\end{split}
\end{align}
where we introduced the rescaled spectral parameter $v=u\,\xi$. The factors $A$, 
where
$A(v)=\frac{v^2}{1+v^2}$,
are due to the R-matrix normalisation as discussed in section~\ref{Rmatrixnorm}.
We also worked out the form of the transfer-matrix eigenvalues for $J=1$ case in  Appendix~\ref{J1res} in terms of a few unknown constants. 
We have explicitly verified that all the $\mathbb{T}$-operators for $J=0$ and $J=1$ commute between themselves and with the charges $\Delta,\,S,\,H$ as expected.

In the next section we will extend these results to the general $J$ case.

\subsubsection{Eigenvalues of the transfer matrices}\la{Tprops}
Here we deduce the general form of the eigenvalues of the transfer matrices. First, one can notice explicitly that for $J=0$ and $J=1$
case they are even functions of the spectral parameter. In Appendix~\ref{Appendixparity}
we prove that this is true for any $J$. Some other properties of the transfer matrices are:

\begin{itemize}
\item ${\mathbb T}^{\bf 4}$ is a polynomial of degree $2J+2$ in $v$, as it follows from its definition \eq{def:Tq}.
\item ${\mathbb T}^{\bf 6}$ is a rational function with two poles at $v=\pm i/2$, coming from the normalisation factor $A(2v)$. Another potential pole at $v=0$, coming from the boundary reflection operators, is cancelled by the same $A(2v)$.
At large $v$ it behaves as $\sim v^{4J+4}$.
\item Previously we noticed that
${\mathbb K}^{\bf 6}(u=+i/\xi)=0$
and
$\bar {\mathbb K}^{\bf 6}(u=-i/\xi)=0$, therefore
we can see that
$\mathbb{T}^6$ should have a prefactor of $v^2+1=\xi^2 u^2+1$. 
\item Finally, in section~\ref{sec:ham}
we have showed that $\mathbb{T}^{\bf 6}(0)=4$ due to \eq{BT}.
\item The properties of $\mathbb{T}^{\bar{\bf 4}}$
are very similar to those of 
$\mathbb{T}^{{\bf 4}}$, apart from the trivial factors of $A$'s and additional trivial factors coming from ${\mathbb L}^{\bf\bar{4}}$
and ${\mathbb K}^{\bf\bar{4}}$.
\item Finally, 
$\mathbb{T}^{\bar{\bf 1}}$ (the quantum determinant) contains only trivial factors and can be computed explicitly for any $J$.
\end{itemize}
Basing on these observations we can write the transfer matrices in terms of the polynomials $P^{\lambda}_{k}(v^2)$ as:
\begin{align}
    \begin{split}
        \label{baxterTP}
{\mathbb T}^{\bf 1}(v)&=1\;,\\
{\mathbb T}^{\bf 4}(v)&\equiv\frac{P^{\bf 4}_{J+1}(v^2)}{\xi^{2J+2}}\;,\\
{\mathbb T}^{\bf 6}(v)&\equiv 
A(2v)\frac{v^2+1}{v^2}
\frac{P^{\bf 6}_{2J+2}(v^2)}{\xi^{4J+4}}\;,\\
{\mathbb T}^{\bar{\bf 4}}(v)&=
A(2v)A(2v+i)A(2v-i)
\frac{(v^2+\tfrac{9}{4})
(v^2+\tfrac{1}{4})^{2J+1}
P^{\bf \bar 4}_{J+1}(v^2)}{\xi^{6J+6}}\;,\\
{\mathbb T}^{\bar{\bf 1}}(v)&=
A^2(2v)A(2v+i)A(2v-i)
A(2v+2i)A(2v-2i)
\frac{
(v^2+4)(v^2+1)^{2J+2}v^{4J+2}}{\xi^{8J+8}}\;.
    \end{split}
\end{align}
Here, $P^{\lambda}_{k}$ is a polynomial of degree $k$, labelled by the representation $\lambda$ in the auxiliary space.
The eigenvalues of the conserved charges of the system are the coefficients of the powers of $v^2$ in these polynomials.
We will denote them as (defining $w\equiv v^2$):
\begin{align}\label{eqn:polycoeffs}
    \begin{split}
    P^{\bf 4}_{J+1}(w) &= \sum_{i = 0}^{J + 1} a_{i} w^{-i+J+1} \;, \\
    P^{\bf 6}_{2J+2}(w) &= \sum_{i = 0}^{2J + 2} b_{i} w^{-i+2J+2} \;,\\
    P^{\bf \bar 4}_{J+1}(w) &= \sum_{i = 0}^{J + 1} c_{i} w^{-i+J+1} \;.
     \end{split}
\end{align}
In our definition, $a_0$ will represent the coefficient of the highest power in $v^2$ in $P^{\bf4}_{J+1}$,  $a_{1}$ the second highest etc. The leading coefficients are easy to compute explicitly directly from the definition
\beq\label{eqn:leadinglargeutwist}
        a_{0} = c_0 = 4\cos\varphi\;\;,\;\;
        b_{0} = 2\cos 2\varphi + 4\;.\\
\eeq
They just give the twisted (or q-)dimension of the corresponding representations.
Since the leading coefficients are trivial,
in total we get $4J+4$ non-trivial coefficients in the polynomials $P$.
As our system has $4J+2$ degrees of freedom it may suggest that there are $2$ more relations between the coefficients of the polynomials $P$. Indeed, in the $J=0$
and $J=1$ cases we found them by computing the differential operators explicitly, but it is rather hard to deduce the general relations. In $J=1$ case we found exactly $6$ independent operators guarantees integrability of the system.

We found that the global charges $\Delta$ and $S$ are encoded into the sub-leading coefficients in the following way:
\begin{align}\label{eqn:sublead}
    c_{1} - a_{1} &= 8\, i\, S\, \Delta \,\sin\varphi\;, \\
    \nonumber\left(a_{1} + c_{1} \right)\cos\varphi - b_{1} 
    &=
    2\,\left(2\,S^2 + 2\,\Delta^2 + J\right)\,\sin^2\varphi + 2\,\cos^2\varphi\;.
\end{align}
These relations are also quite hard to derive in general, but we explicitly verified the first relation up to $J = 3$ and the second up to $J=2$. 

Finally, the condition ${\mathbb T}^{\bf 6}(0)=4$
implies:
\beq
b_{2J+2}={\xi^{4J+4}}=16\hat g^{4J+4}\;.
\eeq

In order to find the eigenvalues of all coefficients of the transfer matrices we will have to develop a numerical procedure. For that we will first build the TQ-relations in the next section.

\section{Baxter equation}
\label{Baxtersection}
In this section we follow the derivation of \cite{Gromov:2019jfh} to deduce the general simplified form of the TQ-relations and deduce asymptotic of the Q-functions.
The starting point is the TQ-relation~\cite{Baxter:1982zz,BAXTER19731,Bazhanov:1996dr}:
\beq
\label{Baxter}
Q(v+2i)+
{\mathbb T}^{\bf 4}(v+i/2)
Q(v+i)
+
{\mathbb T}^{\bf 6}(v)
Q(v)
+
{\mathbb T}^{\bar{\bf 4}}(v-i/2)
Q(v-i)
+
{\mathbb T}^{\bar{\bf 1}}(v-i)
Q(v-2i)=0\;.
\eeq
As we discussed above the transfer matrices have a number of trivial factors.  In order to remove these fixed, non-dynamical factors, we perform the following {\it gauge} transformation of the Q-function
\beq
Q(v)=q(v)\frac{e^{\pi  (J+1) v}  \Gamma (-i v) \xi ^{2 i (J+1) v}
   \Gamma (i v+1)^{-2 J-1}}{\Gamma \left(-i
   v-\frac{1}{2}\right) \Gamma (i v+2)}\,,
\eeq
which brings \eq{Baxter} to a simpler
and more symmetric form:
\beqa \label{eqn:Bax}
\frac{P^{\bf 6}_{2J+2}(v^2)}{v^{2J+3}}q(v)&=&-(v+i)^{2J+1}q(v+2i)
-\frac{v+\tfrac i2}{
v(v+i)
}P^{\bf 4}_{J+1}\((v+\tfrac i2)^2\)q(v+i)
\\ \nonumber
&&-(v-i)^{2J+1}q(v-2i)
-\frac{v-\tfrac i2}{
v(v-i)
}
P^{\bf \bar 4}_{J+1}\((v-\tfrac i2)^2\)q(v-i)\;.
\eeqa
As a test of this equation we can compare with the case $J=0$, studied as a ladder limit of QSC in ${\cal N}=4$ SYM. For $J=0$, by plugging in the explicit form of the polynomials \eq{eqn:J0T} into
\eq{eqn:Bax}, we obtain:
\begin{multline}
\la{J0Bax}
q(v) \left(\frac{2 \left(8 \hat{g}^2 v^2 \cos (\varphi )+8 \hat{g}^4+v^4 (\cos (2 \varphi
   )+2)\right)}{v^3}-\frac{4 \Delta ^2 \sin ^2(\varphi )}{v}\right)\\
+\frac{2 (2 v-i) q(v-i)
   \left(2 \hat{g}^2+v (v-i) \cos (\varphi )\right)}{v (v-i)}+\frac{2 (2 v+i) q(v+i) \left(2
   \hat{g}^2+v (v+i) \cos (\varphi )\right)}{v (v+i)}\\
   +(v-i) q(v-2 i)+(v+i) q(v+2 i)=0\;.
\end{multline}
This is the same as what was
found in \cite{Gromov:2016rrp,Cavaglia:2018lxi} for a cusped Wilson line in the ladders limit, as expected (see detailed comparison in Appendix~\ref{sec:J0T}). 

The equation \eq{eqn:Bax} for general $J$ is one of our main result. As we show in section~\ref{sec:numerics} it lets us evaluate  numerically the spectrum.

\subsection{Large $v$ asymptotic of Q-functions}
For the numerical evaluation, which we describe in the next section,
it is important to have the large $v$ asymptotics under control.
As the leading and partially subleading coefficients in the polynomials $P$ are known from \eq{eqn:leadinglargeutwist} and \eq{eqn:sublead}, we can deduce that the $4$ linearly independent solutions of the equation \eq{eqn:Bax} should have the following large $v$ asymptotic expansion:
\begin{align}\label{eqn:asym}
    \begin{split}
        q_{1} &= e^{+ \phi v} v^{+\Delta - S - J} \left(1 + \frac{c_{1,1}}{v} + \dots\right) \;,\\
        q_{2} &= e^{- \phi v} v^{+\Delta + S - J} \left(1 + \frac{c_{2,1}}{v} + \dots\right) \;,\\
        q_{3} &= e^{+ \phi v} v^{-\Delta + S - J} \left(1 + \frac{c_{3,1}}{v} + \dots\right) \;, \\
        q_{4} &= e^{- \phi v} v^{-\Delta - S - J} \left(1 + \frac{c_{4,1}}{v} + \dots\right) \;.
    \end{split}
\end{align}
where $\phi=\pi-\varphi$.
The above asymptotics suggest the following relation to the QSC Q-functions of~\cite{Gromov:2015dfa}:
\beq
q_i(v)\sim \frac{Q_i(v)}{v^{J+1/2}}\;, 
\eeq
which is similar to the relations found in the fishnet model~\cite{Gromov:2017cja}.
Subleading coefficients in $1/v$ can be found systematically in terms of the coefficients of the polynomials $P$, i.e.
$a_i,\;b_i$ and $c_i$, by plugging the expansion \eq{eqn:asym} into \eq{eqn:Bax}.
In order to fix the coefficients of the polynomials $P$ one has to use the gluing (or quantisation) condition, which we describe in the next section.

\section{Numerical solution}\label{sec:numerics}
After having established the key properties of the Baxter equation we can solve them numerically 
and fix the remaining coefficients $a_i,\;b_i$
and $c_i$. The method we implement is essentially the one of \cite{Gromov:2015wca}
which was adopted and simplified to the current type of problems in \cite{Gromov:2016rrp,Gromov:2017cja,Cavaglia:2020hdb,Gromov:2019jfh}.
The 4th order finite difference equation \eq{eqn:Bax} has $4$ linearly independent solutions with the asymptotic \eq{eqn:asym}. The way to find them numerically is first finding the asymptotic solution at large $v$, where \eq{eqn:Bax}
reduces to a linear problem for the asymptotic expansion coefficients. The truncated asymptotic
series gives a very good approximation at sufficiently large $|{\rm Im}\, v|$.
In order to bring ${\rm Im}\, v$
to a finite value, we can simply use 
\eq{eqn:Bax} itself, as it allows to find $q(v)$
in terms of $q(v+i n),\;n=1,\dots,4$ (or 
$q(v-i n),\;n=1,\dots,4$). Using \eq{eqn:Bax}
as a recursion relation, we can gradually decrease $|{\rm Im}\, v|$. By doing this there are two options: starting from $+i\infty$ or from $-i\infty$. Correspondingly, we will find 
$4$ analytic solutions in the upper-half plane, $q^{\downarrow}_i$, and other $4$ analytic in the lower-half plane, $q^{\uparrow}_i$.
Since the Baxter equation is a fourth order equation, we can have only four independent solutions, meaning that the ${{q}}_{i}^\uparrow$ and ${{q}}_{i}^\downarrow$ should be related by a linear transformation. We should therefore have:
\begin{align}
    q_{i}^{\uparrow}(v)=\Omega_{i}^{\;j}(v) q_{j}^{\downarrow}(v), \quad \Omega_{i}^{\;j}(v+i)=\Omega_{i}^{\;j}(v)\;,
\end{align}
where
\begin{align}
    \Omega_{i}^{\;j}(v)=\frac{\epsilon^{j\, j_{1}\, j_{2}\, j_{3}}}{3 !} \frac{\operatorname{det}_{n=0, \ldots, 3}\left\{q_{i}^{\uparrow}(v-i n), q_{j_{1}}^{\downarrow}(v-i n), q_{j_{2}}^{\downarrow}(v-i n), q_{j_{3}}^{\downarrow}(v-i n)\right\}}{\operatorname{det}_{n=0, \ldots, 3}\left\{q_{1}^{\downarrow}(v-i n), q_{2}^{\downarrow}(v-i n), q_{3}^{\downarrow}(v-i n), q_{4}^{\downarrow}(v-i n)\right\}}\;.
\end{align}
$\Omega_{i}^{\;j}(v)$ is an $i$-periodic function which can have poles at $v=i n$
of order no higher than $q_i(v)$'s
themselves.
From the Baxter equation \eq{eqn:Bax}  it is easy to see that $q_i(v)$ only has poles at $v=in $
of maximal order $2J+2$, which implies that 
$\Omega_{i}^{\;j}(v)$ is a trigonometric rational function of the form:
\begin{align}
    \Omega_{i}^{\;j}(v)=\frac{\sum\limits_{n=0}^{2J+2} C_{\quad i}^{(n)\;j} e^{2\, \pi\, n\, u}}{\left(1-e^{2\, \pi\, u}\right)^{2J+2}} \;,
\end{align}
The quantisation condition can be obtained by comparing with the QSC description of the cusped Wilson line \cite{Gromov:2015dfa}, where one defines an antisymmetric matrix $\omega_{ik}$,
related to $\Omega_{i}^{\;j}$
in the following way:
\beq\la{om}
\omega_{ik} = \Omega_{i}^{\;j} \Gamma_{jk}\,,
\eeq
where the so-called {\it gluing} matrix is:
\beq
\Gamma_{jk}=
 \left(
\begin{array}{cccc}
 0 & \gamma_1 \sinh (2 \pi  v) & 0 & \gamma_3 \\
 \gamma_2 \sinh (2 \pi  v) & 0 & \gamma_4 & 0 \\
 0 & \gamma_5 & 0 & 0 \\
 \gamma_6 & 0 & 0 & 0 \\
\end{array}
\right)\,,
\eeq
where $\gamma_i$ are some constants.
All we need to know, from QSC, is that $\omega$ in  \eq{om} is anti-symmetric i.e. $
\Omega_{i}^{\;j} \Gamma_{jk}=-
\Omega_{k}^{\;j} \Gamma_{ji}
$, which, in particular, implies:
\beq\la{quant}
\Omega_{41}=\Omega_{32}=0\,.
\eeq
As each component of $\Omega(u)$ is a nontrivial function parametrised in terms of $2J+3$ constants $C^{(n)}_{ij}$,
imposing \eq{quant} is usually sufficient to fix
$4J+4$ unknown constants, contained in the Baxter equations.

\begin{figure}[ht]%
\centering
\includegraphics[scale = 0.8]{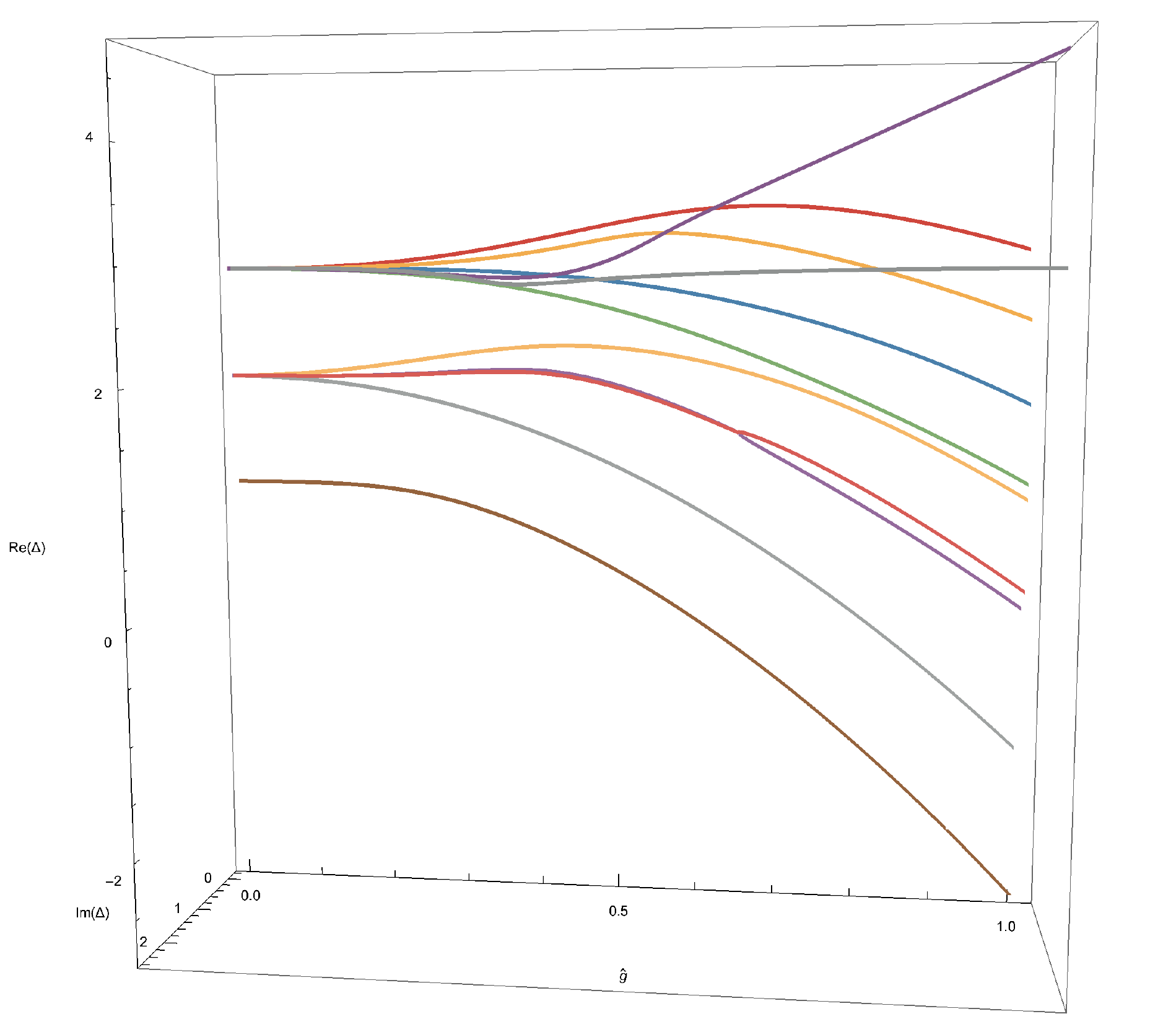}
\caption{Numerical spectrum with excited states for $J = 1$ and $S = 0$. The lowest curve (starting at $\Delta = 1$ at zero coupling) corresponds to the case with a single insertion of $Z$ at the cusp. The curves which begin at higher integers at zero coupling correspond to excited states of the solution of the Baxter equation, which correspond to additional insertions of $\Phi_2$ and $\Phi_1$ 
at the cusp
(see \cite{Cavaglia:2018lxi} for some explicit examples).
Whereas for the ground state the dimension $\Delta$ is real, excited states could appear in complex conjugate pairs.
}%
\label{fig:numspecJ1S0}%
\end{figure}
\begin{figure}
    \centering
    \includegraphics{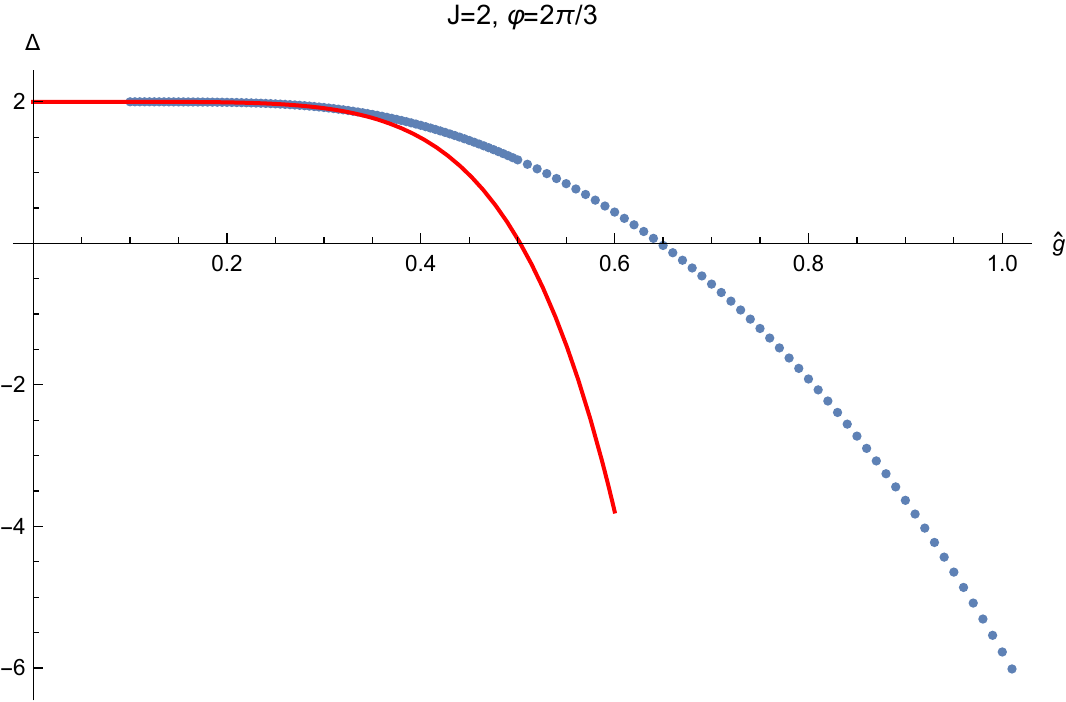}
    \caption{Numerical data (dots) for the ground state of length-2 chain with $\varphi=2\pi/3$. Red solid line shows the L\"uscher formula prediction of \cite{Correa:2012hh}.}
    \label{fig:J2}
\end{figure}

\paragraph{Tests.}
By applying the numerical method we 
studied the spectrum for $J=1$ and $J=2$ cases.
For $J=1$ we also found a large number of excited states (see figure~\ref{fig:numspecJ1S0}), corresponding to additional insertions of $\Phi_2$ and $\Phi_1$ fields at the cusp, as discussed in \cite{Cavaglia:2018lxi}.
We tested our results against the weak coupling result of \cite{Correa:2012hh},
which in our notations reads
\beq
\Delta=J+\hat g^{2J+2}\frac{(-1)^J 2^{4 J+3} \pi ^{2 J+1}  \csc (\varphi ) B_{2
   J+1}\left(\frac{\varphi }{2 \pi }\right)}{\Gamma (2 J+2)}+{\cal O}(\hat g^{4J+4})\,,\la{Lu}
\eeq
which agreed with high precision (of more than $15$ digits) with our numerical data for $J=0,\;J=1$ and $J=2$.
For example for $J=2$ and $\varphi=2\pi/3$
we get the following fit for the numerical data on figure~\ref{fig:J2}:
\beq
\Delta=2-124.08839542210 \hat{g}^6+23271.513371517
   \hat{g}^{12}+\dots\;
\eeq
in agreement with \eq{Lu}, which for $J=2$ gives
\beq
2+\hat g^6 \frac{8}{45} \varphi  \left(3 \varphi ^4-15 \pi  \varphi ^3+20 \pi ^2 \varphi ^2-8 \pi
   ^4\right) \csc \varphi =
   2-\hat g^6 \frac{512 \pi ^5 }{729 \sqrt{3}}\;.
\eeq
The states $Z^J$ for the cases $J=1$ and $J=2$ do not behave classically at large $\xi$, i.e. $\Delta$ decreases faster than linear. Like in \cite{Gromov:2017cja} we expect the classical regime to describe the highly excited states. 

\section{Conclusion} \label{sec:conclusion}

In this paper we show that the cusped Wilson-Maldacena loop with insertions of $J$ orthogonal scalars is an integrable system, and also 
describe its strong coupling dual description in terms of a classical
chain of particles with nearest neighbour interactions, sitting on the lightcone in $6$-dimensional Minkowski spacetime --- the open fishchain. We compute the transfer matrices of this model, which gives us all the conserved charges of the system. Moreover, we obtain a Baxter equation which can be solved numerically for any $J$ and find the spectrum of dimensions $\Delta$ non-perturbatively (for $J=0,1,2$). This lets us find the $Q$-functions of the system, which are a crucial quantity in a quantum integrable systems.  
There is a number of future directions of our work.
We outline some of them below.

Since we have the spectrum under control, the natural next step is to compute correlation functions. In \cite{Cavaglia:2018lxi,McGovern:2019sdd}, the first steps towards this is taken, where the authors calculate the three point functions of three cusped Wilson lines in the ladders limit. Remarkably, they observe that the structure constants can be expressed as overlaps between the same $q$-functions that appear in the QSC that describes the system. This is one of many results \cite{Gromov:2016itr,Gromov:2018cvh,Maillet:2018bim,Ryan:2018fyo,Giombi:2018hsx,Derkachov:2018rot,Cavaglia:2019pow,Gromov:2019wmz,Derkachov:2019tzo,Gromov:2020fwh,Ryan:2020rfk,Derkachov:2020zvv,Cavaglia:2020hdb}
in the rapidly developing separation of variables (SoV) program. 
Another possibility is to 
first probe the SoV structure at strong coupling, using the
dual fishchain description which becomes classical in this limit.

In general 
it would be interesting to get away from the ladders/fishnet limit both in our set-up and in deformed ${\cal N}=4$ SYM. Such an exploration could give some clues as to how to develop a first principles holographic derivation for the full theory. 

It would be also interesting to establish links with the bootstrap program, in particular in  the straight line limit, where the Wilson line defines a defect CFT. 
Our Baxter equation allows to more easily generate the spectrum here than in the case of the full ${\cal N}=4$ SYM, which could give a better understanding on how the conformal bootstrap works in this case.

Another direction of exploration would be to try to expand our construction to Wilson loops in the ABJM theory~\cite{Drukker:2019bev} where there is already some evidence of integrability~\cite{Bai:2017jpe} and further, it admits treatment from a defect CFT point of view~\cite{Correa:2019rdk,Bianchi:2020hsz}. Here too, a fishnet limit exists with Feynman graphs which look like a triangular lattice~\cite{Caetano:2016ydc} and one can envisage the definition and study of a similar CFT wavefunction like the one we studied in this paper. 

Finally it would be interesting to look at other examples of integrable boundary conditions in ${\cal N} = 4$ SYM. These include determinant operators~\cite{Bajnok:2012xc,Bajnok:2013wsa,Jiang:2019xdz,Jiang:2019zig} (and references therein) and giant Wilson loops~\cite{Giombi:2020amn}.

\section*{Acknowledgements}
We are especially grateful to A.~Cavaglia for
numerous discussions and for carefully reading the manuscript prior to its publication. 
N.G. is also grateful to F.~Levkovich-Maslyuk,
 A.~Sever, E.~Sobko
 and A.~Tumanov for discussions on various stages of this project.
The work of N.G. was supported by European Research Council (ERC) under the European Union’s Horizon 2020 research and innovation programme (grant agreement No. 865075)  EXACTC. \appendix

\section{Proof of Poisson-commutativity of $\mathbb{T}(u)$}
\label{Appendixcommutes}
In this appendix, we prove that the classical transfer matrix \eqref{Tfund} forms a family of mutually Poisson-commuting functions for any value of the spectral parameter and for any $J$.
We start by the $J=0$ case. We have that:
\begin{equation}
    \begin{split}
        \{\mathbb{T}(u),\mathbb{T}(v)\}=&\{\mathbb{K}_{ab}(u),\mathbb{K}_{\alpha\beta}(v)\}\bar{\mathbb{K}}^{ba}(u)\bar{\mathbb{K}}^{\beta\alpha}(v)+\mathbb{K}_{ab}(u)\mathbb{K}_{\alpha\beta}(v)\{\bar{\mathbb{K}}^{ba}(u),\bar{\mathbb{K}}^{\beta\alpha}(v)\}\,.
    \end{split}
\end{equation}
Using \eqref{KPB} and its analog for $\bar{\mathbb{K}}$:
\begin{equation}
\begin{split}
    \{\mathbb{T}(u),\mathbb{T}(v)\}=&-\frac{\bar{\mathbb{K}}^{ba}(u)\bar{\mathbb{K}}^{\beta\alpha}(v)}{\xi(u+v)}\left[\mathbb{K}_{\beta b}(u)\mathbb{K}_{\alpha a}(v)-\mathbb{K}_{b\beta}(v)\mathbb{K}_{a\alpha}(u) \right]+\\&+\frac{\bar{\mathbb{K}}^{ba}(u)\bar{\mathbb{K}}^{\beta\alpha}(v)}{\xi(u-v)}\left[\mathbb{K}_{a \beta}(u)\mathbb{K}_{\alpha b}(v)-\mathbb{K}_{a \beta}(v)\mathbb{K}_{\alpha b}(u) \right]-\\&
    -\frac{\mathbb{K}_{ab}(u)\mathbb{K}_{\alpha\beta}(v)}{\xi(u+v)}\left[\bar{\mathbb{K}}^{\alpha a}(u)\bar{\mathbb{K}}^{\beta b}(v)-\bar{\mathbb{K}}^{a\alpha}(v)\bar{\mathbb{K}}^{b\beta}(u) \right]+\\&
    +\frac{\mathbb{K}_{ab}(u)\mathbb{K}_{\alpha\beta}(v)}{\xi(u-v)}\left[\bar{\mathbb{K}}^{b\alpha}(u)\bar{\mathbb{K}}^{\beta a}(v)-\bar{\mathbb{K}}^{b\alpha}(v)\bar{\mathbb{K}}^{\beta a}(u) \right]\,.
    \end{split}
\end{equation}
By relabelling indices appropriately, all terms cancel as expected.

For the rest of this section, we will use the shorthand notation:
\begin{equation}
    (\mathbb{L})^a_{\,\,b}(u)\equiv(\mathbb{L}_1(u).\mathbb{L}_2(u)\dots\mathbb{L}_J(u))^a_{\,\,b}\,,\qquad (\bar{\mathbb{L}})_a^{\,\,b}(u)\equiv(\mathbb{L}_{-J}(u).\mathbb{L}_{-(J-1)}(u)\dots\mathbb{L}_{-1}(u))_a^{\,\,b}\,.
\end{equation}
It is easy to see that these matrices follow the same Poisson brackets as individual $\mathbb{L}$-matrices, i.e. \eqref{PB1}. Thus, for general $J$ we have that:
\begin{equation}
\begin{split}
    \{\mathbb{T}(u),\mathbb{T}(v)\}=&\{\mathbb{K}_{ab}(u),\mathbb{K}_{\alpha\beta}(v)\}(\bar{\mathbb{L}})_c^{\,\,a}(u)(\bar{\mathbb{L}})_{\gamma}^{\,\,\alpha}(v)\bar{\mathbb{K}}^{dc}(u)\bar{\mathbb{K}}^{\delta\gamma}(v)(\mathbb{L})^b_{\,\,d}(u)(\mathbb{L})^{\beta}_{\,\,\delta}(v)+\\&
    \mathbb{K}_{ab}(u)\mathbb{K}_{\alpha\beta}(v)(\bar{\mathbb{L}})_c^{\,\,a}(u)(\bar{\mathbb{L}})_{\gamma}^{\,\,\alpha}(v)\{\bar{\mathbb{K}}^{dc}(u),\bar{\mathbb{K}}^{\delta\gamma}(v)\}(\mathbb{L})^b_{\,\,d}(u)(\mathbb{L})^{\beta}_{\,\,\delta}(v)+\\&
    +(\text{Poisson brackets between  }\mathbb{L})=\\&
    =(\bar{\mathbb{L}})_c^{\,\,a}(u)(\bar{\mathbb{L}})_{\gamma}^{\,\,\alpha}(v)(\mathbb{L})^b_{\,\,d}(u)(\mathbb{L})^{\beta}_{\,\,\delta}(v)\\&\Bigg\{-\frac{1}{\xi(u+v)}\bigg[\mathbb{K}_{\beta b}(u)\mathbb{K}_{\alpha a}(v)\bar{\mathbb{K}}^{dc}(u)\bar{\mathbb{K}}^{\delta\gamma}(v)-\mathbb{K}_{b\beta }(v)\mathbb{K}_{a\alpha }(u)\bar{\mathbb{K}}^{dc}(u)\bar{\mathbb{K}}^{\delta\gamma}(v) +\\&
    +\mathbb{K}_{a b}(u)\mathbb{K}_{\alpha \beta}(v)\bar{\mathbb{K}}^{\gamma c}(u)\bar{\mathbb{K}}^{\delta d}(v)-\mathbb{K}_{a b}(u)\mathbb{K}_{\alpha \beta}(v)\bar{\mathbb{K}}^{c\gamma}(v)\bar{\mathbb{K}}^{d\delta }(u) \bigg]+
    \\&+\frac{1}{\xi(u-v)}\bigg[\mathbb{K}_{a\beta}(u)\mathbb{K}_{\alpha b}(v)\bar{\mathbb{K}}^{dc}(u)\bar{\mathbb{K}}^{\delta\gamma}(v)-\mathbb{K}_{a\beta }(v)\mathbb{K}_{\alpha b}(u)\bar{\mathbb{K}}^{dc}(u)\bar{\mathbb{K}}^{\delta\gamma}(v) +\\&
    +\mathbb{K}_{a b}(u)\mathbb{K}_{\alpha \beta}(v)\bar{\mathbb{K}}^{d\gamma }(u)\bar{\mathbb{K}}^{\delta c}(v)-\mathbb{K}_{a b}(u)\mathbb{K}_{\alpha \beta}(v)\bar{\mathbb{K}}^{d\gamma}(v)\bar{\mathbb{K}}^{\delta c}(u) \bigg]\Bigg\}+\\&
   +(\text{Poisson brackets between  }\mathbb{L})\,.
    \end{split}
\end{equation}
We now relabel indices in order to collect the boundary reflection matrices as:
\begin{equation}
\begin{split}
    \{\mathbb{T}(u),\mathbb{T}(v)\}=&\mathbb{K}_{a b}(u)\mathbb{K}_{\alpha \beta}(v)\bar{\mathbb{K}}^{dc}(u)\bar{\mathbb{K}}^{\delta\gamma}(v)\\&\Bigg\{-\frac{1}{\xi(u+v)}\bigg[(\bar{\mathbb{L}})_c^{\,\,\beta}(u)(\bar{\mathbb{L}})_{\gamma}^{\,\,\alpha}(v)(\mathbb{L})^b_{\,\,d}(u)(\mathbb{L})^{a}_{\,\,\delta}(v)-\\&
    -(\bar{\mathbb{L}})_{\delta}^{\,\,a}(u)(\bar{\mathbb{L}})_{\gamma}^{\,\,\alpha}(v)(\mathbb{L})^b_{\,\,d}(u)(\mathbb{L})^{\beta}_{\,\,c}(v)+\\& +(\bar{\mathbb{L}})_c^{\,\,a}(u)(\bar{\mathbb{L}})_{d}^{\,\,\alpha}(v)(\mathbb{L})^b_{\,\,\gamma}(u)(\mathbb{L})^{\beta}_{\,\,\delta}(v)-\\&
    -(\bar{\mathbb{L}})_c^{\,\,a}(u)(\bar{\mathbb{L}})_{\gamma}^{\,\,b}(v)(\mathbb{L})^{\alpha}_{\,\,d}(u)(\mathbb{L})^{\beta}_{\,\,\delta}(v) \bigg]+
    \\&+\frac{1}{\xi(u-v)}\bigg[(\bar{\mathbb{L}})_c^{\,\,a}(u)(\bar{\mathbb{L}})_{\gamma}^{\,\,\alpha}(v)(\mathbb{L})^{\beta}_{\,\,d}(u)(\mathbb{L})^{b}_{\,\,\delta}(v)-\\&
    -(\bar{\mathbb{L}})_c^{\,\,\alpha}(u)(\bar{\mathbb{L}})_{\gamma}^{\,\,a}(v)(\mathbb{L})^b_{\,\,d}(u)(\mathbb{L})^{\beta}_{\,\,\delta}(v)+\\&+(\bar{\mathbb{L}})_{\gamma}^{\,\,a}(u)(\bar{\mathbb{L}})_{c}^{\,\,\alpha}(v)(\mathbb{L})^b_{\,\,d}(u)(\mathbb{L})^{\beta}_{\,\,\delta}(v)-\\&
    -(\bar{\mathbb{L}})_c^{\,\,a}(u)(\bar{\mathbb{L}})_{\gamma}^{\,\,\alpha}(v)(\mathbb{L})^b_{\,\,\delta}(u)(\mathbb{L})^{\beta}_{\,\,d}(v) \bigg]\Bigg\}+\\&
    +(\text{Poisson brackets between  }\mathbb{L})\,.
    \end{split}
\end{equation}
The Poisson Brackets between $\mathbb{L}$-matrices give:
\begin{equation}
    \begin{split}
         \{\mathbb{T}(u),\mathbb{T}(v)\}=&\text{(Poisson brackets between  }\mathbb{K})+\\&+\mathbb{K}_{a b}(u)\mathbb{K}_{\alpha \beta}(v)\bar{\mathbb{K}}^{dc}(u)\bar{\mathbb{K}}^{\delta\gamma}(v)\\&
         \Bigg[\{(\bar{\mathbb{L}})_c^{\,\,a}(u),(\bar{\mathbb{L}})_{\gamma}^{\,\,\alpha}(v)\}(\mathbb{L})^b_{\,\,d}(u)(\mathbb{L})^{\beta}_{\,\,\delta}(v)+\\&
        +(\bar{\mathbb{L}})_{\gamma}^{\,\,\alpha}(v)\{(\bar{\mathbb{L}})_c^{\,\,a}(u),(\mathbb{L})^{\beta}_{\,\,\delta}(v)\}(\mathbb{L})^b_{\,\,d}(u)+\\&
         +(\bar{\mathbb{L}})_c^{\,\,a}(u)(\bar{\mathbb{L}})_{\gamma}^{\,\,\alpha}(v)\{(\mathbb{L})^b_{\,\,d}(u),(\mathbb{L})^{\beta}_{\,\,\delta}(v)\}+\\&
        +(\bar{\mathbb{L}})_c^{\,\,a}(u)\{(\mathbb{L})^b_{\,\,d}(u),(\bar{\mathbb{L}})_{\gamma}^{\,\,\alpha}(v)\}(\mathbb{L})^{\beta}_{\,\,\delta}(v)
         \Bigg]\,.
    \end{split}
\end{equation}
Using the Poisson brackets \eqref{PB1} and their analogues:
\begin{equation}
\label{PB2}
     \xi\,\{(\mathbb{L}_{-n})_{a}^{\,\,b}(u),(\mathbb{L}_{-m})_{c}^{\,\,d}(v)\}=\dfrac{(\mathbb{L}_{-n})_{a}^{\,\,d}(u)(\mathbb{L}_{-n})_{c}^{\,\,b}(v)-(\mathbb{L}_{-n})_{c}^{\,\,b}(u)(\mathbb{L}_{-n})_{a}^{\,\,d}(v)}{u-v}\,\delta_{nm}\,,
 \end{equation}
\begin{equation}
\label{PB3}
    \xi\,\{(\mathbb{L}_{-n})_{a}^{\;\; b}(u),(\mathbb{L}_{m})_{\;\; d}^{c}(v)\} = \frac{(\mathbb{L}_{-n})_{a}^{\;\; c}(u)(\mathbb{L}_{n})_{\;\; d}^{b}(v)-(\mathbb{L}_{-n})_{d}^{\;\; b}(u)(\mathbb{L}_{n})_{\;\; a}^{c}(v)}{u + v}\delta_{m\,n}\,,
\end{equation}
we get:
\begin{equation}
    \begin{split}
     \{\mathbb{T}(u),\mathbb{T}(v)\}=&\text{(Poisson brackets between  }\mathbb{K})+\\&+\mathbb{K}_{a b}(u)\mathbb{K}_{\alpha \beta}(v)\bar{\mathbb{K}}^{dc}(u)\bar{\mathbb{K}}^{\delta\gamma}(v)\\&
         \Bigg\{
       \frac{1}{\xi(u-v)}\bigg[
       (\bar{\mathbb{L}})_c^{\,\,\alpha}(u)(\bar{\mathbb{L}})_{\gamma}^{\,\,a}(v)(\mathbb{L})^b_{\,\,d}(u)(\mathbb{L})^{\beta}_{\,\,\delta}(v)-\\&
       -(\bar{\mathbb{L}})_{\gamma}^{\,\,a}(u)(\bar{\mathbb{L}})_{c}^{\,\,\alpha}(v)(\mathbb{L})^b_{\,\,d}(u)(\mathbb{L})^{\beta}_{\,\,\delta}(v)+\\&
       +(\bar{\mathbb{L}})_{c}^{\,\,a}(u)(\bar{\mathbb{L}})_{\gamma}^{\,\,\alpha}(v)(\mathbb{L})^b_{\,\,\delta}(u)(\mathbb{L})^{\beta}_{\,\,d}(v)-\\&
       -(\bar{\mathbb{L}})_{c}^{\,\,a}(u)(\bar{\mathbb{L}})_{\gamma}^{\,\,\alpha}(v)(\mathbb{L})^{\beta}_{\,\,d}(u)(\mathbb{L})^{b}_{\,\,\delta}(v)
       \bigg]  +\\&
       +\frac{1}{\xi(u+v)}\bigg[
        (\bar{\mathbb{L}})_c^{\,\,\beta}(u)(\bar{\mathbb{L}})_{\gamma}^{\,\,\alpha}(v)(\mathbb{L})^b_{\,\,d}(u)(\mathbb{L})^{a}_{\,\,\delta}(v)-\\&
         -(\bar{\mathbb{L}})_{\delta}^{\,\,a}(u)(\bar{\mathbb{L}})_{\gamma}^{\,\,\alpha}(v)(\mathbb{L})^b_{\,\,d}(u)(\mathbb{L})^{\beta}_{\,\,c}(v)-\\&
         -(\bar{\mathbb{L}})_c^{\,\,a}(u)(\bar{\mathbb{L}})_{\gamma}^{\,\,b}(v)(\mathbb{L})^{\alpha}_{\,\,d}(u)(\mathbb{L})^{\beta}_{\,\,\delta}(v)+\\&
         +(\bar{\mathbb{L}})_c^{\,\,a}(u)(\bar{\mathbb{L}})_{d}^{\,\,\alpha}(v)(\mathbb{L})^{b}_{\,\,\gamma}(u)(\mathbb{L})^{\beta}_{\,\,\delta}(v)
       \bigg]
\Bigg\}\,.
    \end{split}
\end{equation}
It is easy to verify that the terms from the Poisson Brackets of $\mathbb{L}$-matrices cancel exactly the ones from the Poisson brackets of $\mathbb{K}$-matrices. Therefore, the transfer matrices form a family of functions in convolution between themselves:
\begin{equation}
    \{\mathbb{T}(u),\mathbb{T}(v)\}=0\,.
\end{equation}

\section{Parity of quantum transfer matrices}
In this appendix we will prove explicitly the parity of the quantum transfer matrices in all the antisymmetric representations of the auxiliary space.
\label{Appendixparity}
\subsection*{Parity of $\mathbb{T}^4$}
We need to evaluate:
\begin{equation}
\mathbb{\hat{T}}^{\bf4}(-u)=\text{Tr}(
{\hat{\bar{\mathbb{L}}}_J^{\bf{4}}}(u).{\hat{\bar{\mathbb{L}}}_{J-1}^{\bf{4}}}(u)\dots{\hat{\bar{\mathbb{L}}}_1^{\bf{4}}}(u).{\mathbb{\hat{K}}^{\bf 4}}(-u).
{\mathbb{\hat{L}}_{1}^{\bf 4}}(-u).{\mathbb{\hat{L}}_{2}^{\bf 4}(-u)\dots{\mathbb{\hat{L}}_{J}^{\bf 4}}}(-u).G^{\bf{4}}.\hat{\bar{\mathbb{K}}}^{\bf 4}(-u).G^{{\bf{4}}\,T})\,.
\end{equation}
Transposing inside of the trace:
\begin{equation}
\begin{split}
\mathbb{\hat{T}}^{\bf4}(-u)=&\text{Tr}(G^{\bf{4}}.\hat{\bar{\mathbb{K}}}^{{\bf 4}\,T}(-u).G^{{\bf{4}}\,T}.
\mathbb{\hat{L}}_J^{{\bf{4}}\,T}(-u).\mathbb{\hat{L}}_{J-1}^{{\bf{4}}\,T}(-u)\dots \mathbb{\hat{L}}_1^{{\bf{4}}\,T}(-u).\\&.\mathbb{\hat{K}}^{{\bf{4}}\,T}(-u).\hat{\bar{\mathbb{L}}}_{1}^{{\bf{4}}\,T}(u).\hat{\bar{L}}_{2}^{{\bf{4}}\,T}(u)\dots\hat{\bar{L}}_{J}^{{\bf{4}}\,T}(u)) \,.
\end{split}
\end{equation}
Now since $\mathbb{\hat{L}}_j^{{\bf{4}}\,T}(-u)=-\hat{\bar{\mathbb{L}}}_j^{\bf{4}}(-u)$ and $\hat{\bar{\mathbb{L}}}_{j}^{{\bf{4}}\,T}(u)=-\mathbb{\hat{L}}_{1}^{\bf{4}}(u)$ we get:
\begin{equation}
\label{eqn:Tparity1}
\mathbb{\hat{T}}^{\bf4}(-u)=\text{Tr}(
{\hat{\bar{\mathbb{L}}}_J^{\bf{4}}}(-u).{\hat{\bar{\mathbb{L}}}_{J-1}^{\bf{4}}}(-u)\dots{\hat{\bar{\mathbb{L}}}_1^{\bf{4}}}(-u).{\mathbb{\hat{K}}^{{\bf 4}\,T}}(-u).
{\mathbb{\hat{L}}_{1}^{\bf 4}}(u).{\mathbb{\hat{L}}_{2}^{\bf 4}(u)\dots{\mathbb{\hat{L}}_{J}^{\bf 4}}}(u).G^{\bf{4}}.\hat{\bar{\mathbb{K}}}^{{\bf 4}\,T}(-u).G^{{\bf{4}}\,T})\,.
\end{equation}
We can now insert a pair of $\bar{S}$-matrices near the $\mathbb{\hat{K}}$-operator using the unitarity condition $\bar{S}(2u)\bar{S}(-2u)=I$ and then commute $\bar{S}(2u)$ through the $\mathbb{\hat{L}}$-operators using the Yang-Baxter equation, obtaining:
\begin{equation}
\begin{split}
\mathbb{\hat{T}}^{\bf4}(-u)=&\text{Tr}(
{\hat{\bar{\mathbb{L}}}_J^{\bf{4}}}(-u).{\hat{\bar{\mathbb{L}}}_{J-1}^{\bf{4}}}(-u)\dots{\hat{\bar{\mathbb{L}}}_1^{\bf{4}}}(-u).{\mathbb{\hat{K}}^{{\bf 4}\,T}}(-u).\bar{S}(-2u).\\&.
{\mathbb{\hat{L}}_{1}^{\bf 4}}(u).{\mathbb{\hat{L}}_{2}^{\bf 4}(u)\dots{\mathbb{\hat{L}}_{J}^{\bf 4}}}(u).G^{\bf{4}}.\bar{S}(2u).\hat{\bar{\mathbb{K}}}^{{\bf 4}\,T}(-u).G^{{\bf{4}}\,T})\,.
\end{split}
\end{equation}
Using the following identities:
\begin{equation}\label{eqn:KtomK}
    \bar{\alpha}(2u)\bar{R}^{c\,d}_{\,a\,b}(2u)\mathbb{\hat{K}}^{\bf{4}}_{cd}(-u)=\mathbb{\hat{K}}^{\bf{4}}_{ba}(u)\,,
\end{equation}
\begin{equation}\label{eqn:KtomK2}
    \bar{\alpha}(-2u)\bar{R}^{a\,b}_{\,d\,c}(-2u)(\hat{\bar{\mathbb{K}}}^{\bf{4}})^{dc}(-u)=(\hat{\bar{\mathbb{K}}}^{\bf{4}})^{ba}(u)\,,
\end{equation}
which in matrix notation are:
\begin{equation}
\label{eqn:SK1}
    \mathbb{\hat{K}}^{{\bf{4}}\,T}(-u).\bar{S}(2u)= \mathbb{\hat{K}}^{\bf{4}}(u)\,,
\end{equation}
\begin{equation}
\label{eqn:SK2}
   \bar{S}(-2u).\hat{\bar{\mathbb{K}}}^{{\bf{4}}\,T}(-u)= \hat{\bar{\mathbb{K}}}^{\bf{4}}(u)\,,
\end{equation}

we obtain that $\mathbb{\hat{T}}^{\bf4}(-u)=\mathbb{\hat{T}}^{\bf4}(u)$, thus seeing that $\mathbb{\hat{T}}^4$ is even for any $J$.

We will now give a more detailed proof of the last passage above.
We will use the YBE in figure~\ref{fig:RLLM} (substituting $R(-u)$ with $\bar{S}(u)$), the unitarity condition \eqref{unitarity} and the identities \eqref{eqn:KtomK} and \eqref{eqn:KtomK2}. Starting from \eqref{eqn:Tparity1} we insert an identity and we use \eqref{eqn:unitarity} to get:
\begin{equation}
\begin{split}
    \hat{\mathbb{T}}^{\bf4}(-u)=&(\hat{\bar{\mathbb{L}}}_J^{\bf{4}}(-u).{\hat{\bar{\mathbb{L}}}_{J-1}^{\bf{4}}}(-u)\dots{\hat{\bar{\mathbb{L}}}_1^{\bf{4}}}(-u))_a^{\,\,\,\gamma}\bar{S}^{\delta\,\,\alpha}_{\,\,\zeta\,\,\eta}(2u)\bar{S}^{\zeta\,\,\eta}_{\,\,\gamma\,\,\beta}(-2u)\hat{\mathbb{K}}^{\bf{4}}_{\alpha\delta}(-u)\\&({\hat{\mathbb{L}}_{1}^{\bf 4}}(u).{\hat{\mathbb{L}}_{2}^{\bf 4}(u)\dots{\hat{\mathbb{L}}_{J}^{\bf 4}}}(u))^{\beta}_{\,\,h}(G^{\bf{4}}.\hat{\bar{\mathbb{K}}}^{{\bf 4}\,T}(-u).G^{{\bf{4}}\,T})^{ha}\,.
    \end{split}
\end{equation}
We can now use \eqref{eqn:KtomK} to get:
\begin{equation}
\begin{split}
    \mathbb{\hat{T}}^{\bf4}(-u)=&({\hat{\bar{\mathbb{L}}}_J^{\bf{4}}}(-u).{\hat{\bar{\mathbb{L}}}_{J-1}^{\bf{4}}}(-u)\dots{\hat{\bar{\mathbb{L}}}_1^{\bf{4}}}(-u))_a^{\,\,\,\gamma}\bar{S}^{\zeta\,\,\eta}_{\,\,\gamma\,\,\beta}(-2u)\mathbb{\hat{K}}^{\bf{4}}_{\zeta\eta}(u)\\&({\mathbb{\hat{L}}_{1}^{\bf 4}}(u).{\mathbb{\hat{L}}_{2}^{\bf 4}(u)\dots{\mathbb{\hat{L}}_{J}^{\bf 4}}}(u))^{\beta}_{\,\,h}(G^{\bf{4}}.\hat{\bar{\mathbb{K}}}^{{\bf 4}\,T}(-u).G^{{\bf{4}}\,T})^{ha}\,.
    \end{split}
\end{equation}
We now use YBE in figure~\ref{fig:RLLM} (substituting $R(-u)$ with $\bar{S}(u)$) to commute the remaining S-matrix through all the $\mathbb{\hat{L}}_i$ and $\hat{\bar{\mathbb{L}}}_i$, obtaining for $i=1$:
\begin{equation}
\begin{split}
    \mathbb{\hat{T}}^{\bf4}(-u)=&({\hat{\bar{\mathbb{L}}}_J^{\bf{4}}}(-u).{\hat{\bar{\mathbb{L}}}_{J-1}^{\bf{4}}}(-u)\dots{\hat{\bar{\mathbb{L}}}_2^{\bf{4}}}(-u))_{a}^{\,\,\,\omega}\,\bar{S}^{\gamma\,\,\beta}_{\,\,\epsilon\,\,\omega}(-2u)\mathbb{\hat{L}}^{\bf4\,\,\eta}_{1\,\,\,\,\,\,\gamma}(u)\mathbb{\hat{L}}^{\bf4\,\,\,\,\,\,\zeta}_{1\,\,\beta}(-u) \mathbb{\hat{K}}^{\bf{4}}_{\zeta\eta}(u)\\&({\mathbb{\hat{L}}_{2}^{\bf 4}(u)\dots{\mathbb{\hat{L}}_{J}^{\bf 4}}}(u))^{\epsilon}_{\,\,h}(G^{\bf{4}}.\hat{\bar{\mathbb{K}}}^{{\bf 4}\,T}(-u).G^{{\bf{4}}\,T})^{ha}\,.
    \end{split}
\end{equation}
Continuing this process for $\forall i=2...J$ we obtain:
\begin{equation}
\begin{split}
   \mathbb{\hat{T}}^{\bf4}(-u)=&({\hat{\bar{\mathbb{L}}}_J^{\bf{4}}}(-u).{\hat{\bar{\mathbb{L}}}_{J-1}^{\bf{4}}}(-u)\dots{\hat{\bar{\mathbb{L}}}_1^{\bf{4}}}(-u).\mathbb{\hat{K}}^{\bf{4}}(u).{\mathbb{\hat{L}}_{1}^{\bf 4}}(u).{\mathbb{\hat{L}}_{2}^{\bf 4}(u)\dots{\mathbb{\hat{L}}_{J}^{\bf 4}}}(u))_{\epsilon\,\omega}\\&\bar{S}^{\epsilon\,\,\omega}_{\,\,h\,\,a}(-2u)(G^{\bf{4}}.\hat{\bar{\mathbb{K}}}^{{\bf 4}\,T}(-u).G^{{\bf{4}}\,T})^{ha}\,.
    \end{split}
\end{equation}
Finally we use \eqref{eqn:KtomK2} and get: 
\begin{equation}
\begin{split}
    \mathbb{\hat{T}}^{\bf4}(-u)=&({\hat{\bar{\mathbb{L}}}_J^{\bf{4}}}(-u).{\hat{\bar{\mathbb{L}}}_{J-1}^{\bf{4}}}(-u)\dots{\hat{\bar{\mathbb{L}}}_1^{\bf{4}}}(-u).\mathbb{\hat{K}}^{\bf{4}}(u).{\mathbb{\hat{L}}_{1}^{\bf 4}}(u).{\mathbb{\hat{L}}_{2}^{\bf 4}(u)\dots{\mathbb{\hat{L}}_{J}^{\bf 4}}}(u))_{\epsilon\,\omega}\\&(G^{\bf{4}}.\hat{\bar{\mathbb{K}}}^{\bf 4}(u).G^{{\bf{4}}\,T})^{\omega\,\epsilon}=\mathbb{\hat{T}}^{\bf4}(u)\,.
    \end{split}
\end{equation}
\subsection*{Parity of $\mathbb{T}^{\bf6}$}
Remembering that $\mathbb{\bar{L}}^{\bf6}_i(u)=\mathbb{L}^{\bf6}_i(-u)$, we write:
\begin{equation}
    \mathbb{\hat{T}}^{\bf6}(u)=\text{Tr}\left(\mathbb{\hat{L}}_J^{\bf6}(u)\dots\mathbb{\hat{L}}_1^{\bf6}(u).\mathbb{\hat{K}}^{\bf6}(u).\mathbb{\hat{L}}_1^{\bf6}(u)\dots\mathbb{\hat{L}}_J^{\bf6}(u).G^{\bf6}.\hat{\bar{\mathbb{K}}}^{\bf6}(u).G^{6\,T} \right)\,.
\end{equation}
Hence we have that:
\begin{equation}
    \mathbb{\hat{T}}^{\bf6}(-u)=\text{Tr}\left(\mathbb{\hat{L}}_J^{\bf6}(-u)\dots\mathbb{\hat{L}}_1^{\bf6}(-u).\mathbb{\hat{K}}^{\bf6}(-u).\mathbb{\hat{L}}_1^{\bf6}(-u)\dots\mathbb{\hat{L}}_J^{\bf6}(-u).G^{\bf6}.\hat{\bar{\mathbb{K}}}^{\bf6}(-u).G^{{\bf6}\,T} \right)\,.
\end{equation}
Taking a transpose inside the trace and noticing from definition \eqref{eqn:L6} that $\mathbb{\hat{L}}^{{\bf6}\,T}_i(-u)=\mathbb{\hat{L}}^{\bf6}_i(u)$:
\begin{equation}
\label{eqn:T6pari1}
    \mathbb{\hat{T}}^{\bf6}(-u)=\text{Tr}\left(\mathbb{\hat{L}}_J^{\bf6}(u)\dots\mathbb{\hat{L}}_1^{\bf6}(u).\mathbb{\hat{K}}^{6\,T}(-u).\mathbb{\hat{L}}_1^{\bf6}(u)\dots\mathbb{\hat{L}}_J^{\bf6}(u).G^{\bf6}.\hat{\bar{\mathbb{K}}}^{{\bf6}\,T}(-u).G^{{\bf6}\,T} \right)\,.
\end{equation}
 We now need identities analogous to \eqref{eqn:KtomK} for $\bf{6}$ irrep.
First, we need the $\bar{S}^{\bf6}$-matrix. Making the ansatz that it is formed by all compatible indices structures, we can fix the relative coefficients by requiring that it satisfies the Yang-Baxter equation:
\beq
\hat{\mathbb{L}}_{\;\;\;\;\,E}^{{\bf6}\,B}(u)\hat{\bar{\mathbb{L}}}^{{\bf6}\,\;\;D}_{\;\;F}(-v)\bar{S}_{\;\;\,\,A\,C}^{{\bf6}E\,F}(u+v)=\bar{S}_{\;\;\,\,F\,E}^{{\bf6}B\,D}(u+v)\hat{\bar{\mathbb{L}}}^{{\bf6}\,\;\;E}_{\;\;C}(-v)\mathbb{\hat{L}}_{\;\;\;\;\,A}^{{\bf6}\,F}(u)\,.
\eeq
We get that:
\beq
\bar{S}^{{\bf6}\,A\,\,\,C}_{\;\;\;\;\,B\,\,\,D}(u)=
c(u) \left(\delta^{A}_{\,B}\, \delta^C_{\,D}-\frac{i}{\,\xi\,  u}\delta^A_{\,D}\, \delta^C_{\,B}+\frac{i}{\,\xi  \left(u-\frac{2i}{\xi}\right)}\eta^{AC}\, \eta_{BD}\right)\,.
\eeq
 The overall coefficient $c(u)$ is fixed by unitarity,
$
 \bar{S}^{{\bf6}\,A\,\,\,C}_{\;\;\;\;\,B\,\,\,D}(u)\bar{S}^{{\bf6}\,B\,\,\,D}_{\;\;\;\;\,E\,\,\,F}(-u)=\delta^A_{\,E}\,\delta^C_{\,F}$, as:
 \beq
 c(u)=\frac{u \left(u-\frac{2i}{\xi }\right)}{\frac{2}{\xi^2}+u\left(u+\frac{i}{\xi}\right) }\,.
 \eeq
The identities we need are:
\beqa
&\hat{\mathbb{K}}^{{\bf{6}}\,T}(-u).\bar{S}^{\bf6}(2u)=\hat{\mathbb{K}}^{{\bf{6}}}(u)\,,\\&
\bar{S}^{\bf6}(-2u).\hat{\bar{\mathbb{K}}}^{{\bf{6}}\,T}(-u)=\hat{\bar{\mathbb{K}}}^{{\bf{6}}}(u)\,.
\eeqa
Hence, inserting into \eqref{eqn:T6pari1} a pair of $\bar{S}$-matrices via unitarity and repeating the passages of the section above, it is easy to prove that:
\beq
\hat{\mathbb{T}}^{\bf6}(-u)=\hat{\mathbb{T}}^{\bf6}(u)\,.
\eeq

\subsection*{Parity of $\mathbb{T}^{\bf\bar{4}}$}
Using the definitions of section~\ref{T4barstuff} one can rewrite $\mathbb{\hat{T}}^{\bf{\bar{4}}}(u)$ in terms of $\mathbb{\hat{L}}^{\bf4}$-operators and $\mathbb{\hat{K}}^4$-operators as: 
\begin{equation}
\mathbb{\hat{T}}^{\bf{\bar{4}}}(u)=\hat{\beta}(u)\,\text{Tr}(
\hat{\bar{\mathbb{L}}}_1^{\bf{4}}(u)\dots \hat{\bar{\mathbb{L}}}_J^{\bf{4}}(u).(G^{\bf4})^{-T}.\mathbb{\hat{K}}^{\bf 4}(-u).(G^{\bf4})^{-1}.
\mathbb{\hat{L}}_{J}^{\bf 4}(-u)\dots\mathbb{\hat{L}}_{1}^{\bf 4}(-u).\hat{\bar{\mathbb{K}}}^{\bf 4}(-u))\,.
\end{equation}
Where $\hat{\beta}(u)$ is an operator which is an even polynomial in $u$, composed by all the prefactors appearing in the definitions of the $\bf\bar{4}$ operators. Following the same passages used for the parity of $\mathbb{T}^{\bf4}$, it is then easy to prove that:
\begin{equation}
    \mathbb{T}^{\bf\bar{4}}(-u)=\mathbb{T}^{\bf\bar{4}}(u)\,.
\end{equation}
\subsection*{Parity of $\mathbb{T}^{\bf\bar{1}}$}
From the definitions of section~\ref{1barstuff}, it is evident that the $\mathbb{L}^{\bf\bar{1}}$-operators are even polynomials in $u$.
Also, since the $\mathbb{K}^{\bf\bar{1}}$-operators are proportional to the identity operator, we can move them together through all $\mathbb{L}^{\bf\bar{1}}$. Their product is:
\begin{equation}
    \mathbb{K}^{\bf\bar{1}}(u)\mathbb{\bar{K}}^{\bf\bar{1}}(u)=\left(u^2+\frac{4}{\xi^2} \right)\left(u^2+\frac{1}{\xi^2} \right)^2\,u^2\,.
\end{equation}
Hence, $\mathbb{T}^{\bf\bar{1}}(u)$ is even as it is a product of even functions. 

\section{Explicit form of transfer matrices}
\subsection*{$J=0$ case}\label{sec:J0T}
The polynomials $P$ that enter the transfer matrices for the $J = 0$ case in \eq{eqn:J0T} are
\begin{align}
\begin{split}
    P^{\bf 4} = P^{\bf \bar{4}} &= 4\,\cos\varphi\,v^2 - 8\,h + \cos\varphi\;,\\
    P^{\bf 6} &= (2 \cos2\varphi + 4) v^4 -(4\,\Delta^2\sin^2\varphi+16\,h\,\cos\varphi)v^2 + 16\,h^2\;,
\end{split}
\end{align}
where $h =- \hat{g}^{2}\,B^{-1}$. The above expressions lead to the Baxter equation \eq{J0Bax}.

In order to make a comparison with 
\cite{Gromov:2016rrp,Cavaglia:2018lxi}
we introduce the notation
for the final difference operators $\hat O_\pm$
\beq
\hat O_\pm q\equiv q(u) \left(4 \hat{g}^2-2 u^2 \cos (\phi )\pm 2 \Delta  u \sin (\phi )\right)+u^2 q(u-i)+u^2
   q(u+i)\;.
\eeq
The second order equations used in 
\cite{Gromov:2016rrp,Cavaglia:2018lxi} was of the form $\hat O_\pm q = 0$. At the same time the forth order equation \eq{J0Bax} 
can we written as
\beq
-\frac{1}{u^2}\hat O_+ \frac{1}{u} \hat O_- q =
-\frac{1}{u^2}\hat O_- \frac{1}{u} \hat O_+ q = 0\;.
\eeq
We see that the four independent solutions
of the two second order equations $\hat O_\pm q = 0$ are the $4$ solutions of 
\eq{J0Bax}, which indeed demonstrates their equivalence.

\subsection*{$J=1$ case}\label{sec:J1T}
\label{J1res}
For the $J = 1$ case we explicitly built all 
$3$ transfer matrices as differential operators acting on the CFT wavefunction of $6$ variables $s,t,\vec x_1$. 
We verified the general analytic properties outlined in the text in section~\ref{Tprops}.
Furthermore, we found some additional relations  between the coefficients as shown below
\beq
\begin{split}
   & P^{\bf 4}=4\cos{\varphi}\,v^4+a_1\,v^2+\frac{a_1}{4}+8h-\frac{1}{4}\cos{\varphi}\,, \\&
P^{\bf{6}}=(2\cos{2\varphi}+4)\,v^8+2\bigg[\cos{\varphi}\,\frac{a_1+c_{1}}{2}+\cos{2\varphi}\,\Delta^2-(\Delta^2+1)\\&-2S^2\sin^2{\varphi}\bigg]\,
v^6+b_2\,v^4+b_3\,v^2+16\,h^2\,,\\&
    P^{\bar{\bf 4}}=4\cos{\varphi}\,v^4+c_1\,v^2+\frac{c_1}{4}+8h-\frac{1}{4}\cos{\varphi}\,.
\end{split}
\eeq
Here, $h=-\hat g^4 B^{-1}\simeq -\hat g^4$ so we can see that the relation \eq{BT} does hold indeed. The coefficients $a_1, c_1, b_2, b_3$ and $h$ are complicated differential operators whose explicit form can be provided upon request. 
There are no further simple relations we found between them except for $c_1-a_1=8\,i\,S\,\Delta\, \sin{\varphi}$,
agreeing with \eq{eqn:sublead}. This implies that under spin flipping $S \rightarrow -S$,  $\mathbb{T}^{\bf4}(u)$ interchanges with $\mathbb{T}^{\bf\bar{4}}(u)$ up to the trivial explicit prefactor.    
We see that in total we have $6$ independent commuting operators $a_1,b_2,b_3,h,S,\Delta$,
which equates the number of degrees of freedom in $J=1$ case.

The limit of straight line $\varphi\to\pi$ is especially interesting as 1D Conformal symmetry gets restored. The space naturally decomposes into a $1D$ line and the $3D$ space orthogonal to it.
The corresponding symmetry is thus $SO(3)\times SO(2,1)$, and its representations are parametrised by the spin $S$ of $SO(3)$ and the conformal weight $\Delta$ of $SO(2,1)$.
Fixing $\Delta$ and $S$ removes two variables in our CFT wavefunction out of $6$.
Furthermore, we can restrict ourselves to Highest Weight states w.r.t. to both subgroups, which imposes on the wave function $2$ more conditions $K\psi=0$ and $S^+\psi=0$, which can be used to further reduce
the number of variables from $4$  to $2$. In this reduced system $B^{-1}$ and $b_3$ remain two non-trivial
differential operators, whereas all others can be expressed explicitly in terms of $\Delta$ and $S$. In particular, $a_1$ becomes $2(P_1 K_1+\Delta^2-\Delta+1)$, where $K_\mu$ is the special conformal transformation generator and $P_\mu$
is the generator of translations, and thus simplifies considerably for the primary operators in the 1D defect CFT, for which by definition $K_1=0$.

In the simplified case $S=0$ we get
the following relations
\beq
\begin{split}
   & P^{\bf 4}=-4 v^4+\left(2\Delta ^2-2\Delta +2\right)v^2+\frac{1}{4} \left(2 \Delta ^2-2 \Delta+3\right) -8\hat g^4\,,
    \\&
P^{{\bf 6}}=+6 v^8
-(4 (\Delta -1) \Delta +6)v^6
+v^4 \left((\Delta -1)^2 \Delta ^2+16 \hat g^4\right)
+b_3 v^2+16 \hat g^8\,,\\&
    P^{\bar{\bf 4}}=-4 v^4+\left(2\Delta ^2-2\Delta +2\right)v^2+\frac{1}{4} \left(2 \Delta ^2-2 \Delta+3\right) -8 \hat g^4\,,
\end{split}
\eeq
so there are only two non-trivial functions $\Delta(g)$ and $b_3(g)$, which  can only be deduced numerically.

\section{Generalisation: addition of impurities}
To generalise our setup, it is possible to introduce impurities in the spin chain. This is done by introducing a dependence on some parameters $\{\theta_i\},\;i=1\dots J$ in the rapidities of the bulk particles. To preserve parity in the argument $u$ of the $T$-operators, the correct choice (up to a normalisation of the $\theta_i$) amounts to:
\begin{align}
    &\mathbb{\hat{T}}^{\lambda}_{\theta}(u)=\text{Tr}\left(\hat{\bar{\mathbb{L}}}_J^{\lambda}(-u-\theta_J)\dots\hat{\bar{\mathbb{L}}}_1^{\lambda}(-u-\theta_1).\mathbb{\hat{K}}^{\lambda}(u).\mathbb{\hat{L}}^{\lambda}_1(u-\theta_1)\dots\mathbb{\hat{L}}^{\lambda}_J(u-\theta_J).G^{\lambda}.\hat{\bar{\mathbb{K}}}^{\lambda}(u).G^{\lambda\,t} \right)\;,
\end{align}
where $\lambda ={\bf \{4,\bar{4},6,\bar{1}\}}$. These transfer matrices form a family of mutually commuting operators: this was verified explicitly up to the case $J=1$. However, they do not commute with the original Hamiltonian $H$: this is expected, as introducing impurities changes the physical system and thus the Hamiltonian as well.

The next step is to introduce the polynomials $P^{\lambda}_k$ and to write the Baxter equation. In this case, the polynomials will acquire a $\{\theta_i\}$ dependence. Moreover, the prefactors appearing in equations \eqref{baxterTP} will also be modified. We obtain that: 
\beq
\begin{split}
&{\mathbb T}^{\bf 1}_{\theta}(v,\{\zeta_i\})=1\;,\\&
{\mathbb T}^{\bf 4}_{\theta}(v,\{\zeta_i\})\equiv\frac{P^{\bf 4}_{J+1}(v^2,\{\zeta_i\}
)}{\xi^{2J+2}}\;,\\&
{\mathbb T}^{\bf 6}_{\theta}(v,\{\zeta_i\}
)\equiv 
A(2v)\frac{v^2+1}{v^2}
\frac{P^{\bf 6}_{2J+2}(v^2,\{\zeta_i\}
)}{\xi^{4J+4}}\;,\\&
{\mathbb T}^{\bar{\bf 4}}_{\theta}(v,\{\zeta_i\})=
A(2v)A(2v+i)A(2v-i)
\frac{(v^2+\tfrac{9}{4})
(v^2+\tfrac{1}{4})\prod_{i=1}^J\left(\zeta_i^2+\left(v^2-\zeta_i^2+\frac{1}{4}\right)^{2}\right)
P^{\bf \bar 4}_{J+1}(v^2,\{\zeta_i\})}{\xi^{6J+6}}\;,\\&
{\mathbb T}^{\bar{\bf 1}}_{\theta}(v,\{\zeta_i\})=
A^2(2v)A(2v+i)A(2v-i)
A(2v+2i)A(2v-2i)\\&\qquad\qquad\qquad
\frac{
(v^2+4)(v^2+1)^{2}v^{2}\prod_{i=1}^J\left(\left(v^2-\zeta_i ^2\right)^{2} \left(4 \zeta_i ^2+\left(v^2-\zeta_i ^2+1\right)^2\right)\right)}{\xi^{8J+8}}\;,
\end{split}
\eeq
where $v\equiv u\,\xi$ and $\zeta_i\equiv\theta_i\, \xi$. We can now rewrite the Baxter equation \eqref{Baxter}. Defining $\zeta_0\equiv 0$, $\zeta_{-i}\equiv -\zeta_i$ and identifying:
\begin{equation}
    Q(v)\to \frac{\Gamma (-i v) \exp (\pi  (J+1) v)\, q(v)\, \xi ^{2 i v(J+1)} }{\Gamma \left(-i v-\frac{1}{2}\right) \Gamma (i v+2)}\prod_{i=-J}^J\left(\Gamma (i (v+ \zeta_i) +1)^{-1}\right)\;,
\end{equation}
we obtain:
\beqa \nonumber
\dfrac{P^{\bf 6}_{2J+2}(v^2)}{v^2\prod_{i=-J}^J(v-\zeta_i)}q(v)&=&-\prod_{i=-J}^J(v+i-\zeta_i)\, q(v+2i)
-\frac{v+\tfrac i2}{
v(v+i)
}P^{\bf 4}_{J+1}\((v+\tfrac i2)^2\)q(v+i)
\\ \nonumber
&&-\prod_{i=-J}^J(v-i-\zeta_i)\, q(v-2i)
-\frac{v-\tfrac i2}{
v(v-i)
}
P^{\bf \bar 4}_{J+1}\((v-\tfrac i2)^2\)q(v-i)\;.
\eeqa
A similar construction with inhomogeneities in the closed fishchian is introduced in~\cite{fishnetSoV}.

\bibliographystyle{JHEP.bst}
\bibliography{references}

\end{document}